\documentclass[lettersize,journal]{IEEEtran}

\usepackage{amsmath,amsfonts,amssymb}
\usepackage{algorithmic}
\usepackage{algorithm}
\usepackage{array}
\usepackage[caption=false,font=normalsize,labelfont=sf,textfont=sf]{subfig}
\usepackage{textcomp}
\usepackage{stfloats}
\usepackage{url}
\usepackage{verbatim}
\usepackage{graphicx}
\usepackage{cite}
\usepackage{multirow}
\usepackage{academicons}
\usepackage{mathtools}
\usepackage[thinc]{esdiff}
\usepackage{color,soul}
\usepackage{titlesec}
\usepackage[table]{xcolor}

\begin{document}

\title{An Overview of Electromagnetic Illusions: Empowering Smart Environments with Reconfigurable Metasurfaces}

\author{Hamidreza Taghvaee,~\IEEEmembership{Member,~IEEE,}
Mohsen Khalily,~\IEEEmembership{Senior Member,~IEEE} 
Gabriele Gradoni,~\IEEEmembership{Senior Member,~IEEE} and Rahim Tafazolli,~\IEEEmembership{Senior Member,~IEEE}

\thanks{H. Taghvaee, M. Khalily, G. Gradoni, and R. Tafazolli are with the Institute for Communication Systems (ICS), Home of the 5G and 6G Innovation Centre, University of Surrey, Guildford GU2 7XH, U.K.}
\thanks{Manuscript received April 00, 0000; revised August 00, 0000.}}

\markboth{Proceedings of the IEEE
,~Vol.~00, No.~0, August~0000}%
{Shell \MakeLowercase{\textit{et al.}}: An Overview of Electromagnetic Illusions: Empowering Smart Environments with Reconfigurable Metasurfaces}


\maketitle

\begin{abstract}
Metasurfaces, the two-dimensional counterparts of metamaterials, represent cutting-edge technology in electromagnetic (EM) engineering. 
Comprising thin planar subwavelength structures that boast arbitrary EM responses when arranged in specific sequences. 
The result is the ability to craft engineered impedance, providing exceptional capabilities to manipulate EM waves. 
These metasurfaces emerge as ideal candidates for realizing ultra-compact and highly efficient EM devices, offering pivotal advantages as wireless communications and defence technologies continue to advance. In this evolving landscape, the demand for innovative solutions to address challenges such as unwanted scattering and enhancing stealth capabilities becomes increasingly urgent, positioning metasurfaces as a key technological solution in meeting these critical needs.
This study delves into the innovative landscape of metasurfaces, with a particular focus on their role in achieving EM illusion (EMI) — a facet of paramount significance. 
The control of EM waves assumes a pivotal role in mitigating issues such as signal degradation, interference, and reduced communication range. 
Furthermore, the engineering of waves serves as a foundational element in achieving invisibility or minimized detectability. 
This survey unravels the theoretical underpinnings and practical designs of EMI coatings, which have been harnessed to develop functional metasurfaces. 
EMI, practically achieved through engineered coatings, confers a strategic advantage by either reducing the radar cross-section of objects or creating misleading footprints. 
In addition to illustrating the outstanding achievements in reconfigurable cloaking, this study culminates in the proposal of a novel approach, suggesting the emergence of EMI without the need for physically coating the device to conceal and thus proposing the concept of a smart EMI \emph{environment}. 
This groundbreaking work opens a new way for engineers and researchers to unlock exotic and versatile designs that build on reconfigurable intelligent surfaces (RIS). 
Crucially the designs enabled by the proposed approach, present a wide array of applications, encompassing camouflaging, deceptive sensing, radar cognition control, and defence security, among others.
In essence, this research stands as a beacon guiding the exploration of uncharted territories in wave control through smart EMI environments, with profound implications spanning basic academic research in RIS through advanced security technologies and communication systems.

\end{abstract}


\section{Introduction}

\begin{figure*}[!h]
  \centering
  \includegraphics[width=140mm]{ 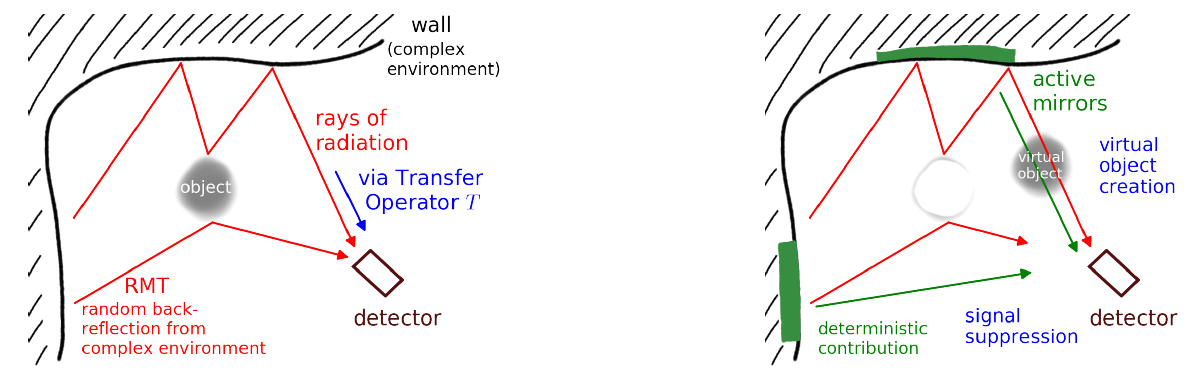}
  \caption{Demonstrating EMI within smart environments featuring tailored surfaces. The left panel presents a situation in which a sensing device captures reflected signals and utilizes random matrix theory to interpret the surrounding environment. Conversely, the right panel shows a smart environment outfitted with dynamic mirrors that modify the scattering profile, effectively crafting an illusion to deceive the detecting device.}
  \label{fig:cart}
\end{figure*}

The ability to detect the presence and shape of arbitrary objects is
of paramount importance in imaging technologies as well as in localization for the future generation of wireless networks. 
This capability currently stands on sophisticated digital signal processing algorithms that process and somehow invert the signal back-scattered from a target. 
Of particular importance, are emerging integrated sensing and communication (ISAC) technologies, whose objective is to localize both network terminal and passive objects within the surrounding environment. 
While most of the research effort has to date focused on improving the accuracy and efficiency of these algorithms, a key question that remains is whether we are (or will be) able to hide an object from imaging/localization technologies.
Inherently, there is a strong need to protect electronic devices from external attacks in the context of EM compatibility, and data sniffing in the context of wireless communications. 
The answer to this need has been provided by extensive research in optics and electromagnetics. 
The main idea stands on the fact that physical cloaking of objects via engineered `mantles' constitutes the way forward to hide objects. 
However, this technology never achieves perfect concealing due to basic physics limits \cite{monticone2016invisibility}. 
Furthermore, the calculation of the re-radiated fields of the cloak 
assumes free-space conditions, while the influence of the back-radiation from the environment (walls, surrounding objects, etc) onto the surface current densities needs to be factored in in design procedures. 
More importantly, real-life protection technologies should be capable of manipulating the complex EM environment to form a prescribed response at selected receiving areas. 
After reviewing the current literature on cloaking and illusion across multiple disciplines, we discuss a paradigm-shift idea that stands on smart EM environments enabled by reconfigurable metasurfaces.
In other words, a system of engineered mirrors creates the illusion of object displacement by shaping the EM
wavefront as depicted in Fig. \ref{fig:cart}.

\emph{Why is cloaking not enough?} 
Dreaming to achieve optical invisibility, a series of studies across the physics and engineering communities have devised an artificial cloak, known as meta-material, which can be wrapped around the object to suppress the wave scattering and restore the incident wavefront \cite{schurig2006metamaterial}.
An observer detecting the wave processed by the perfect cloak would not notice a difference between the presence and absence of the object. 
More recently, researchers have shown that only imperfect cloaks can be realized without violating both causality \cite{miller2006perfect} and conservation of energy \cite{schurig2006metamaterial}.
While imperfect cloaking is still useful, as it still leads to a reduced scattering cross section, meta-material design has focused on processing monochromatic waves and pulses propagating in free space. 
The imperfection in cloaking, stemming from the absence of causality in the time domain response to a pulse excitation, can be improved with active cloaks, achieving near-perfection. However, it's essential to note that current demonstrations of this capability have only been realized in free space.
However, to effectively hide from radar imaging and advanced localization technologies, it's crucial to achieve cloaking of objects within confined propagation environments such as rooms and streets.
This scenario is radically different from free space with regards to: 
i) Wave physics, where the EM response of the meta-material is changed by waves back-reflected from the environment to the cloak, e.g., caused by proximity to a wall;
ii) Mathematical modelling, for which the field incident on the cloak is not a simple monochromatic plane wave but needs to be found as a solution to the boundary-value problem. 
The two aspects are intertwined, necessitating the resolution of the boundary-value problem in a self-consistent manner that accounts for the environment, object, and cloak altogether.
The solutions present in the literature are still not enough to understand how to hide an object within a confined environment. 
Inherently, a real environment cannot be considered static: people moving, for example, within an office space change the interaction of EM waves with the environment. 
This would result in some variability of the wave scattering that can hardly be modeled deterministically as it would not provide unique boundary conditions to the mathematical formulation
of the wave problem.
Novel wave modeling grounded on statistical electromagnetics is necessary to understand wave control in complex environments. 

\emph{How does object illusion work?} In the context of wave-based imaging and localization, a probing EM field that enters the environment would be transmitted and reflected after progressive scattering through anomalous mirrors as well as passive objects.
The wave would then develop a complex space-frequency behaviour that can be manipulated by wall-mounted reflectors, whose wavefront shaping acting on ensembles of waves redistributes the energy in a prescribed
way. 
From a mathematical point of view, the problem of object illusion in chaotic environments poses extraordinary challenges. 
Being a cascaded scattering process, it configures as an interior \emph{stochastic} boundary-value problem that naturally involves random matrix theory (RMT) within the interaction between the environment and reflectors. 
The scientific literature has seen a series of rigorous studies to understand the physical limits of free space EM wave illusion.
In particular, waves reflected/transmitted through engineered surface meta-films were shown to achieve holograms of a virtual prescribed object \cite{liu2017source,smy2020surface}. 
Again, no back-reflection from the environment has been accounted for \cite{jiang2011radar}.
Developing on modern metasurface research, the task of object illusion in complex environments finds a way forward through reconfigurable metasurfaces, which are thus capable of manipulating wavefronts far from the object (and are typically referred to as the third generation of metasurfaces \cite{barbuto2021metasurfaces}).

\emph{Route to wave modeling.}
Modeling the progressive scattering between metasurfaces and environment as driven by multiple reflections through walls and objects is grounded on boundary integral equations. 
Methods based on RMT have been developed through the years to capture the statistics of wave scattering within complex environments with stochastic variability \cite{stockmann2002effective,kuhl2013microwave}. 
However, the wave modelling of environments with engineered boundaries leads to more challenging boundary-value problems. 
While methods based on the transfer operator \cite{creagh2016propagating} have been proposed to solve boundary-value in the high-frequency asymptotics regime, solving the interior scattering problem of confined domains with nonuniform boundaries remains a challenging task. 
Some notable exceptions can be found that go beyond solving the exterior problem of curved in-homogeneous boundaries \cite{sandeep2023application}, and propose wave dispersion engineering via quantum graphs and spectral formalisms \cite{lawrie2022quantum}.
More specifically, an interior problem of large rooms assisted by metasurfaces has been solved \cite{karamanos2023topology}.
Statistical methodologies have also been proposed that help solve boundary-value problems involving large and irregular environments with non-homogenous boundary conditions from a circuital perspective \cite{gradoni2014predicting,gradoni2015statistical}. 
This physics-based view has been recently embraced to evaluate the performances of wireless links operating within extremely confined environments \cite{singh2023shannon}.
Furthermore, despite the scientific community still lacks a self-consistent model of metasurface-assisted cavity environments, the technological application of smart illusions is getting ahead of the curve and already spans from acoustic, e.g., isospectral resonant systems \cite{lenz2023transformation}, through wireless communications, e.g., creation of dark regions within indoor environments \cite{encinas2023rishield} and reconfigurable boundary modulation \cite{dong2023wireless}.

\subsection{Illusion and camouflage in Nature}
In the intricate tapestry of nature, every object exhibits a distinctive pattern of scattering EM waves. Sensing systems, whether embodied in human senses or advanced sensing devices, rely on the discernment and identification of objects based on these unique scattering signatures. The intentional or inadvertent distortion of collected data by any sensor introduces an illusion—a concept foundational to the perceptual understanding of the surrounding world. Biological sensors showcase various natural illusions. Take, for example, the phenomenon of Fata Morgana (a phenomenon where a layer of warm air rests above a layer of cooler air), a captivating natural illusion resulting from thermal inversion that gives rise to a superior mirage visible above the horizon \cite{10.3389/fnhum.2018.00120}. In the realm of optical illusions, the vibrant colouration of a Peacock's feather \cite{1650059} is achieved through a combination of microscopic structures called photonic crystals and pigments. The photonic crystals are arranged in layers, creating a structural colouration effect that results in iridescent blue, green, and gold. Additionally, melanin pigments in the central part of the feather absorb excess light and enhance the contrast of the structural colours. This interplay of structural and pigmentary elements gives the peacock feather its distinctive and stunning visual display. Similarly, the intricate patterns on a Beetle's wing \cite{1412810111} scale and diffract light, producing structural colouration. Additionally, pigments in the wing absorb and reflect specific wavelengths of light, contributing to the overall colour and pattern. The interplay of these structural and pigmentary elements results in the visually striking and intricate patterns observed on a beetle's wing. The spectrum of illusions extends beyond the visual domain to acoustics, where artificially induced false perceptions of real sounds or external stimuli constitute acoustic illusions \cite{1981302}. In the commercial realm, the application of cancelling earphones stands as a pragmatic use case of acoustic illusion. Within the domain of human perception, illusions manifest as both artificial effects and artefacts stemming from intricate brain functions, as exemplified by the intriguing McGurk Effect \cite{MCGURK1976} (a perceptual phenomenon where individuals perceive a third, distinct sound when confronted with conflicting audio and visual speech cues). This effect underscores the intricate relationship between auditory and visual information in speech perception. This exploration into the multifaceted nature of illusions sets the stage for a deeper examination of EM illusion (EMI); artificial manipulations of EM waves that hold promise for diverse applications. As one delves into this realm, it unravels the intricate interplay between perception, manipulation, and the rich landscape of EM phenomena.

\begin{figure}[!h]
  \centering
  \includegraphics[trim={3cm 4cm 4cm 5cm},clip,width=80mm]{ 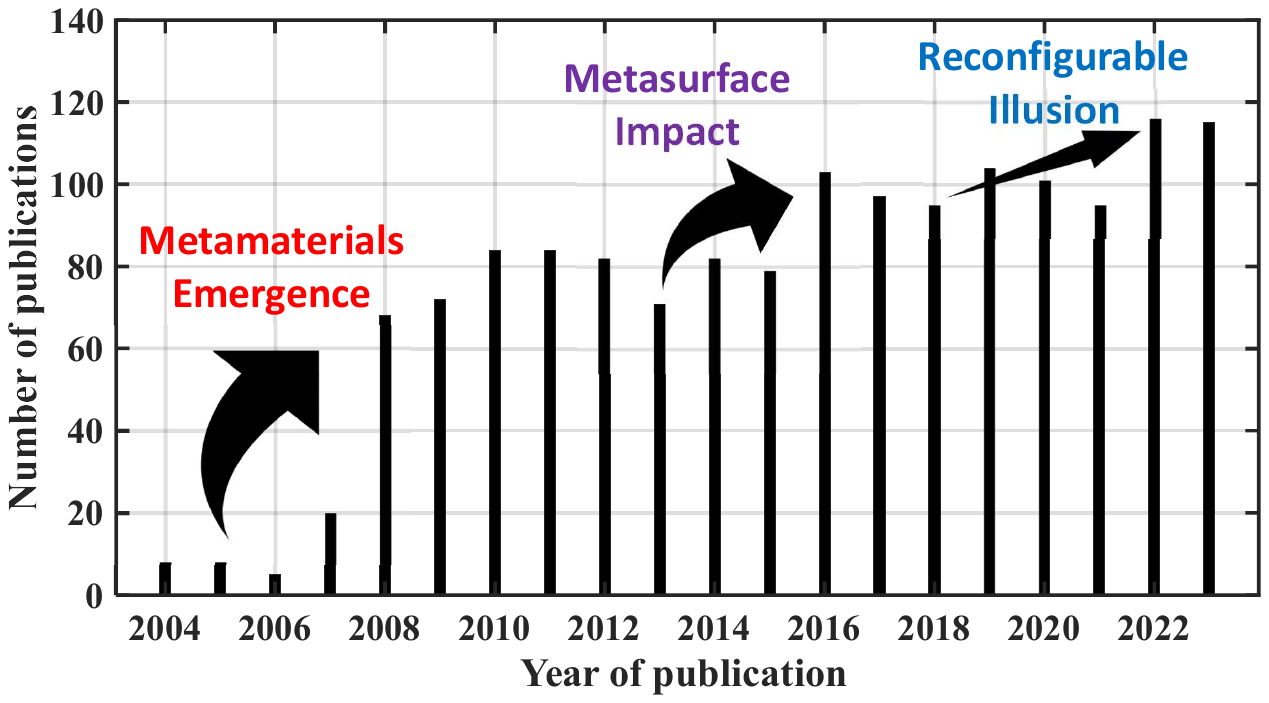}
  \caption{This chart illustrates the growth in scholarly output related to EMI from 2003 to 2023. Every bar in the diagram quantifies the annual publication count, highlighting the escalating engagement and investigative efforts in this area throughout the years. Upward arrows illustrate the progressive rise in the volume of publications, underscoring the expanding interest within the scientific community. (Information derived from Scopus).}
  \label{fig:pubs}
\end{figure}

In this paper, the focus is on artificially generated illusions stemming from external or environmental conditioning, rather than those arising from human perceptual effects. A specific manifestation of this concept is the creation of EMI, strategically designed to confound sensors or receiving antennas operating in the microwave, millimetre-wave, and optical regimes. The integration of object illusions has found widespread applications in various real-world scenarios, including security, communications, and warfare. Figure \ref{fig:pubs} illustrates the number of publications related to EMI and cloaking, emphasizing the growing interest and relevance in these fields. The conceptual roots trace back to the advent of double-negative (DNG) materials, initially sparking the idea of transparent materials. Subsequently, scientists pushed the boundaries, ushering in a new era of engineered composites, namely Metamaterials, and their 2D successors, Metasurfaces. Figure \ref{fig:pubs} underscores the instrumental role these engineered materials have played in advancing the field. While the introduction of these materials predates the depicted timeline (DNG in 1968 \cite{Veselago_1968}, Metamaterials in 2001 \cite{Smith2001}, and Metasurfaces in 2013 \cite{1232009}), their lasting impact has rippled through subsequent research endeavours. Therefore, the arrows in Fig. \ref{fig:pubs} signify the impact of each technology on the surge of the publication. This paper delves into the evolution and applications of EMI within this broader context, offering insights into their development and strategic implementation in various technological domains.

The remainder of this section serves as a foundational exploration for theoretical studies on EM manipulation and light bending. Following this groundwork, a concise historical overview of EMI is presented, tracing its evolution alongside practical designs facilitated by the advent of metamaterials. The final subsection endeavours to motivate the unique contribution of this paper, elucidating the specific areas of focus and innovation. A clear outline of the subsequent sections is provided to guide the reader through the organization of the paper, enhancing comprehension and navigation.

\subsection{Early works on the theory of electromagnetic wave control}
According to Fermat's principle, EM waves inherently choose the path that minimizes the travel time between two points. In essence, it posits that light, or any EM radiation, follows the path that requires the least amount of time to traverse, considering all possible routes. Exploiting Fermat's principle offers a strategic avenue for controlling and manipulating the propagation of EM waves. By intentionally engineering inhomogeneous materials, one can influence EM waves within these materials. This intentional manipulation of the medium allows for precise control over the trajectory of the waves. Inhomogeneous materials, characterized by variations in their composition or structure, introduce spatial gradients that can effectively guide the EM waves along desired paths. In practical terms, this means that by carefully tailoring the properties of the material through which the waves propagate, one can exert control over the direction, speed, and ultimately, the behaviour of EM waves. So, artificial structures can be designed to have properties difficult or impossible to find in nature like simultaneous negative permittivity ($\epsilon$) and permeability ($\mu$), discussed by V. Veselago in 1968 \cite{Veselago_1968}. However, pursuing this goal was not favourable for a long time during which the only record of EM manipulation was the use of absorbing screens and antireflection coatings to diminish the scattering or the reflection from objects that were common for decades \cite{8632,WARD1972369}. 

In 1996, J. Pendry continued Veselago's work and applied DNGs to engineer artificial materials \cite{PhysRevLett.76.4773,798002}. In 2000, J. Pendry proposed to use a perfect lens based on DNG \cite{PhysRevLett.85.3966}. In 2003, Alù and Engheta suggested pairing an epsilon-negative slab with a $\mu$-negative slab that was later called DNG materials to implement a transparent component \cite{1236073}. In 2005, by adopting a standard Mie expansion, for the first time, lossless and passive elements were proposed to reduce the scattering from spherical and cylindrical objects without requiring high dissipation \cite{PhysRevE.72.016623}. These artificial components owe their unusual properties to subwavelength inclusions of composition rather than to their chemical structure which is why they are called metamaterials. In 2006, Pendry et al \cite {10.1126/science.1125907} used analytical calculations to show metamaterials can drag EM fields into almost any desired and arbitrary configuration. Later, Leonhardt \cite{doi:10.1126/science.1126493} developed a general recipe to bend the light around an object as if nothing were there. In the same year, Schurig implemented the first practical realization of cloaking with split-ring resonators (SRRs) \cite{10.1126/science.1133628}. Later more attention has been attracted to new fascinating ideas of cloaking. Among the notable advancements in cloaking, \cite{PhysRevLett.101.203901} introduces the intriguing idea of employing a dielectric background rather than a free-scattering medium. 
Furthermore, \cite{10.1126/science.1166949,Valentine2009} put forth a broadband design featuring extremely low-loss elements. In a distinct approach, \cite{PhysRevLett.102.233901} employed a plasmonic coating to cloak a sensor without perturbing the impinging field, while \cite{PhysRevB.80.245115} introduced an angle-independent cloaking technique.

\subsection{Toward realization of electromagnetic illusions}

Embarking on the journey toward the realization of EMI, it delves into technical approaches that have reshaped the landscape of wave manipulation. In 2006, the distortions of coordinate transformation between the original Cartesian mesh and the distorted mesh were adopted for EM wave manipulation \cite{10.1126/science.1125907}. This method was further developed for comprehensive EMI applications within the framework of transformation optics (TO), allowing arbitrary objects to appear as other objects \cite{PhysRevLett.102.253902}. In 2010, metamaterials emerged as a key element for generating virtual objects \cite{10.1063/1.3371716, XiangJiang:10}. Later, in \cite{PhysRevLett.105.233906}, the authors experimentally demonstrated the first metamaterial illusion device using a transmission-line medium. In 2011, the utilization of a perfect lens introduced a paradigm shift, creating the illusion of multiple objects appearing as a singular entity in the far field—a groundbreaking concept in illusion optics \cite{Xu_2011}. Subsequent works swiftly emerged, pioneering the design of new illusion applications, including radar and ghost illusions \cite{PhysRevE.83.026601,10.1002/adfm.201203806}.

\begin{figure*}[!h]
  \centering
  \includegraphics[trim={1.5cm 4cm 2cm 3cm},clip,width=180mm]{ 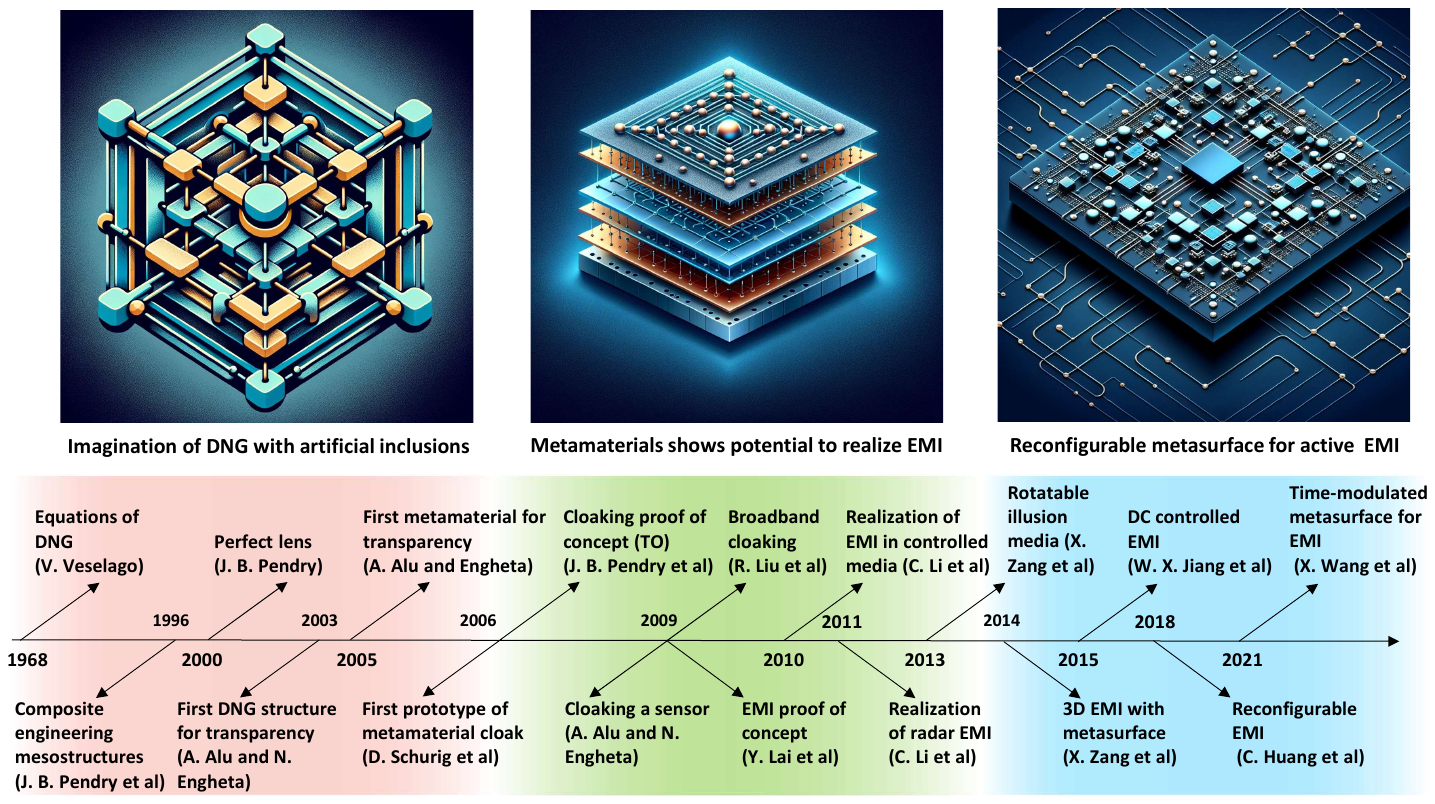}
  \caption{This figure outlines the progression of Electromagnetic Illusion research, starting from the seminal exploration of Double-Negative (DNG) material equations by Veselago. It tracks the burgeoning interest sparked by the advent of metamaterials, paving the way for the development of metasurfaces as a more compact and practical iteration. The timeline culminates with the advent of reconfigurable metasurfaces, heralding a new era of versatile applications in the field, showcasing the dynamic evolution and expanding possibilities within electromagnetic research.}
  \label{fig:map}
\end{figure*}

A proposal for a rotatable illusion medium, capable of achieving singular parameter-independent illusions, was introduced \cite{Zang:13}. This innovative medium can manipulate terahertz (THz) EM waves. Consequently, additional mediums, either rotating \cite{ZANG2018977} or shifting \cite{Yang:18, Zang:11}, were subsequently designed to explore further possibilities in wave manipulation. In 2014, the theory, and experimental verification of quasi-three-dimensional and angle-tolerant EMI were presented \cite{10.1002/adfm.201401561,6929143}. This was realized by metasurfaces as the successor version of metamaterials and more similar research was followed \cite{7995050}. In 2015, the concept of multifunctional transformation-dc devices was proposed and experimentally verified \cite{https://doi.org/10.1002/adma.201500729}. The research demonstrated that the illusion function of metamaterials can be optically manipulated, thereby opening a new avenue for tunable illusions. Building on the principles of general transformation electromagnetics and utilizing quasi-conformal mapping, various illusion devices and microwave lens antennas have been presented \cite{C2NR31140B, Yang_2015, doi:10.1163/156939310793675664, 6926742}. Moreover, the transformation of source radiation into desired patterns has been achieved \cite{10.1063/1.4913596, Chen_2013, PhysRevLett.119.034301}. In 2017, an experimentally realized reconfigurable carpet cloak utilizing tunable metasurface technology was demonstrated at the microwave frequency \cite{doi:10.1021/acsphotonics.7b01114}. Subsequently, in 2018, a graphene-based metasurface was developed by incorporating graphene ribbons onto a dielectric cavity resonator. The study revealed that by applying electric bias, the wavefront of the field reflected from a triangular bump covered by the metasurface could be tuned to resemble that of a bare plane or a spherical object \cite{PhysRevApplied.9.034021}. Figure \ref{fig:map} illustrates the evolution of EMI research, beginning with the foundational works on DNG by Veselago, which laid the groundwork for future advancements. It highlights the transition towards practical design through the introduction of metamaterials and further expands on the opportunities within EMI research brought about by the integration of reconfigurable architectures.



\subsection{Contribution and Organization}

A survey on EMI is crucial because it provides a comprehensive overview of the existing research landscape, helping researchers and engineers identify key trends, challenges, and opportunities in this evolving field \cite{15011403,5.0048846,2020013}. The importance of such a survey lies in its potential to:
\begin{itemize}

\item \textbf{Knowledge Consolidation:} EMI encompasses a diverse range of techniques and applications. A survey consolidates this scattered knowledge, offering a centralized resource for researchers to gain insights into the current state of the field.

\item \textbf{Identifying Trends and Gaps:} By systematically reviewing existing literature, a survey enables the identification of trends and emerging patterns within EMI research. It sheds light on areas that have received substantial attention and those that might be underexplored.

\item \textbf{Benchmarking and Evaluation:} Researchers can use a survey to benchmark their work against existing studies and evaluate the effectiveness of different techniques. This aids in understanding the strengths and limitations of current approaches.

\item \textbf{Guiding Future Research:} Understanding the missing link in EMI research is pivotal for guiding future investigations. A survey can highlight gaps in knowledge, technological limitations, or unexplored application areas, providing a roadmap for future research endeavours.

\end{itemize}

The gap in EMI research may encompass several aspects, including hurdles in practical implementation, the need for greater scalability of illusion devices, challenges within certain frequency ranges, and the exploration of new applications. Identifying this gap is pivotal for the advancement of the field, guiding researchers to focus their efforts where they can make the most profound impact and further expand the capabilities of EMI technology. In the sections that follow, this paper presents a curated selection of seminal papers, examining their significant contributions in detail. It also provides a comparative analysis of various approaches, highlighting their advantages and disadvantages, with a particular emphasis on metasurface-based devices.

The prevailing research trend on this topic often focuses on the application of object coatings to facilitate illusions. However, this strategy may not always be suitable for creating illusion devices, especially when it involves active devices whose functionality could be compromised by such coatings. For instance, a communication device requiring signal transmission and reception might experience signal absorption by the coating material. Additionally, direct access to the object or device for coating application is not always possible. Previous studies typically assume environments devoid of obstacles and ignore the effects of back-reflections from surroundings, which is not practical. To effectively cloak an object from detection by imaging and localization technologies within confined spaces like rooms or streets, a shift from reliance on scattering-free environments is necessary. This leads to the recommendation of a new strategy: implementing smart object illusion within a multi-scattering scenario. This approach suggests utilizing the environment itself, equipped with reconfigurable elements that adjust based on the specific scenario, instead of applying coatings directly to the object. While this direction poses its own set of challenges and has seen limited focus in existing research, particularly in tackling the wave boundary-value problems it introduces, it underscores the need for developing a comprehensive mathematical framework to accurately model the interactions between the object, its reflections, and the environment. Addressing this gap represents a crucial step forward in the field.

\begin{figure*}[!t]
  \centering
\subfloat[Without protection]{\includegraphics[height=5cm]{ 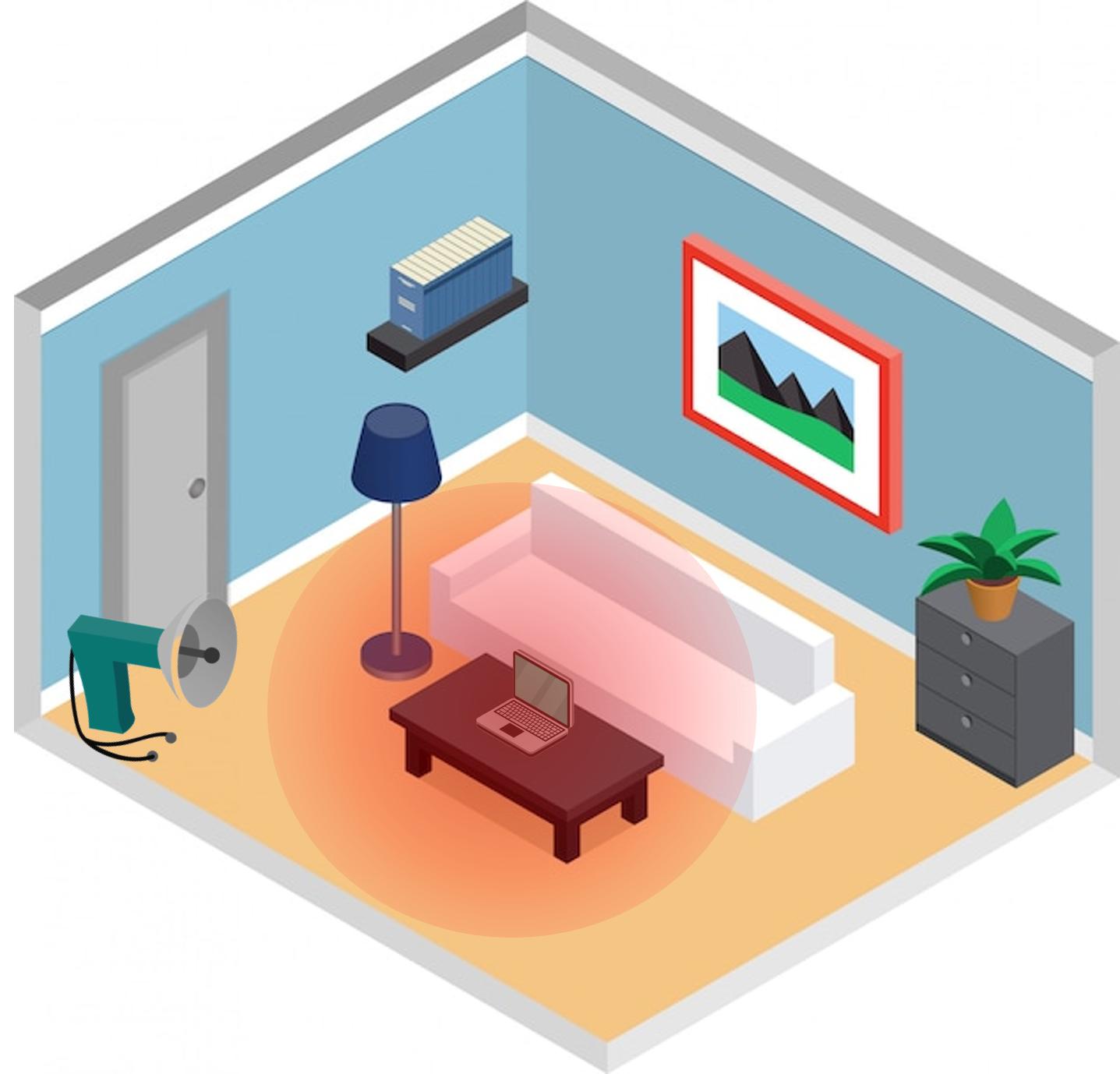}}
\subfloat[Coating solution]{\includegraphics[height=5cm]{ 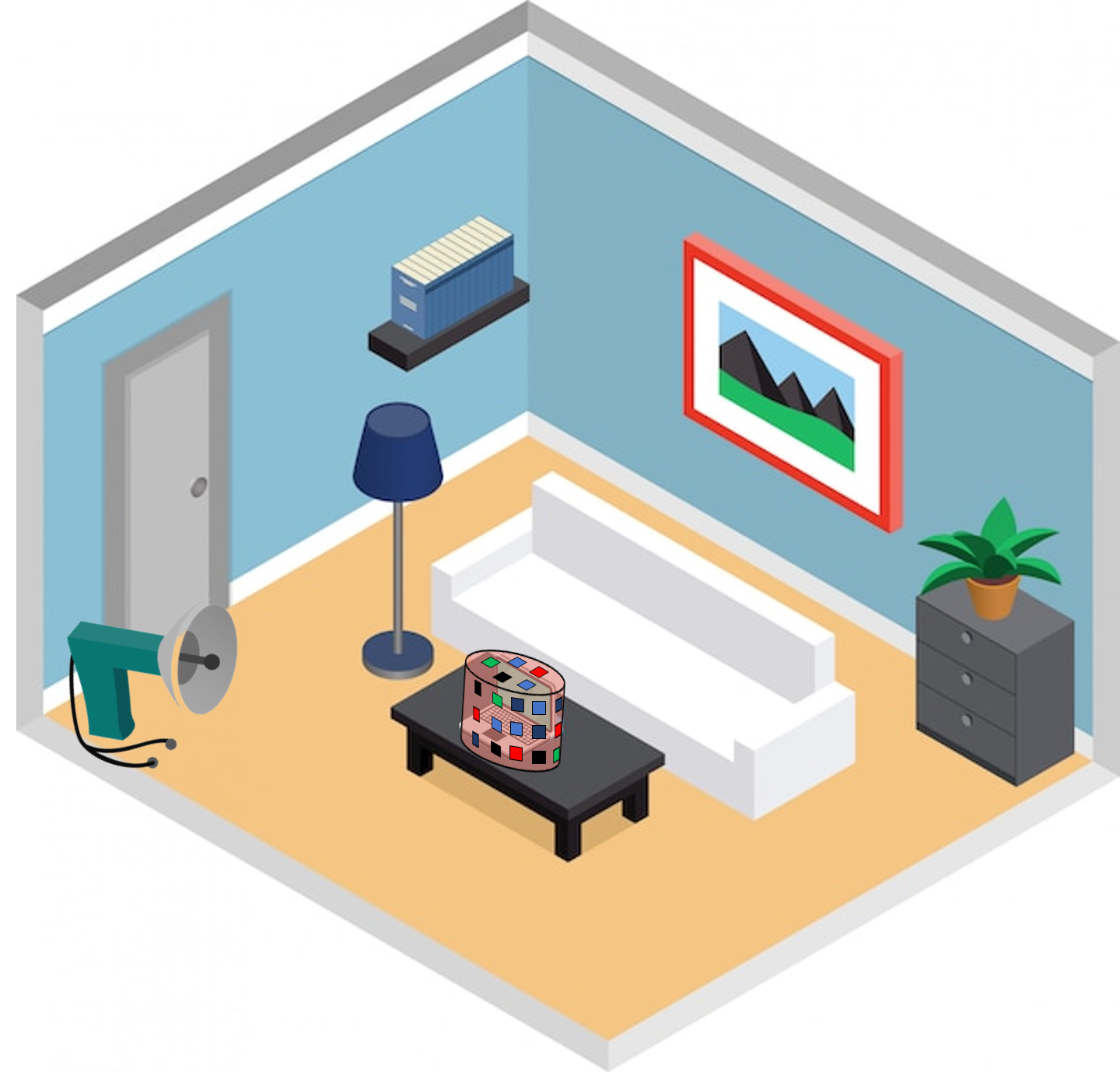}}
\subfloat[Smart environment]{\includegraphics[height=5cm]{ 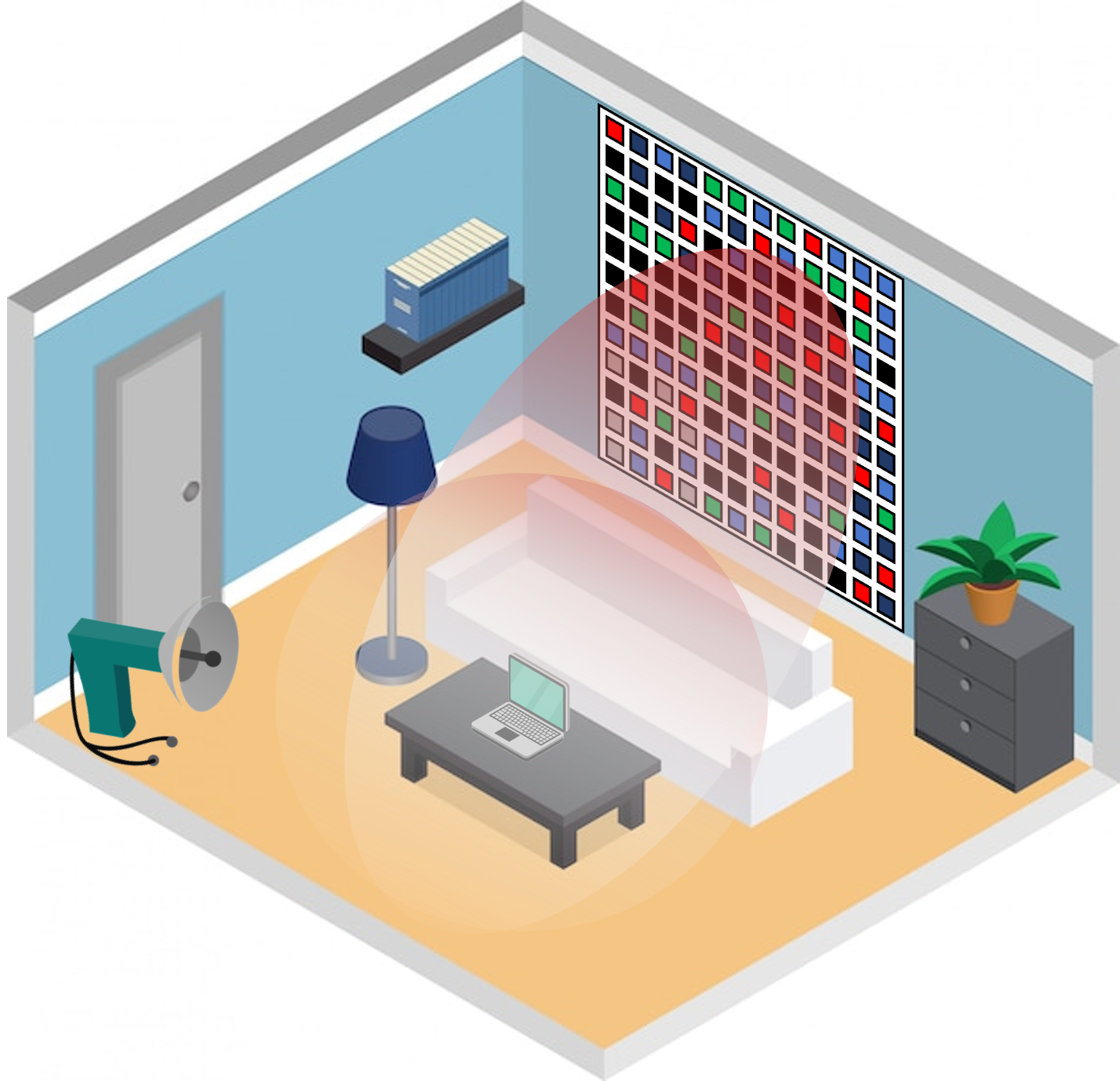}}
  \caption{Comparison of different object illusions in multi-reflective environment. In the cloaking case, the object is coated by conformal metasurfaces to engineer the backscattering but in the smart environment, the metasurface on the wall cancels out the scattering of the device.}
  \label{fig:comp}
\end{figure*}

As illustrated in Fig.~\ref{fig:comp}, the scattering environment (indoor) significantly diverges from the physics of waves and mathematical modelling associated with a scattering-free system. These intertwined aspects necessitate a consistent resolution of the boundary-value problem, accounting comprehensively for the interaction among the environment, object, and cloak. Our contribution aims to revolutionize existing solutions presented in the state of the art, focusing on achieving object cloaking within confined environments. From a mathematical perspective, the problem of smart illusion presents exceptional challenges. It involves a cascaded scattering process that can be effectively modelled using products and sums of transfer matrices of arbitrary dimensions. Our proposed approach paves the way for advanced and futuristic research in this domain, offering a conceptual breakthrough. As a proof of concept, it has considered a one-dimensional (1D) case of the problem, laying the groundwork for further exploration.

mathematical framework to accurately model the interactions between the object, its reflections, and the environment

The structure of this paper unfolds as follows: Section II traces the development of metamaterials and metasurfaces leading up to their application in EMI. Section III examines current EMI methods, providing a theoretical framework and assessing their strengths and weaknesses. Section IV proposes a smart environment approach and establishes a mathematical model to describe the interaction between objects and their surroundings. Section V discusses a reconfigurable system for achieving advanced EMI, including a demonstration of the concept through a 1D example. Section VI categorizes EMI use cases, detailing their technological prerequisites and exploring their potential influence across various aspects of daily life. Finally, Section VII concludes the paper with a summary of key points and findings.


\section{Metamaterials and metasurfaces}
With the emergence of metamaterials, the ability to tailor the wavefront arbitrarily has become achievable. This section delves into the background of metamaterials and their successors, metasurfaces. The prefix "meta" signifies "beyond," indicating that metamaterials exhibit characteristics that surpass natural standards. These materials are artificially crafted compositions that derive their properties from small inclusions rather than traditional chemical bonds. Although the concept was introduced some time ago \cite{Veselago_1968}, the realization of its potential applications has gained momentum in recent decades \cite{6230714,798001}. Notably, metamaterials have found applications in various fields, including invisibility cloaks, superlens, transceivers, imaging and, energy harvesting.

\begin{figure}[htbp]
  \centering
  \includegraphics[width=81mm]{ 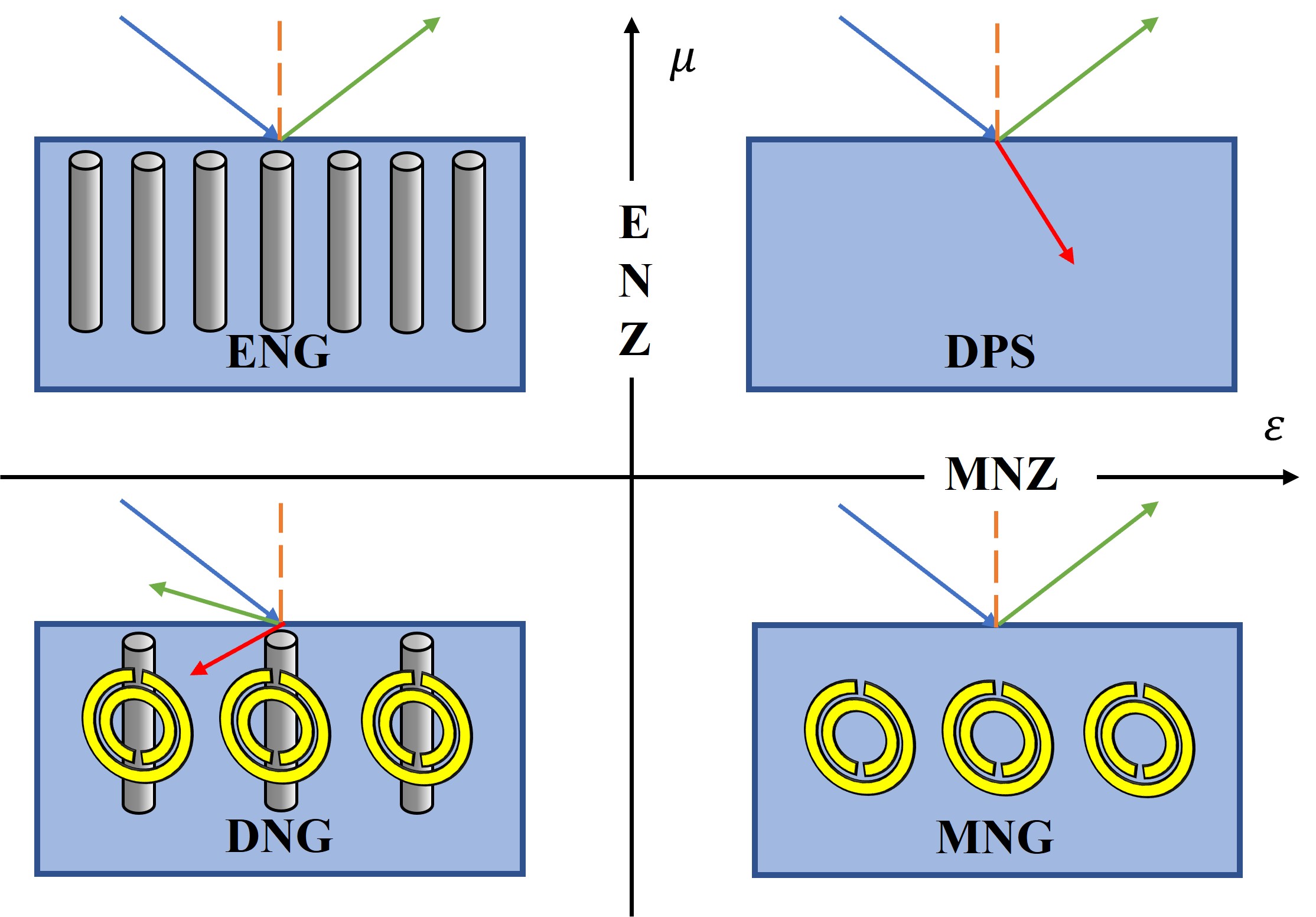}
  \caption{This diagram visually categorizes four unique types of metamaterials, differentiated by the signs of their permittivity ($\epsilon$) and permeability ($\mu$). These groups are identified as "DPS" (Double-positive), "ENG" (epsilon-negative), "MNG" (mu-negative), and "DNG" (Double-negative), each demonstrating exceptional electromagnetic traits that surpass traditional material capabilities, facilitating novel approaches to wave control. Moreover, the illustration emphasizes areas adjacent to the axes, designated as "ENZ" (Epsilon Near Zero) and "MNZ" (Mu Near Zero), indicating specific regions that possess distinctive EM features.}
  \label{fig:metacat}
\end{figure}

The fundamental structure of metamaterials comprises periodic units crafted from metallic or dielectric materials to attain specific properties and functionalities. These subwavelength structural units serve as the foundational elements of metamaterials and can be precisely customized in terms of shape and size. The lattice constant and interatomic interaction within these units can be manipulated to adjust the EM response. In essence, through the deliberate arrangement of these small unit cells into the desired architecture or geometry, one can finely tune the effective permittivity or permeability of the metamaterial to exhibit positive, near-zero, or negative values \cite{0471754323,6230714}. Figure \ref{fig:metacat} illustrates how materials are categorized based on their effective permittivity ($\epsilon$) and permeability ($\mu$) \cite{7880594}.

Natural materials, such as dielectrics, exhibit both positive permittivity ($\epsilon$) and positive permeability ($\mu$), classifying them as double-positive (DPS) materials. Epsilon-negative materials (ENG) manifest a negative permittivity ($\epsilon$) while maintaining positive permeability ($\mu$). For instance, noble metals like gold or silver function as ENG materials in the infrared and visible spectrum. The realization of ENG can involve using conducting rods, resulting in an imaginary refractive index and evanescent propagation within the medium. On the other hand, mu-negative (MNG) materials feature positive permittivity ($\epsilon$) but negative permeability ($\mu$). MNG materials can be created using SRRs. Meanwhile, double-negative (DNG) materials possess both negative permittivity ($\epsilon$) and negative permeability ($\mu$). DNG materials can be constructed using a combination of conducting rods and SRRs, leading to a negative refractive index ($n$). This unique property causes EM waves to travel in the backward direction within the medium, a phenomenon not observed in natural materials.

DNG metamaterials are sometimes referred to as "left-handed materials (LHMs)" because electric and magnetic fields are related through left-hand orientation to the wave vector. In DNG materials, both reflection and refraction deviate from normal behaviour, and Snell’s law does not apply. This characteristic makes DNG metamaterials distinct and introduces novel possibilities not achievable with natural materials. Precisely, a material can be considered a metamaterial if its effective relative permittivity is positive but less than 1 and/or its effective relative permeability is positive but less than 1. Thus, another property that cannot normally be found in nature but can be achieved with metamaterials is that of a near-zero refractive index. This could be Epsilon near-zero (ENZ) or Mu near-zero (MNZ) materials, as shown in Fig. \ref{fig:metacat}.

Metamaterials involve intricate structures such as complex metallic wires, posing challenges in fabrication and assembly. The extension of 3-D metamaterials can be achieved by arranging electrically small scatters or holes into a 2-D pattern at a surface or interface, giving rise to a metasurface. The term "metasurface" refers to this 2D counterpart of metamaterial. Metasurfaces offer a practical alternative to full 3-D metamaterial structures in various applications. They share the unique properties of their 3D counterparts while overcoming challenges associated with bulkiness, fabrication complexity, and cost. Unlike 3-D metamaterials, metasurfaces occupy less physical space, making them advantageous in scenarios where size constraints are critical.

Additionally, metasurfaces address other issues inherent in traditional metamaterials, such as bulkiness, dispersive properties, and the complexity of tensor analysis. The analytical challenges associated with metamaterials, where constitutive relations must be followed, are alleviated with metasurfaces. Instead, the analysis primarily revolves around boundary conditions, simplifying the understanding and utilization of these engineered surfaces in various EM applications.

Hence, metasurfaces have garnered more attention compared to 3D metamaterials and have progressed rapidly over the past decade. Throughout this period, numerous specialized periodic surface structures have been developed, including frequency-selective surfaces (FSS), high-impedance surfaces (HIS), and electromagnetic bandgap (EBG) structures. However, they are named based on their functional characteristics. Consequently, metasurfaces often intersect with various research domains involving conventional 2D periodic structures. For instance, FSSs can be realized using metasurfaces rather than metamaterials, depending on the specific design. Generally, conventional FSSs are not considered metamaterials, whereas HISs and EBG structures are typically classified as such.

While metasurfaces have been typically static and resonant-based structures, thus responding to very specific functionalities or conditions, recent works have proven that the behaviour of metasurfaces can be tuned during and after deployment with global or local reconfigurability \cite{Oliveri2015}. Moreover, in recent years, there has been a rise in programmable metasurfaces. These metasurfaces integrate local tunability and digital logic, enabling convenient reconfiguration of electromagnetic behaviour from external control. This can involve the integration of components such as varactors, diodes, or memristors within the unit cells \cite{10.1002.adom.202000783} or the use of materials enabling thermal \cite{Lewi2019}, electrical \cite{ju2011graphene}, optical \cite{Zhao2015, Fan2013}, magnetic tuning \cite{Chen2017, Yang2017} and waveform-dependent response\cite{nakasha2021pseudo}. These metasurfaces enabled functionalities such as absorption, beam steering, focusing, polarisation control, power-limiter, waveform-filtering and radar cross-section (RCS) reduction across the spectrum, from microwaves \cite{Huang2017, Li2017, Tcvetkova2018} to THz \cite{Qu2015, Hosseininejad2019, Liu2016a, Taghvaee_2017, Qu2017}, or optical frequencies \cite{ChenXZ2012, Li2015}.

Two main approaches have been proposed for the implementation of programmable metasurfaces, namely, (i) by interfacing the tunable elements through a control unit such as Field-Programmable Gate Array (FPGA) \cite{Cui2014, Wan2016}, or (ii) by integrating sensors, control units, and actuators within the metasurface structure \cite{AbadalACCESS, Tasolamprou8788546, Liaskos2018a, PhysRevApplied.11.044024}. Programmable metasurfaces have opened the door to disruptive paradigms such as reconfigurable intelligent surface (RIS), leading to the interconnection of metasurfaces, the use of machine learning and, eventually, the implementation of software-driven distributed intelligence on EM control \cite{AbadalACCESS, Liaskos2015, di2019smart, huang2019reconfigurable}. This has the potential to exert a disruptive impact in a plethora of application domains, including but not limited to holographic displays \cite{Li2017}, stealth technology \cite{ma2019smart}, active control of surface plasmons \cite{nwz148} or wireless communications \cite{Liaskos2018a, HsfNetworkTNET.2019}.

\section{Electromagnetic Illusion approaches}
The introduction section reviewed the early works on EMI and described the progress toward advanced designs with metasurfaces. This section highlights pivotal and recent advancements in EMI, providing insights into the advantages and disadvantages associated with each approach.

\subsection{Transformation optics (TO)}
To stands as one of the pioneering tools for designing intricate EM structures. This approach involves a spatial coordinate transformation, enabling the creation of an arbitrary wave propagation. In essence, by strategically configuring radiating elements in space, TO can alter the radiation pattern and power flow in free space or other media. The fundamental application of this concept involves the virtual modification of the spatial location of a radiating EM source. In practical terms, an observer perceives the emitter as being situated elsewhere, contrary to its actual position. A notable analogy can be drawn from \cite{PhysRevLett.85.3966}, where a perfect lens demonstrates this principle. In the context of a slab with a refractive index of $n=-1$, the perfect lens creates two focusing points outside the slab, effectively delocalizing the source. Within the realm of TO, the perfect lens can be conceptualized as a medium where space is entirely suppressed and folded outside, a manipulation termed embedded transformation \cite{PhysRevLett.100.063903}. Additional instances of perfect lenses, particularly those with cylindrical geometries, have been employed as magnifying devices using linear transformations \cite{PhysRevB.78.125113, Pendry:03}. 

\begin{figure*}[ht]
    \centering
  \includegraphics[trim={0cm 6cm 0cm 4cm},clip,width=0.75\linewidth]{ 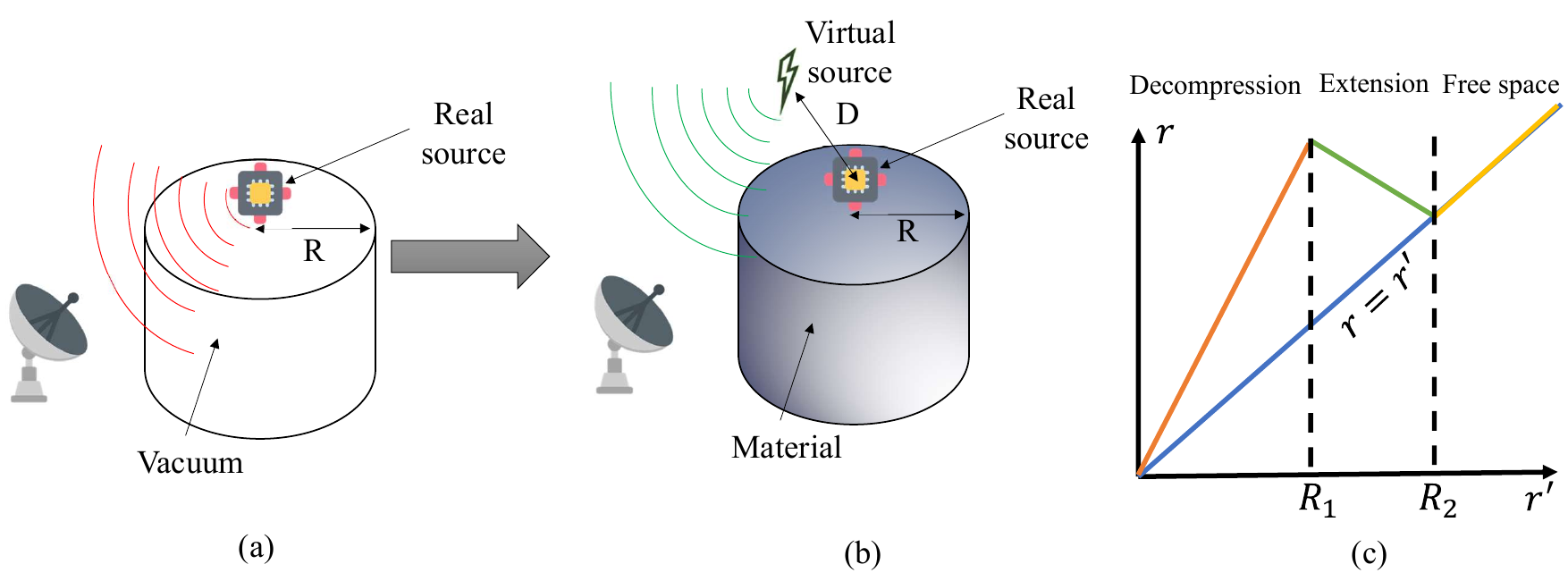} 
    \caption{(a, b) Conceptualization of the transformation process: A detector registers electromagnetic radiation appearing to emanate from a direction other than the true source, which is ingeniously concealed within a specialized material shell. (c) The pivotal transformation required for fabricating this device is accomplished utilizing linear transformation methodologies.}
    \label{fig:tra}
\end{figure*}

A radial transformation employing a sinusoidal function was initially introduced in \cite{6922484}. However, it's noteworthy that in such sinusoidal transformations, material parameter values may become prohibitively high. Even after undergoing parameter reduction processes, achieving these values with conventional metamaterial structures can remain challenging. In a recent study \cite{10.1063/1.4913596}, the authors adopted a radial transformation with a linear form. The fundamental principle of this transformation is depicted in Fig. \ref{fig:tra}(a), showcasing its ability to modify the location and size of the radiation source.

When the radiating element is positioned within the transformed material shell, observers perceive a distinct radiation pattern, creating the illusion that the emission originates from a different location. The essence of this transformation involves stretching a portion of the space within a radius outside the shell and compressing it at the radius $r=R_2$ of the material, as illustrated in Fig. \ref{fig:tra}(b).

The space transformation is divided into two segments: initially, $r$ increases sharply, indicating space compression (derivative $>1$), followed by a segment where $r$ decreases (derivative $<1$), extending space to ensure perfect impedance matching with vacuum. Any source located at a specific distance within the material shell will generate three images at different positions, as depicted by their projection on the $r'$ axis. Here, ($r$, $h$, $z$) denote coordinates in the initial cylindrical space, while ($r'$, $h'$, $z'$) represent coordinates in the final transformed space. The transformation assumes the independence of $h$ and $z$. Mathematically, this transformation is expressed as:
\begin{equation}
    \left\{
    \begin{aligned}
  r’&=f(r)\\
  \theta’&=\theta\\
  z'&=z
\end{aligned}
\right.
\end{equation}
where $f(r)$ corresponds to a perfect radial transformation and takes the form,
\begin{equation}
    f(r)=\left\{
    \begin{aligned}
  \frac{r}{q} \hspace{10mm} (0<r<R_1q)\\
  (ar)+b \hspace{10mm}     (R_1q<r<R_2)
\end{aligned}
\right.
\end{equation}
with $q>1$, $a=\frac{R_2-R_1}{R_2-R_1q}$ and $b=\frac{(1-q)R_2R_1}{R_2-R_1q}$, the transformation used to achieve the engineered material device is defined as
\begin{equation}
   r=\left\{
    \begin{aligned}
  qr' \hspace{10mm} (0<r'<R_1)\\
  \frac{r'-b}{a} \hspace{10mm}     (R_1<r'<R_2)
\end{aligned}
\right.
\end{equation}
This condition assures a perfect impedance matching between the material and vacuum and leaves the boundary at $r=R_2$ unchanged. 



\subsection{Phase gradient engineering}
Discrete systems often produce discontinuous phase profiles, which can introduce phase noise into scattering fields. This section revisits the principles of scattering control through a mechanism that harnesses these interactions, focusing on deriving a continuous gradient phase through the spin-orbit interaction, utilizing sinusoidal metallic strips. This method enables the modification of scattering properties by varying the amplitude and period of the sinusoidal strips, which can significantly influence the EMI characteristics of the object in question. The technique is noted for its continuous phase profile, its independence from polarization, and its simplicity in application.

\begin{figure}[ht]
    \centering
  \includegraphics[width=1\linewidth]{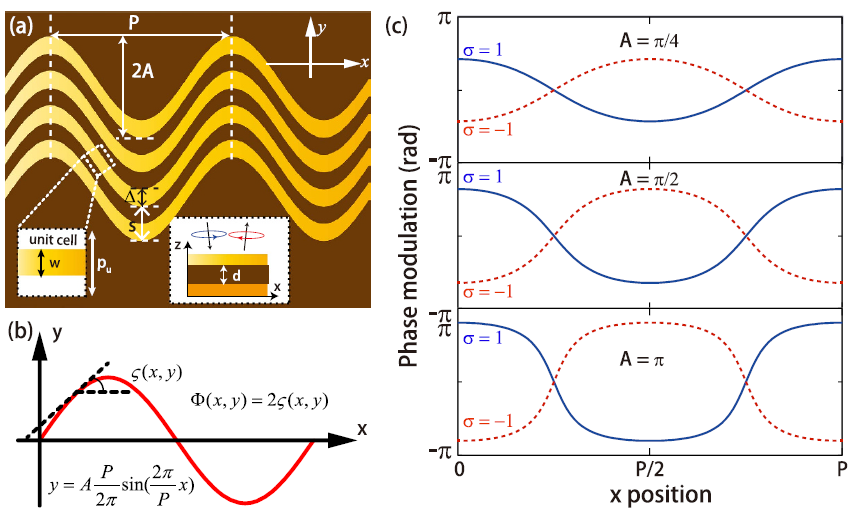} 
    \caption{The concept of continuous gradient phase modulation is rooted in the spin-orbit interaction facilitated by spatially variant half-wave plates. (a) The designed metasurface features an array of sinusoidal metallic strips, which effectively act as a series of metallic patches with varying orientation angles, as shown in the inset. This metasurface, combined with a dielectric spacer and a metallic ground plane, forms space-variant half-wave plates that enable spin-orbit interaction. (b) This is demonstrated through the phase shift observed in a half-wave plate when it is rotated under a circularly polarized wave. (c) The mechanism of Pancharatnam-Berry phase modulation is depicted across a single cycle of the sinusoidal metallic strip, where the amplitude coefficients are varied, as illustrated \cite{Guo2016}. This representation underscores the open-access nature of the article under the Creative Commons CC BY license, allowing for free use, distribution, and reproduction.}
    \label{fig:phs}
\end{figure}

The sinusoidal function is acknowledged as the quintessential oscillatory pattern for EM waves. The Fourier series expansion of perfect triangular or square waveforms is represented as a series of sinusoidal functions. Hence, sinusoids are chosen to facilitate smooth phase transitions. The configuration of the metasurface is shown in Fig. \ref{fig:phs}(a), highlighting four rows and two cycles of sinusoidal metallic strips. Figure \ref{fig:phs}(b) details the metallic strips' spatially varying profile, mathematically modelled by a standard sinusoidal equation \cite{Guo2016}.

\begin{equation}
    y=A\frac{P}{2\pi}\sin{(\frac{2\pi}{P}x)}
    \label{sin}
\end{equation}

In this setup, symbols $A$ and $P$ denote the amplitude coefficient and the period of the sine curve, respectively. The structure includes a metallic ground plane and a dielectric substrate positioned below the metasurface, thereby creating a metamirror, as illustrated in the inset of Fig. \ref{fig:phs}(a) \cite{doi:10.1063/1.4799162,Guo:15}. Tuning the dielectric spacer's thickness optimally leads to the emergence of a space-variant waveplate, composed of the metasurface, the dielectric spacer, and the metallic ground plane. In such a configuration, a half-wave plate inverts the circular polarization (CP) upon reflection. Manipulating the spatial polarization of a beam in this fashion introduces spatially variable phase shifts, known as the Pancharatnam-Berry (P-B) phase \cite{Bomzon:01}. This phase shift is mathematically defined as twice the angle of inclination between the tangent of the curve at any point and the x-axis, as shown in Fig. \ref{fig:phs}(b).
\begin{equation}
    \begin{aligned}
    \Phi(x,y)=2\sigma\zeta(x,y)=2\sigma arc\tan{(\diff{y}{x})}\\=2\sigma arc\tan{(A\cos{(\frac{2\pi}{P}}x))}
    \end{aligned}
    \label{phi}
\end{equation}

In the formula, the variable $\sigma = \pm 1$ is used to represent left-handed circular polarization (LCP) and right-handed circular polarization (RCP), respectively. Applying the equation referenced as Eq. (\ref{phi}), this approach facilitates the examination of phase modulation across a single cycle of the sinusoidal metallic strip, as depicted in Fig. \ref{fig:phs}(c). As expected, the resulting phase profile is continuous, potentially enhancing the metasurface's influence on the laws of reflection and refraction. The amplitude coefficient $A$ determines the extent of the phase shift, which approaches the complete range of ($-\pi, \pi$) as $A$ increases. The deflection angle of the $m$-th order of scattering towards the specular direction is dependent on both frequency and period, approximately adhering to the following relationship:
\begin{equation}
    \theta_m=arc sin(\frac{m\lambda}{P}).
    \label{thetam}
\end{equation}

Adjusting the amplitude coefficient $A$ and period $P$ allows for the generation of diverse scattering angular spectra. Notably, despite the polarization sensitivity of the P-B phase modulation shown in Fig. \ref{fig:phs}(c), the scattering properties of the introduced metasurface are unaffected by the circular polarization's handedness, demonstrating no circular dichroism in its scattering effects. This technique offers the potential for creating reconfigurable optical illusions with the use of adjustable metasurface technologies.

In a study cited as \cite{doi:10.1021/acsphotonics.7b01114}, researchers developed dynamic optical illusions by modifying the EM appearance without altering the physical structure. Their design incorporated several adjustable phase-shifting components that compensated for the phase differences between waves reflected by a reference plane and those at the boundary of an object. This adjustment allowed for the reconstruction of the wavefront of the scattered wave to simulate light interaction with a flat metallic surface. Through this approach, the illusion of an EM shape resembling a flat metallic plate was created. By dynamically adjusting the phase distribution on the surface via a programmable voltage source, the object could be made to present various EM shapes, effectively creating optical illusions to mislead observers.

\subsection{Anisotropic coating}

\begin{figure*}[ht]
    \centering
  \includegraphics[width=0.7\linewidth]{ 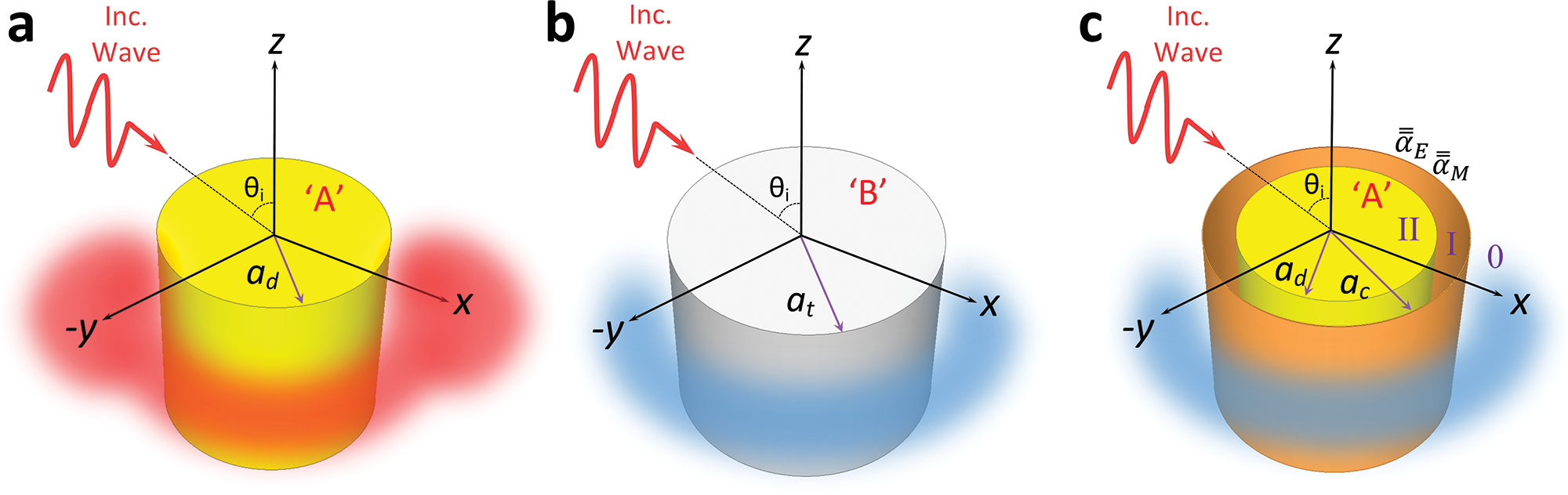} 
    \caption{Schematics illustrating the metasurface-based illusion technique: a) A cylinder made from material $A$, having a radius of $a_d$, produces a unique scattering pattern distinct from b) a cylinder made from material $B$ with a radius of $a_t$. c) By applying a specially tailored metasurface coating around the cylinder in (a), characterized by electric (E) and magnetic (M) surface polarizability tensors at a radius $a_c$, the scattering profile can be manipulated to mirror that of the cylinder in (b). This illustration is presented with the authorization of Wiley \cite{10.1002/adfm.201401561}.}
    \label{fig:cyl}
\end{figure*}

In research documented by \cite{10.1002/adfm.201401561}, the authors both theoretically forecasted and empirically validated the realization of quasi-three-dimensional, angle-resilient EMI effects on objects with a definite length, employing ultrathin, single-layer anisotropic metasurface coatings. This study showed that by finely adjusting the EM characteristics on the surface of a metasurface, it is feasible to alter the scattering profile of a coated dielectric cylinder so that it replicates the behaviour of an uncoated metallic cylinder and the reverse. A key achievement was the optimization of the metasurface coating to manipulate the zeroth- and first-order complex Mie scattering coefficients. This advancement allows for the creation of illusions over a certain angular spectrum for cylinders of finite length and practical diameters up to $0.3\lambda_0$, exceeding the boundaries of the quasi-static limit. The effectiveness of the ultrathin metasurface in creating illusions is illustrated in Fig. \ref{fig:cyl}, showcasing the ability to change the scattering characteristics of an object to resemble those of another object, chosen for its size and material composition.

In the IBCs, all the field components are oriented tangentially. Typically, $Z_e$ and $Z_m$ are denoted as two-dimensional tensors. The GSTCs were originally developed for metafilms, drawing on surface susceptibilities \cite{1236082}, and were subsequently adapted for metasurfaces characterized as isolated scatterers \cite{6230714}. Due to the anisotropic nature of the surface polarizability tensors in the metasurface, the tangential magnetic fields on either side exhibit discontinuities. Hence, these can be precisely modelled using second-order boundary conditions within the cylindrical coordinate framework, as detailed in \cite{10.1002/adfm.201401561}.
\begin{equation}
    \begin{split}
    &\hat{\rho}\times\hat{z}(\vec{H}_z^0-\vec{H}_z^I)|_{\rho=\alpha_c}=\\&j\omega\epsilon_0\alpha_E^\phi(\frac{E_\phi^0+E_\phi^I}{2})|_{\rho=\alpha_c}-\hat{\rho}\times\hat{z}\frac{\partial}{\partial z}(\alpha_M^\rho \frac{H_{\rho}^0+H_{\rho}^I}{2})|_{\rho=\alpha_c}\\
    &\hat{\rho}\times\hat{\phi}(\vec{H}_\phi^0-\vec{H}_\phi^I)|_{\rho=\alpha_c}=\\&j\omega\epsilon_0\alpha_E^z(\frac{E_z^0+E_z^I}{2})|_{\rho=\alpha_c}-\hat{\rho}\times\hat{\phi}\frac{1}{\rho}\frac{\partial}{\partial z}(\alpha_M^\rho \frac{H_{\rho}^0+H_{\rho}^I}{2})|_{\rho=\alpha_c}
    \end{split}
\label{rho}
\end{equation} 

\begin{figure}[h]
    \centering
  \includegraphics[width=1\linewidth]{ 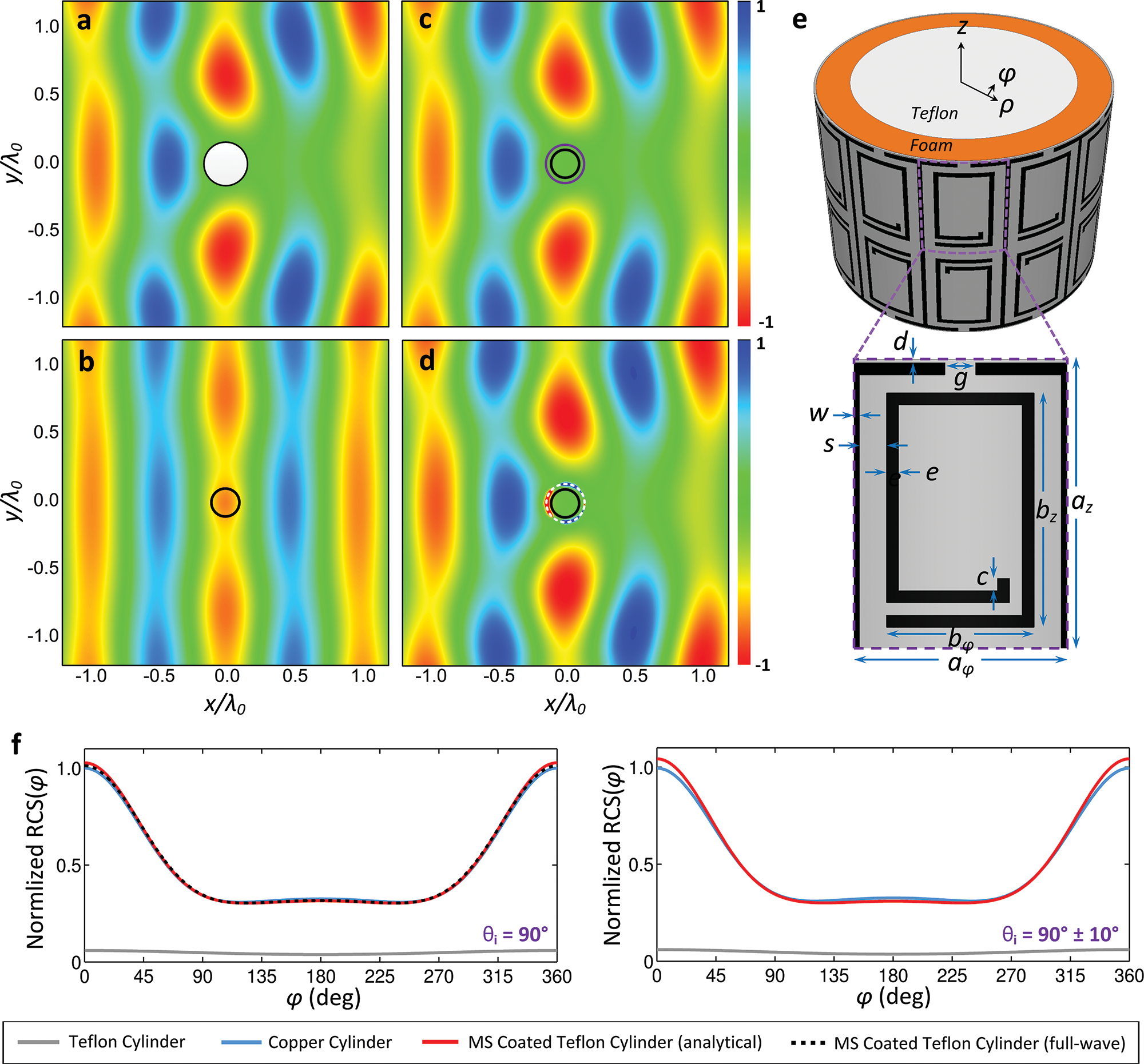} 
    \caption{Altering the scattering profile of an infinitely long dielectric cylinder to match that of an infinitely long conducting cylinder involves a series of analytical and simulated examinations of the total electric (E) field distribution. Specifically, a) the desired scattering pattern of the copper cylinder, b) the natural scattering pattern of the bare Teflon cylinder, and c) the modified scattering pattern of the same Teflon cylinder when enveloped in a specially engineered metasurface are presented through analytical calculations. Furthermore, d) showcases the total E-field distribution for the metasurface-clad Teflon cylinder as determined by comprehensive full-wave simulations. Part e) details the design of the metasurface coating along with its constituent unit cell. Lastly, f) compares the analytically derived and fully simulated far-field Radar Cross Section patterns for the copper target, the uncoated Teflon cylinder, and the Teflon cylinder sheathed in both homogenous and discretely structured metasurfaces under direct and oblique (±10°) incidences. This illustration is reproduced under authorization from Wiley \cite{10.1002/adfm.201401561}.}
    \label{fig:rescyl}
\end{figure}

Applying the mentioned equations along with additional boundary conditions, which guarantee the continuity of tangential electric and magnetic fields at the interface of the inner cylinder, and the electric fields at the boundary of the external metasurface, enables the calculation of Mie scattering coefficients across all scattering harmonic modes. Fig. \ref{fig:rescyl} showcases a comparison of the electric field distributions for three configurations when subjected to plane wave stimulation at a normal incidence angle ($\theta_i = 90^o$): an uncoated copper cylinder, an uncoated Teflon cylinder, and a Teflon cylinder encased in a metasurface layer. Remarkably, the Teflon cylinder with the metasurface coating demonstrates a field pattern that is virtually indistinguishable from that of the copper cylinder, in both visual resemblance and numerical measure.

\subsection{Generalized Sheet
Transition Condition}

\begin{figure*}[ht]
    \centering
  \includegraphics[width=0.7\linewidth]{ 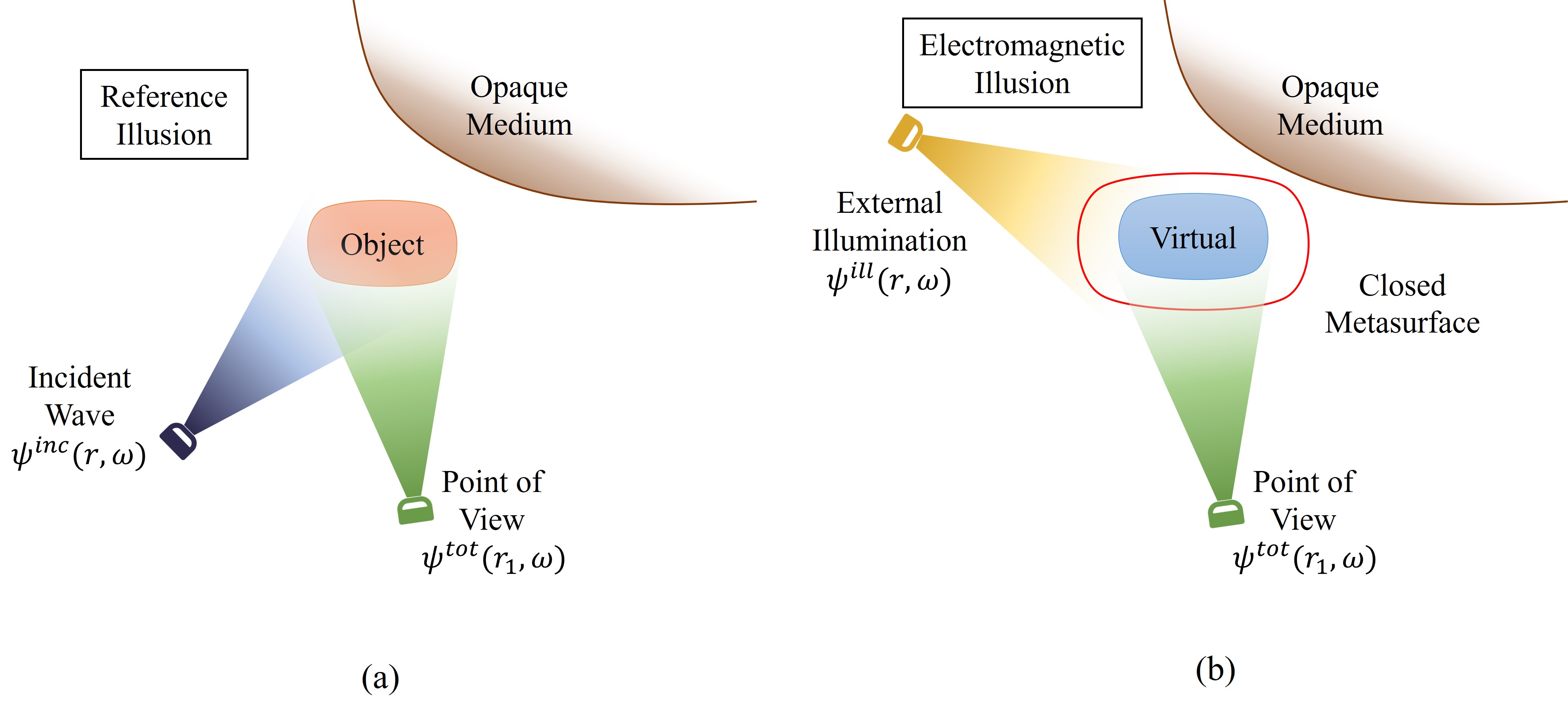}
  \includegraphics[width=0.7\linewidth]{ 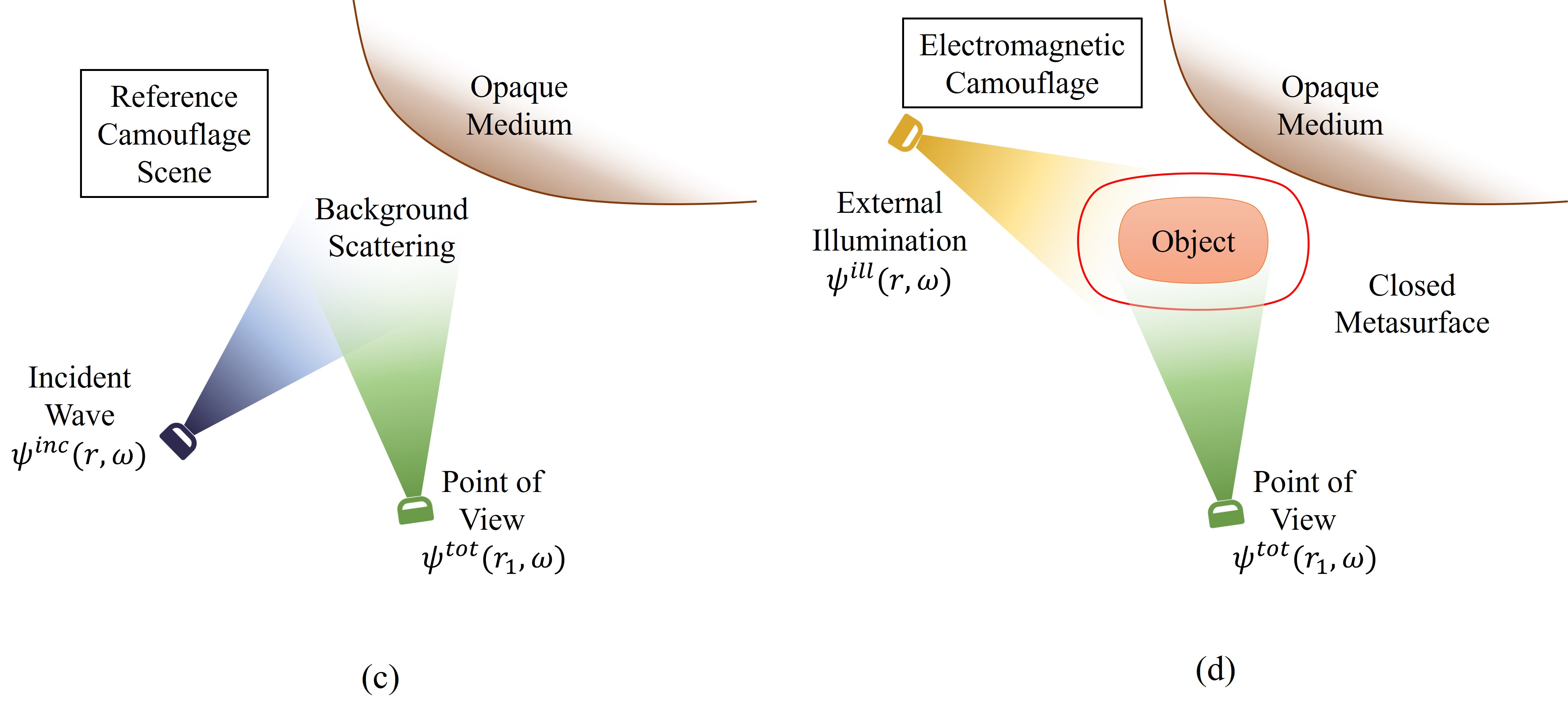}
    \caption{Visual representation of employing a closed metasurface barrier for electromagnetic illusion or camouflaging purposes. a) A reference scenario used to ascertain the required scattered fields from a target object against a backdrop. b) The deployment of a metasurface functioning as a hologram to fabricate an illusion of the object (object is omitted). c) A reference scenario for camouflaging, wherein the observer documents the scene in the absence of the object. d) The application of a metasurface acting as a skin to envelop the object, thereby concealing it within its surroundings.}
    \label{fig:iegbc}
\end{figure*}

This section introduces a comprehensive numerical approach utilizing Integral Equations and Generalized Sheet Transition Conditions (IE-GSTCs) in a two-dimensional framework to design closed metasurface holograms and coatings. These are aimed at generating EMI effects for specific targets, including, as a particular application, camouflaging them against their backgrounds. An example of a field scattering scenario is depicted in Fig. \ref{fig:iegbc}, where a defined incident wave, $\psi^{inc}(r,\omega)$, illuminates an object. This object is situated within a certain setting or Scene, possibly against a background that might be textured and reflective/opaque, or partially reflective/transmissive, thus semi-transparent. The incident wave interacts with various elements within this environment, leading to a multitude of scattered fields from both the object and its backdrop. Within this scenario, an Observer is positioned to view the object from a specific point of view (POV) across a defined field of view (FOV). This perspective involves scanning the scene across the FOV, assessing the intensity of the field as a function of the scanning angle to generate an image of the scenario. Consequently, the Observer captures/measures the overall fields $\psi^{tot}(r_1,\omega)$, which encompass both the incident and scattered fields.

In the most general case of a bianisotropic medium, electric and magnetic surface polarization densities ($P$ and $M$) are related to the averaged polarizabilities and surface susceptibilities for macroscopic description \cite{7088589}.
\begin{equation}
    \begin{split}
P=\epsilon \chi_{ee}E_{av}+\sqrt{\mu\epsilon}\chi_{em}H_{av}\\
M=\chi_{mm}H_{av}+\sqrt{\epsilon/\mu}\chi_{me}E_{av}
    \end{split}
\label{chi}
\end{equation} 

As previously noted, a metasurface can be precisely modelled as a zero-thickness sheet, employing the GSTC alongside defined electric and magnetic surface susceptibility densities. The challenge of generating EMI and camouflage is essentially reduced to identifying the appropriate surface susceptibilities for closed metasurface holograms and coatings. This approach effectively encapsulates the broad wave manipulation capabilities inherent to physical metasurfaces, portraying them as mathematical discontinuities in space with zero thickness \cite{1236082,Idemen,8302458}. In a bid for simplicity, the study in \cite{9298768} narrows the focus to tangential components and adopts scalar susceptibilities within the frequency domain \cite{app9091891,9094677}.
\begin{equation}
    \begin{split}
    &\hat{n}\times\Delta E_T=-j\omega\mu_0(\epsilon\chi_{mm}H_{T,av}+\chi_{me}\sqrt{\epsilon/\mu}E_{T,av})\\
    &\hat{n}\times\Delta H_T=-j\omega(\epsilon\chi_{ee}E_{T,av}+\chi_{em}\sqrt{\epsilon\mu}H_{T,av})
    \end{split}
\label{delta}
\end{equation} 
Section \ref{sec:smrt} simplifies this equation and uses it to develop EMI in a smart environment. 

\subsection{Adaptive illusion}

\begin{figure*}[ht]
    \centering
  \includegraphics[trim={0.5cm 3cm 0cm 2cm},clip,width=140mm]{ 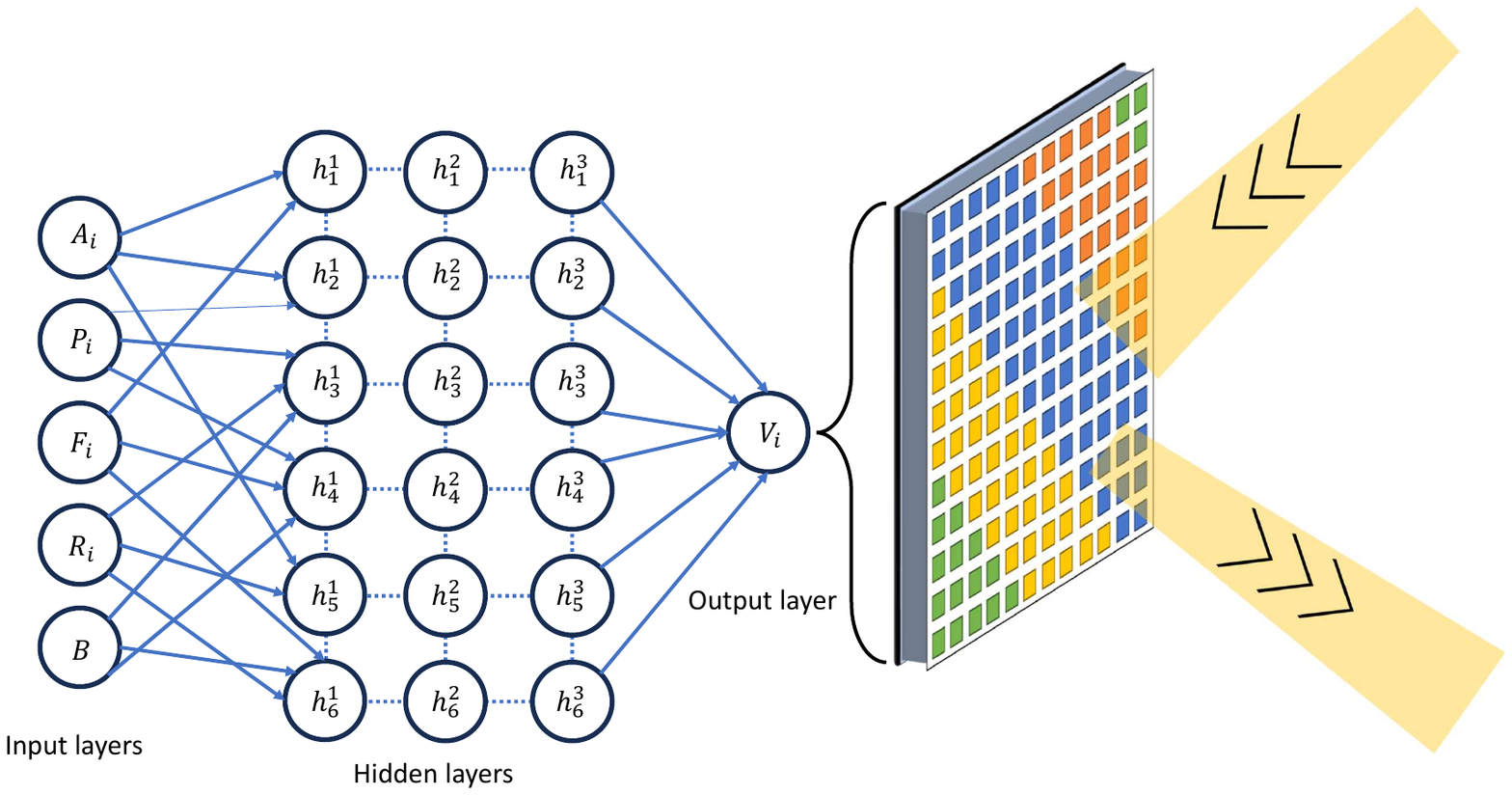}
    \caption{The metasurface features a highly thin layer filled with active unit cells, each embedded with a varactor diode that is individually managed through direct-current (d.c.) bias voltage (depicted in the upper right corner). Upon recognizing an incident wave and evaluating the ambient background, a pre-trained artificial neural network automatically determines the necessary bias voltages, which are then immediately applied via a power supply system. It's important to note that information about the detected background is transformed into the desired reflection spectrum for each unit cell using a wave reconstruction method, which is subsequently provided as input to the pre-trained artificial neural network.}
    \label{fig:dl}
\end{figure*}

For an optimal object illusion, it's vital that it can quickly and autonomously modify its structure in reaction to changes in the external environment or stimuli. This necessitates a deep comprehension of the complex interactions between the object, incident waves, and its surroundings. Reconfigurable Intelligent Surfaces (RIS) offer a solution by interacting with external signals and employing a trial-and-error approach to meet specific user needs, thus significantly enhancing the utility in real-time scenarios. To bypass the need for labour-intensive manual adjustments and foster the creation of illusions that can adapt on their own, the notion of an intelligent or self-adapting illusion has been proposed \cite{Qian2020}. This concept is illustrated through the example of a cloak made from a tunable metasurface. A depiction of this idea is provided in Figure \ref{fig:dl}, showing a deep-learning-powered self-adaptive metasurface. Within this metasurface, the reflective characteristic of each element can be individually adjusted by varying the direct-current (d.c.) bias voltages across a varactor diode at microwave frequencies. With the assistance of an artificial neural network (ANN)—a key deep learning technology—the metasurface cloak is capable of instantly adapting to dynamic incident waves and environmental conditions without human input, with adjustments made in milliseconds. The bias voltages are autonomously determined and applied to the cloak. To evaluate its effectiveness in a practical scenario, a finite-difference time-domain (FDTD) simulation, integrated with the pre-trained ANN, is used to replicate a real-world environment and validate the concept through experimental evidence.

The metasurface cloak is designed to obscure a perfect electrical conductor (PEC) bump or similar objects within a free space environment. Each of its unit cells is adjusted with a reverse direct-current (d.c.) bias voltage that varies from 0 to 20V, which modulates the capacitance of the varactor diode, thereby fine-tuning the reflection spectrum. To demonstrate the capabilities of the smart cloak, an FDTD simulation integrated with a pre-trained ANN is utilized to track the transient behaviour. An antenna array is used for detecting the incident wave, and with a theoretically predefined surrounding, a detailed simulation setup is crafted to replicate an actual situation. In this simulation, a transverse magnetic (TM) plane wave carrying a random cosine-modulated Gaussian profile strikes the cloak. Initially, when all bias voltages are at their minimum, the incident wave interacts with the bump, leading to noticeable disturbances in the scattered waves. Following an effective response period, the detection system gathers information about the incident wave, which is then processed by the ANN. Once activated, the metasurface cloak effectively renders the bump undetectable, showcasing the dynamic operation of the smart cloak in real time.

Expanding on the capabilities of intelligent systems, another example involves the use of deep learning to control a single layer of form-free metasurfaces for creating optical illusions, sidestepping the need for traditional, voluminous metamaterial layers \cite{202109331}. This approach employs a deep learning framework that utilizes two dedicated convolutional neural networks (CNNs) for different optical objectives: a far-field design network (FDN) and a near-field design network (NDN). The research reports success rates of $90\%$ for inverse structural design based on RCS and $87\%$ for designs based on near-field distributions, demonstrating the precision and effectiveness of this innovative approach.
The discussed strategies for EMI have been thoroughly examined. For an overview of further instances of object illusions and camouflage techniques found in scholarly articles, please see Table \ref{Ill}.

\begin{table*}[!htb]
    \centering
    \caption{Overview of various object illusions documented in academic research, categorized by the year of publication. Details include the operating frequency and the methods used to achieve the illusion.}
    \begin{tabular}{|p{2.2cm}|p{3.5cm}|p{3.5cm}|p{3.5cm}|p{1.8cm}|p{0.6cm}|} 
    \hline
           \textbf{Reference} & \textbf{Feature} &  \textbf{Implementation} & \textbf{Evaluation} &  \textbf{frequency} &  \textbf{Year}  \\ \hline
         A. Al\`{u} \cite{PhysRevE.72.016623} & Low-loss and passive elements & Plasmonic coating & Mie theory &  N/A & 2005   \\ \hline
         
         J. B. Pendry \cite{10.1126/science.1125907} & Theoretical EM shielding & TO & Ray approximation & N/A & 2006 \\ \hline
         
         D. Schurig \cite{10.1126/science.1133628} & Realization of camouflage & TO and SRRs & FEM and experiment &  8-12 GHz & 2006 \\ \hline
         
         J. Li \cite{PhysRevLett.101.203901} & Dielectric background & TO & Boundary condition &  N/A & 2008  \\ \hline

         R. Liu \cite{10.1126/science.1166949} & Low-loss and broadband & TO and optimization & Perturbation and experiment & 13-16 GHz & 2009 \\ \hline     
         
        A. Al\`{u} \cite{PhysRevLett.102.233901} & Without perturbation & Plasmonic coating & FEM &  1 GHz & 2009 \\ \hline 

         A. Al\`{u} \cite{PhysRevB.80.245115} & Angle independent & Patterned metallic geometries & Analytical and FEM &  N/A & 2009 \\ \hline 
        
         J. Valentine \cite{Valentine2009} & Low-loss and broadband & TO and optical carpet & FEM and experiment &  166-214 THz & 2009 \\ \hline      
        
        Y. Lai \cite{PhysRevLett.102.253902} & Remote illusion & TO & FEM &  N/A & 2009 \\ \hline   
                 
         W. X. Jiang \cite{10.1063/1.3371716} & Virtual object & TO and SRRs & FEM &  12 GHz & 2010 \\ \hline  

         W. X. Jiang \cite{XiangJiang:10} & Virtual conversion & Layered material shell & FEM &  2 GHz & 2010 \\ \hline

         C. Li \cite{PhysRevLett.105.233906} &  Broadband illusion & Inductor-capacitor network & FDTD &  51 MHz & 2010 \\ \hline

         Y. Xu \cite{Xu_2011} & Overlapped illusion optics & Illusion device & FEM & N/A & 2011 \\ \hline  

         W. X. Jiang \cite{PhysRevE.83.026601} & Radar illusion & Layered SRRs & FEM & 9.8 GHz & 2011 \\ \hline    
         
         W. X. Jiang \cite{10.1002/adfm.201203806} & Ghost illusion & Layered SRRs & FEM & 10 GHz & 2013 \\ \hline
         
         J. J. Li \cite{Li2013} & Shifting illusion media & Inductor-capacitor network & FEM and experiment &  65 MHz & 2013 \\ \hline

         Z. H. Jiang \cite{10.1002/adfm.201401561} & Low profile and angle tolerant & Dipoles and spiral resonators & GSTC and experiment &  2.5 GHz & 2014 \\ \hline

          J. Yi \cite{10.1063/1.4913596} & Virtual emission & TO and layered material shell & FEM &  5 and 10 GHz & 2015 \\ \hline

         X. Ni \cite{10.1126/science.aac9411} & Ultrathin & Nanoantennas & Scanning electron
         microscope &  410 THz & 2015 \\ \hline    
        
          B. Zheng \cite{Zheng2016} & remote cloaking & multi-folded shell & TO and FEM & 3 GHz & 2016 \\ \hline

          Y. Guo \cite{Guo2016} & Polarization-independent & Continues gradient distribution & Snell’s law and experiment & 8-12 GHz & 2016 \\ \hline          

          R. Wang \cite{7995050} & Dual-polarized, wide angle & Patch unit cell & FEM and experiment & 5.8 GHz & 2017 \\ \hline
          
          Z. Hayran \cite{10.1021/acsphotonics.7b01608} & Self-Cloaked & Clear cast acrylic (Plexiglas) & FDTD and experiment & 10.5-12.9 GHz & 2018 \\ \hline    

          T. Nagayama \cite{refId0} & Broadband transmission-line & TO and microstrip-line & SPICE and experiment & 2.60-4.65 GHz & 2019 \\ \hline  
          
          T. J. Smy \cite{9298768} & Opaque background & Susceptibility synthesis & IE-GSTCs and FEM & 60 GHz & 2020 \\ \hline
         
          C. Qian \cite{Qian2020} & Self-adaptive & Deep learning & FDTD and experiment &  6-8 GHz & 2020 \\ \hline         

          D.-H. Kwon \cite{PhysRevB.101.235135} & Passive lossless & Auxiliary evanescent waves & Theoretical and FEM &  300 MHz & 2020 \\ \hline            
          X. Wang \cite{9142383} & Spectrum selective & Time-modulated metasurface & FDTD and experiment &  10 GHz & 2020 \\ \hline   
          
          H.-X. Xu \cite{Xu2021} & Polarization-insensitive & Patterned meta-atoms & FDTD and experiment &  8-18 GHz & 2021 \\ \hline         

          C. Huang \cite{smtd.202000918} & High-Efficiency and dual-band & Graphene and SRR & FDTD and experiment &  5-25 GHz & 2021 \\ \hline

          P. Ding \cite{DING2021111578} & Broadband and wide-angle & Graphene & FEM &  2 THz & 2021 \\ \hline

          P. Ang \cite{0041996} & Active cloaking & Huygens's metasurfaces & HFSS and experiment &  1.2 GHz & 2021 \\ \hline

          H. Taghvaee \cite{9814300} & Smart environment & Impedance Surface & IBC and GSTCs & 10-12 GHz & 2022 \\ \hline

          Y. Jia \cite{202109331} & Global Metasurface & Deep Learning & FDN and NDN & 8 GHz & 2022 \\ \hline

          H. Lee \cite{054012} & Freestanding illusion & Reactance surface & Simulation and experiment & 10 GHz & 2022 \\ \hline

          X. Tian \cite{photonics9030156} & Hybrid Metasurface & phase-change material & FEM & 35 THz & 2022 \\ \hline 

          R. Cacocciol \cite{202101882} & Sandwiched Metasurface & SRR & Simulation and experiment & 4 GHz & 2022 \\ \hline 

          Q. Liang \cite{202201728} &  cloaking and illusion & Sn–Bi metallic rings  & CST and experiment & 4.5 and 15 GHz & 2023 \\ \hline

          Y. Sun \cite{202300318} & Polarization-Insensitive & Passive supercell & CST and experiment &  4.35 GHz & 2023 \\ \hline  
          
          J. Liao \cite{2c21565} & Polarization-Insensitive & Patch connected to Varactor & Simulation and experiment &  4.35 GHz & 2023 \\ \hline          

         Q. Liang \cite{202202020} & Dynamic metasurface  & 4D-printing & CST and experiment & 7.63 GHz & 2023 \\ \hline

         R. Zhu \cite{1289250} & GA-drive  & SRR & CST & 5 GHz & 2023 \\ \hline
        
         J. Budhu \cite{10133421} & Passive and Lossless  & 3D Metasurface & Semi-analytic gradient & 10 GHz & 2023 \\ \hline          
          
   \end{tabular}
    \label{Ill}
\end{table*}

\section{Illusion in smart environment}
\label{sec:smrt}
The core idea behind advancing electromagnetic illusions stands on harnessing the complex scattering generated by an object and interacting with the surrounding environment. 
As anticipated above, the breakthrough comes from surface assisted propagation environments, which can be reconfigured to create anomalous wave effects such as reflection at non-specular angles. 
In rich scattering, e.g., chaotic, environments the waves bouncing around the environment will most certainly reach the wall mounted surfaces, which gives a route to influencing the signal at receiver. 
Whence the idea of creating an illusion by avoiding direct engineering of mantle cloak wrapping the object itself. 
Inherently, an electromagnetic illusion revisits the concept of hiding the object by suppressing its scattered waves (or reduce its radar cross section). 
It rather diverts the receiver by providing the scattering signal of a virtual object located somewhere else in the the environment.
In other words, this process involves modifying the scattering characteristics of the object to mislead the detection capabilities of an observer. 
Further discussions will elaborate on the creation of such illusions by manipulating the susceptibility of the transmissive metasurface or adjusting the surface impedance of the reflective metasurface. The initial step involves outlining the boundary-value problem pertinent to object illusion challenges, incorporating advanced wave-scattering models in complex, constrained settings. Traditional boundary conditions are utilized to illustrate the interaction between field discontinuities at the boundary and the distribution of current sources along it.
\begin{equation}
    \begin{split}
&\hat{z}\times(\Vec{H_2}-\Vec{H_1}) = J_e\\
&(\Vec{E_2}-\Vec{E_1})\times\hat{z}=J_m
    \end{split}
\label{Jem}
\end{equation} 
The discontinuity in fields across EM surfaces can be described through their effective sheet impedances, a concept from which IBCs were formulated, as detailed in \cite{Senior1960}. Thus, IBCs elucidate the correlation between the average fields on surfaces and the discontinuities in fields across these surfaces.
\begin{equation}
    \begin{split}
&\hat{z}\times(\Vec{H_2}-\Vec{H_1}) = E_{av}.Z_e^{-1}\\
&(\Vec{E_2}-\Vec{E_1})\times\hat{z}=H_{av}.Z_m
    \end{split}
\label{Zem}
\end{equation} 

\begin{figure*}[htbp]
  \centering
  \includegraphics[trim={0cm 2cm 0cm 2cm},clip,width=150mm]{ 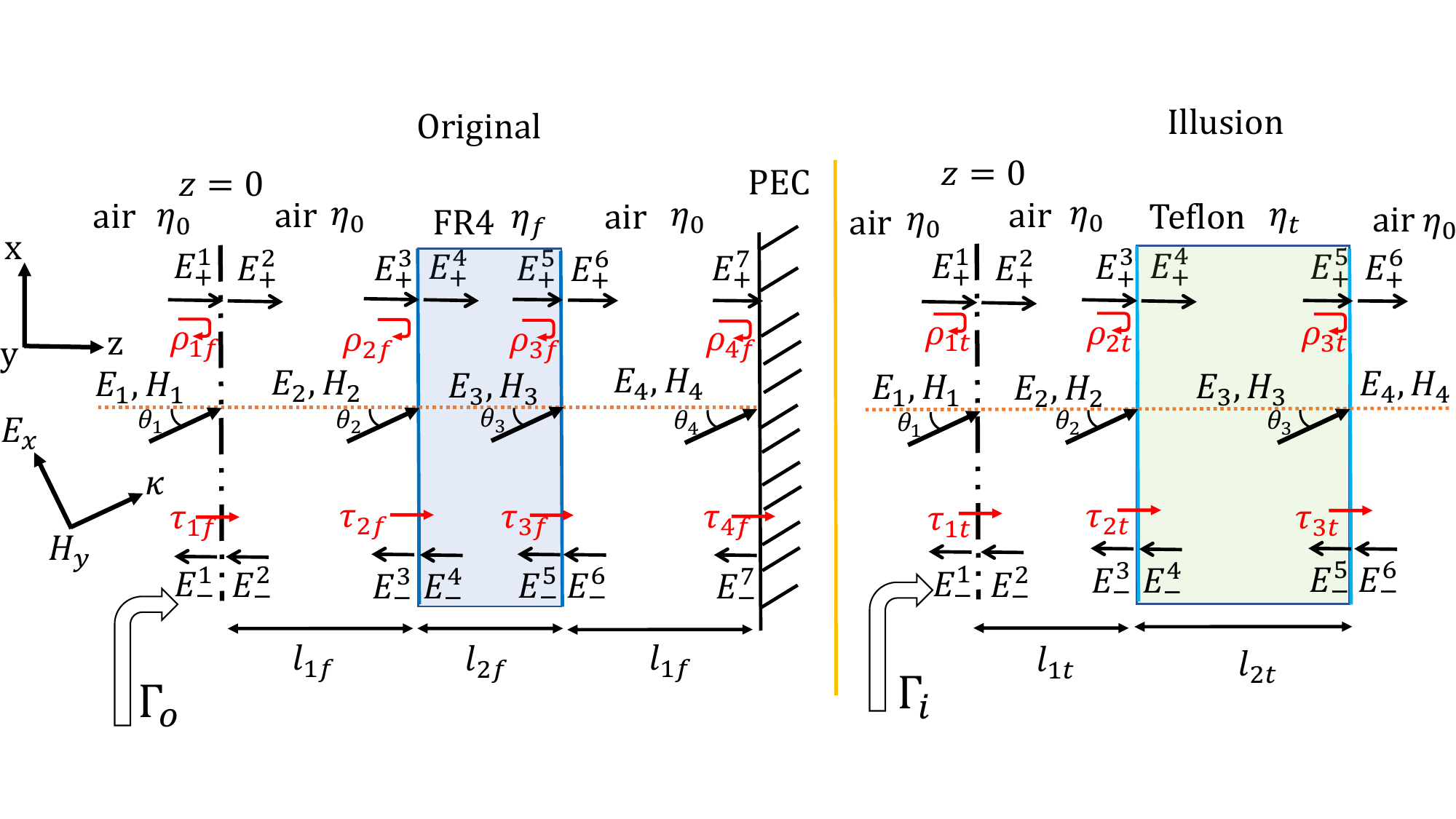}
  \caption{This figure presents a comparative 1D model analysis of an original structure and its illusion counterpart. The left subfigure depicts a slab of FR4 material surrounded by air and backed by a PEC, where analysis of reflections can reveal the slab's thickness and material properties. Conversely, the right subfigure illustrates an illusion, designed to misrepresent the actual thickness and material composition, creating a deceptive perception of the physical attributes.}
  \label{fig:cavity}
\end{figure*}

It's important to recognize that in IBCs, all field components are tangential to the surfaces. IBCs facilitate the calculation of field discontinuities across EM surfaces, provided the sheet impedances, $Z_e$ and $Z_m$, are established. Nonetheless, these impedances are not inherent properties of EM surfaces, and determining their values for various scenarios is not straightforward. In contrast, the defining parameters for 1D interfaces are effective susceptibilities, which remain constant regardless of the external fields applied. These susceptibilities form the basis for deriving the GSTCs for metafilms, as outlined in \cite{app9091891}.
\begin{equation}
  \label{eq:Xe}
  \hat{n}\times(\Vec{H_2}-\Vec{H_1}) = j\omega\epsilon_0\chi_eE_{av}-\hat{n}\times\nabla_t(\chi_mH_{av})
\end{equation}
\begin{equation}
  \label{eq:Xm}
  (\Vec{H_2}-\Vec{H_1})\times\hat{n} = j\omega\mu_0\chi_mH_{av}+\hat{n}\times\nabla_t(\chi_eE_{av})
\end{equation}
In this context, the 1D electric ($\chi_e$) and magnetic ($\chi_m$) tensors denote the effective surface electric and magnetic susceptibilities, respectively. Utilizing these susceptibilities, the transmission and reflection coefficients for the interface can be accurately calculated, as demonstrated in \cite{1580755}.
\begin{equation}
  \label{eq:tau}
  \tau_{M} =\frac{1-(k_0/2)^2\chi_{e}\chi_m}{1+(k_0/2)^2\chi_{e}\chi_m-jk_0/2(\chi_m-\chi_e)}
\end{equation}
\begin{equation}
  \label{eq:rho}
  \rho_{M} =\frac{jk_0/2(\chi_m+\chi_e)}{1+(k_0/2)^2\chi_{e}\chi_m-jk_0/2(\chi_m-\chi_e)}.
\end{equation}
Utilizing IBCs to model the fields allows for the establishment of a relationship between the forward and backward fields on each side of the interface, which is represented through a $2\times2$ transfer matrix. To explore how changes in material properties can affect the slab's behaviour, a metasurface is introduced alongside the PEC wall. In the initial configuration, the slab consists of FR4 material with a permittivity of $\epsilon_f=3.9-0.08i$ and a thickness of $l_{2f}=60mm$, with air gaps on the lateral sides measuring $l_{1f}=l_{3f}=120 mm$. For the illusion setup, the slab is constructed from Teflon, characterized by a permittivity of $\epsilon_t=2.1-0.0006i$ and a thickness of $l_{2t}=120mm$, and the air space on the lateral side is reduced to $l_{1t}=60mm$. Figure \ref{fig:cavity} illustrates the comparison between the original (left figure) and the illusion (right figure) configurations.

\begin{figure}[htbp]
  \centering
  \includegraphics[trim={1cm 6cm 1cm 6cm},clip,width=43mm]{ 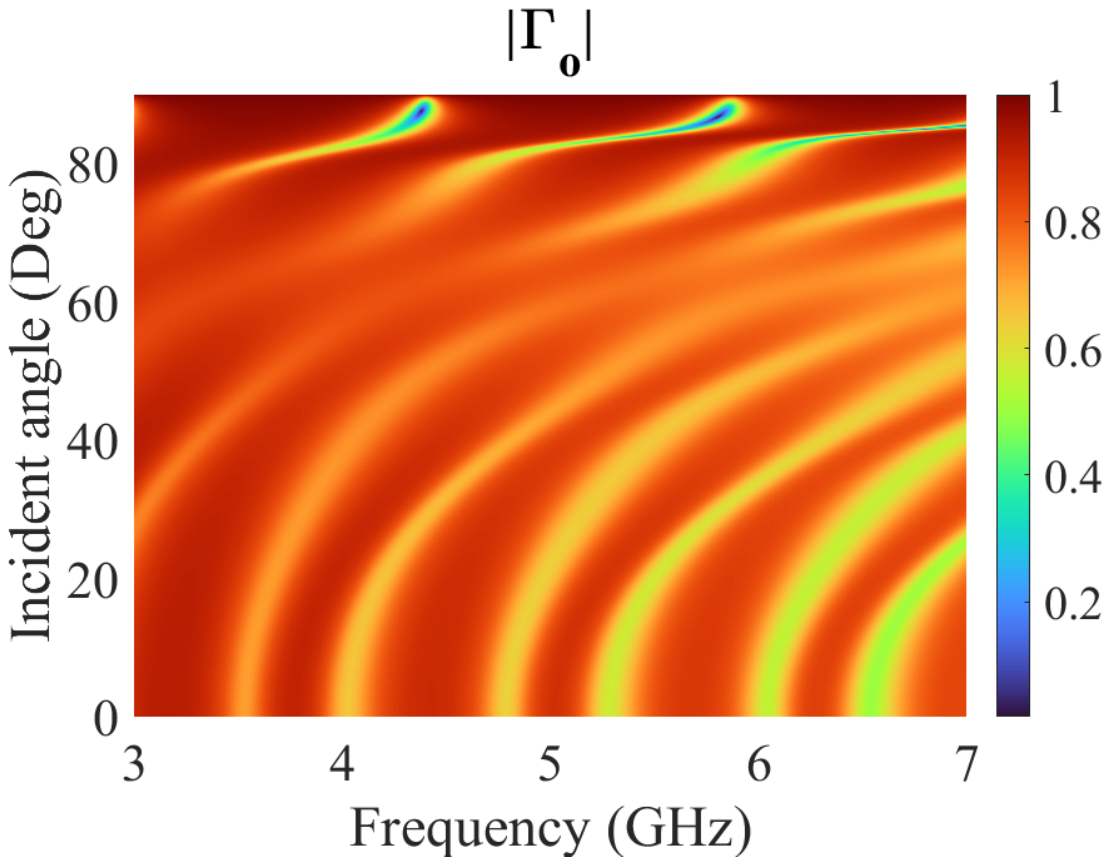}
  \includegraphics[trim={1cm 6cm 1cm 6cm},clip,width=43mm]{ 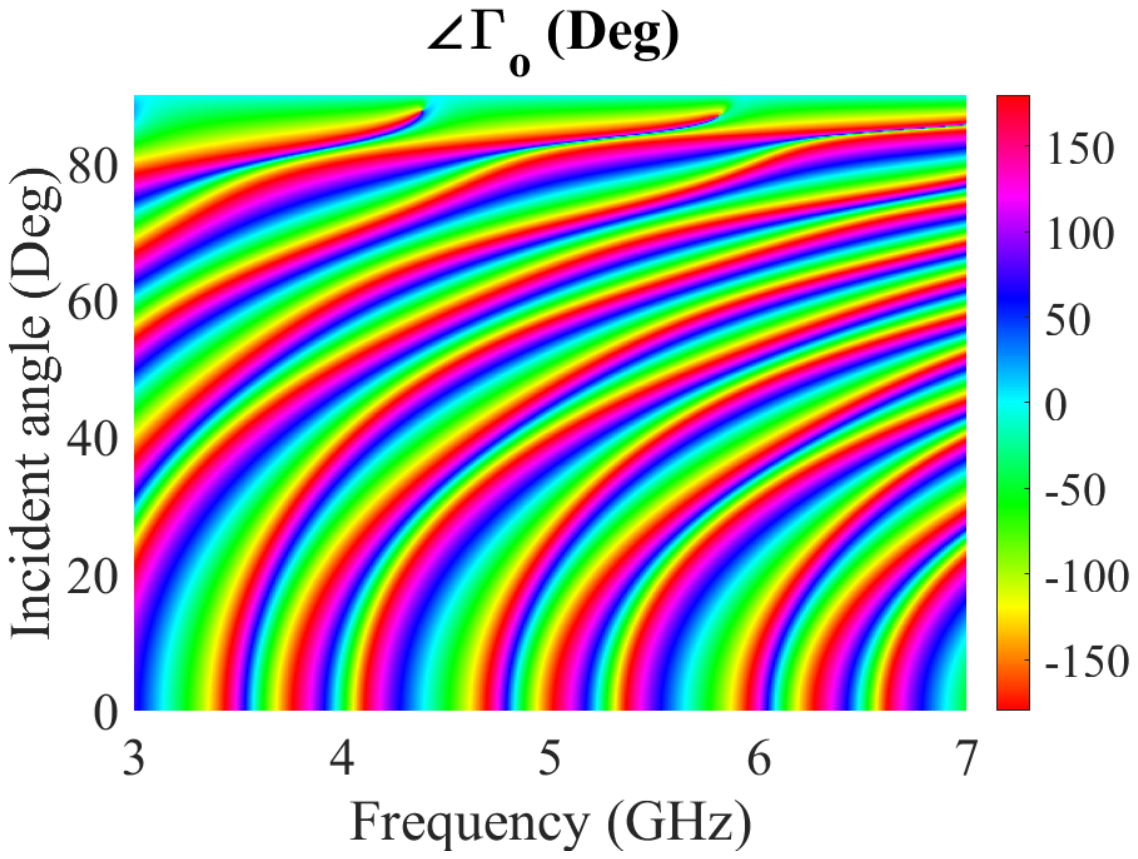}
  \includegraphics[trim={1cm 6cm 1cm 6cm},clip,width=43mm]{ 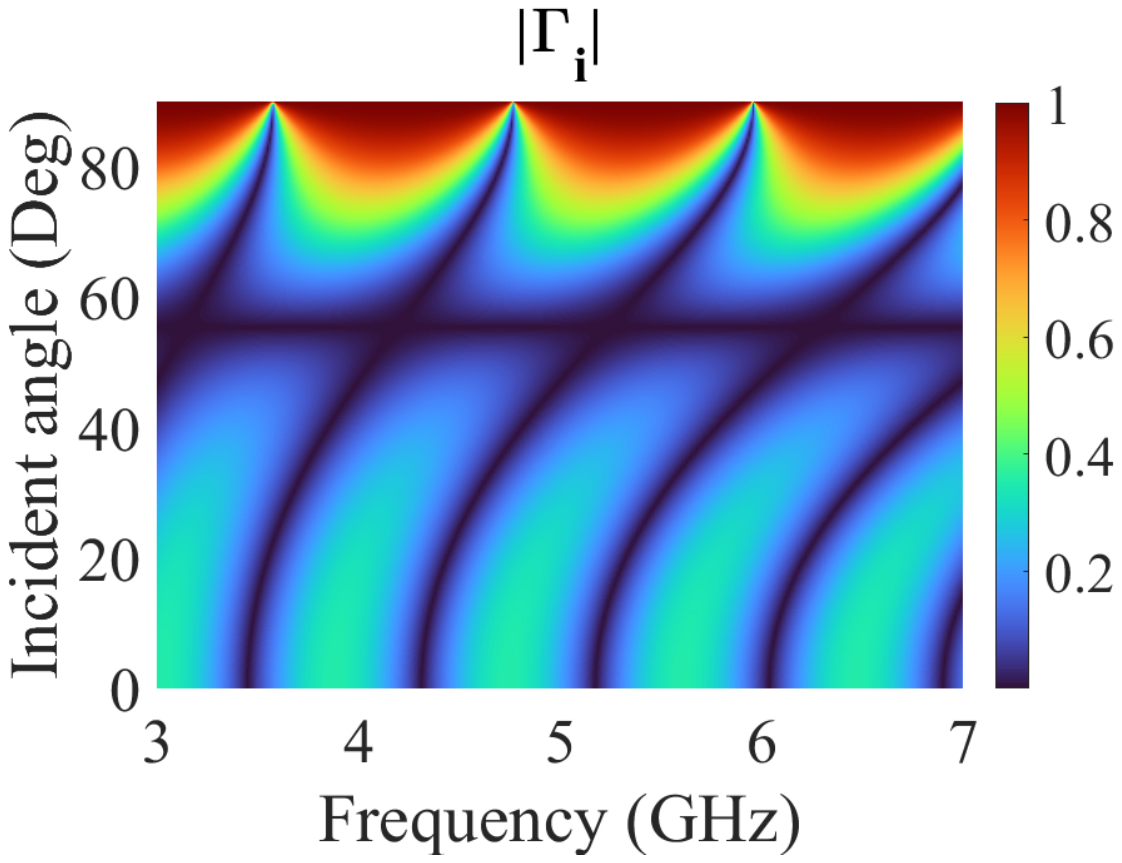}
  \includegraphics[trim={1cm 6cm 1cm 6cm},clip,width=43mm]{ 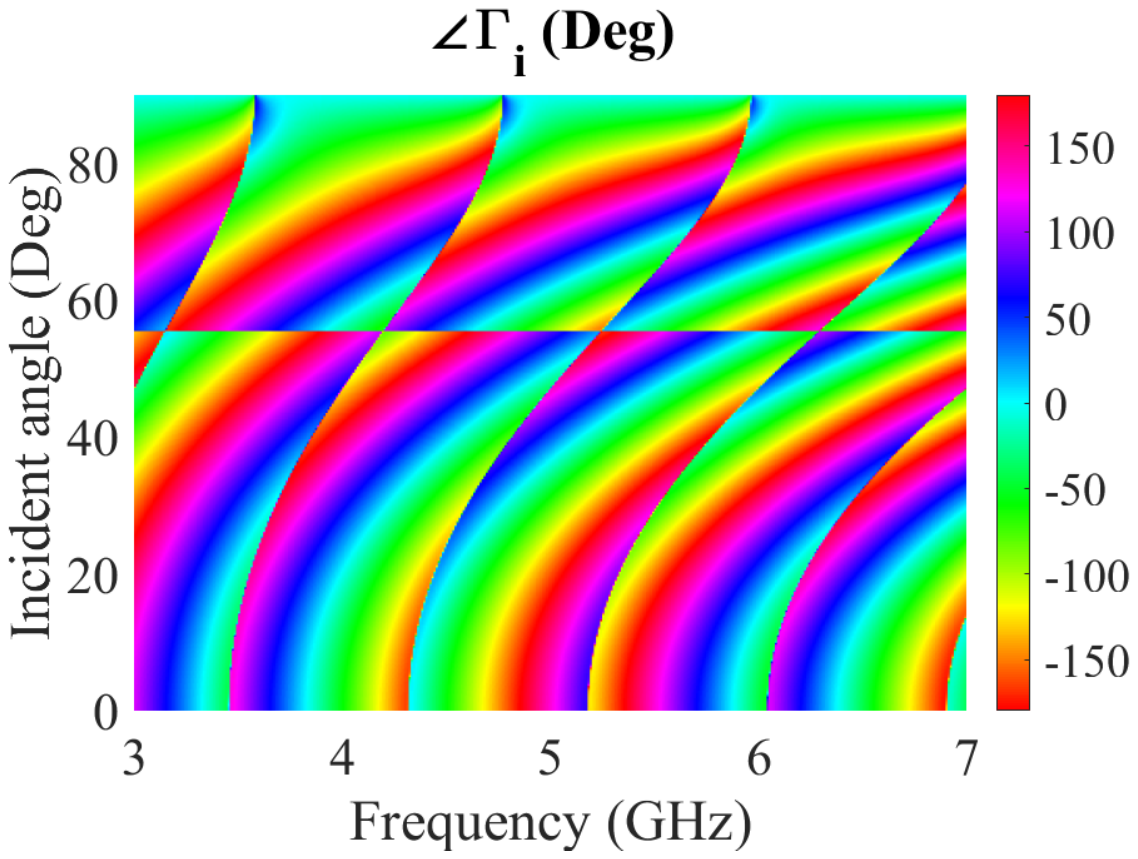}
  \caption{This figure displays a comparison of the reflection coefficients for two scenarios: the original structure and its illusion variant. It consists of four subfigures, each depicting a plot versus incident angles. The plots detail the absolute value and phase of the reflection coefficient for both the original setup, involving an FR4 slab in the air with a PEC background ($\Gamma_o$), and the illusion setup, featuring a Teflon slab in the air without a PEC background ($\Gamma_i$).}
  \label{fig:refslab}
\end{figure}

When a planar oblique wave strikes the slab, part of the wave transmits through and then repeatedly reflects within the slab. The transmitted portion of the wave encounters the PEC and is reflected back. The behaviour of this process is encapsulated by the transfer matrix, which is described as follows:
\begin{multline}
  \label{eq:refmatrix}
  \begin{bmatrix}
E_+^1  \\
E_-^1  
\end{bmatrix}=\frac{1}{\tau_1\tau_2\tau_3}\times
\begin{bmatrix}
Z_1^{-1} & \rho_1Z_1 \\
\rho_1Z_1^{-1}  & Z_1
\end{bmatrix}
\begin{bmatrix}
Z_2^{-1} & \rho_2Z_2 \\
\rho_2Z_2^{-1}  & Z_2
\end{bmatrix}
\\
\begin{bmatrix}
Z_3^{-1} & \rho_3Z_3 \\
\rho_3Z_3^{-1}  & Z_3
\end{bmatrix}
  \begin{bmatrix}
E_+^7  \\
E_-^7  
\end{bmatrix}
\end{multline}
The term $Z_n = e^{-jk_nl_ncos\theta_n}$ is derived from the propagation matrix, where $\rho_n$ and $\tau_n$ represent the local reflection and transmission coefficients, respectively, and are defined as follows:
\begin{multline}
  \label{eq:elementry}
  \rho_n=\frac{\eta_{n+1}cos\theta_{n+1}-\eta_n cos\theta_n}{\eta_{n+1}cos\theta_{n+1}+\eta_n cos\theta_n},\\  \tau_n=\frac{2\eta_{n+1}cos\theta_{n+1}}{\eta_{n+1} cos\theta_{n+1}+\eta_n cos\theta_n}
\end{multline}
The total reflection at the point $z=0$ is calculated using the formula $\Gamma = E_-^1/E_+^1$, where $E_-^1$ and $E_+^1$ represent the amplitudes of the reflected and incident waves, respectively. The figures referenced as~\ref{fig:refslab} display the total reflection amplitude and phase for both FR4 ($\Gamma_o$) and Teflon ($\Gamma_i$) slabs. The resonances observed are due to stationary waves forming within a cavity-like environment, resulting from the interference between the forward-propagating and backwards-propagating wave fields.

\begin{figure}[htbp]
  \centering
  \includegraphics[trim={1cm 6cm 1cm 6cm},clip,width=43mm]{ 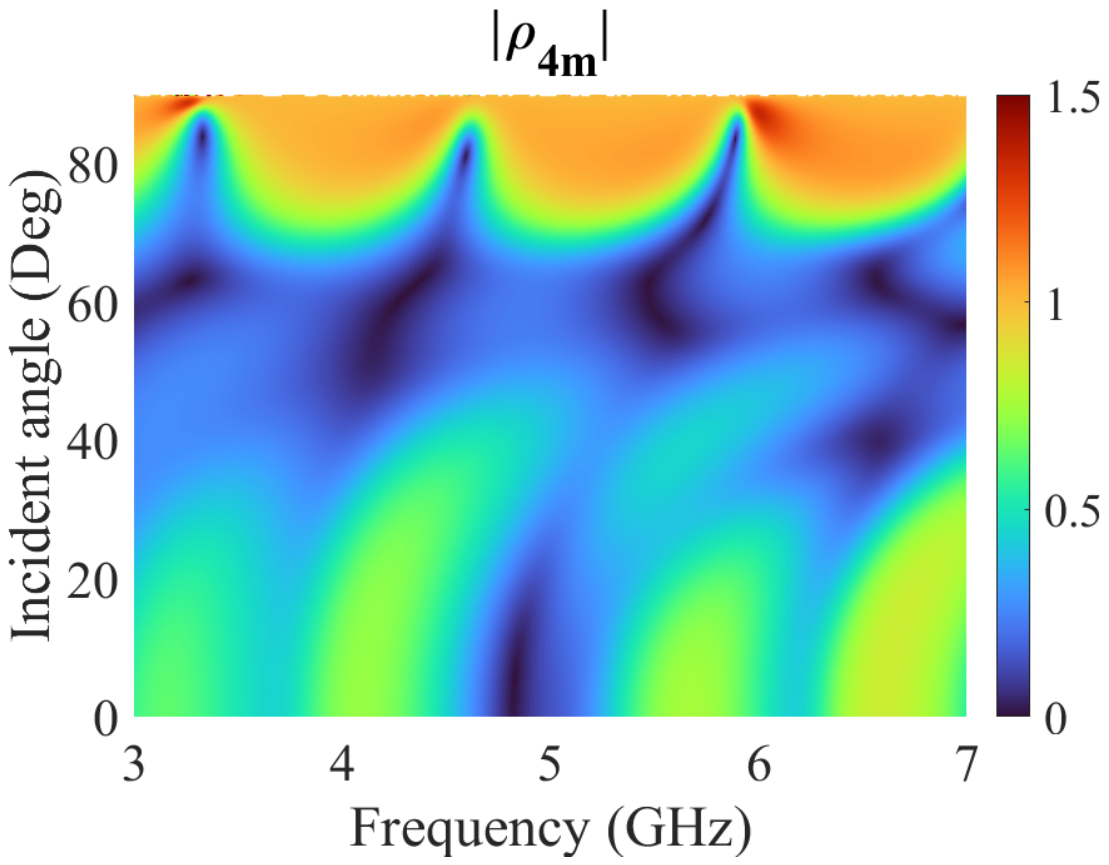}
  \includegraphics[trim={1cm 6cm 1cm 6cm},clip,width=43mm]{ 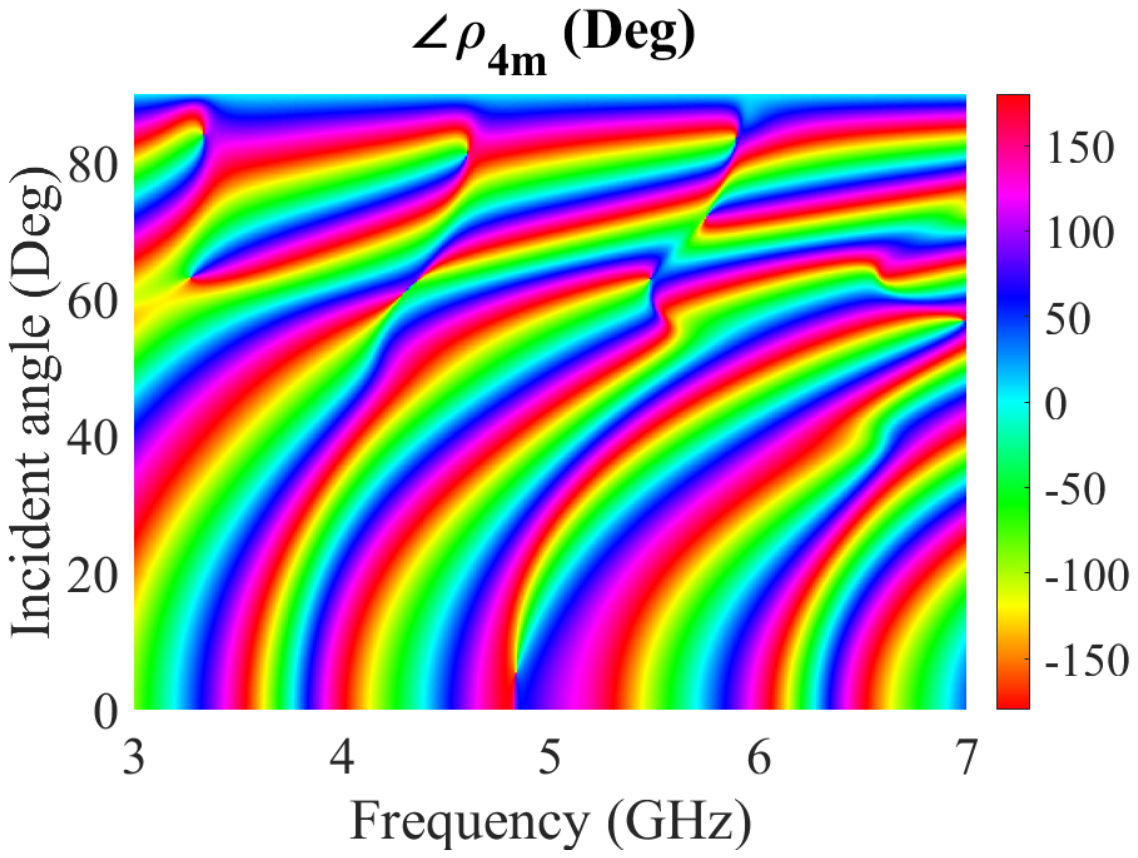}
    \includegraphics[trim={1cm 6cm 1cm 6cm},clip,width=43mm]{ 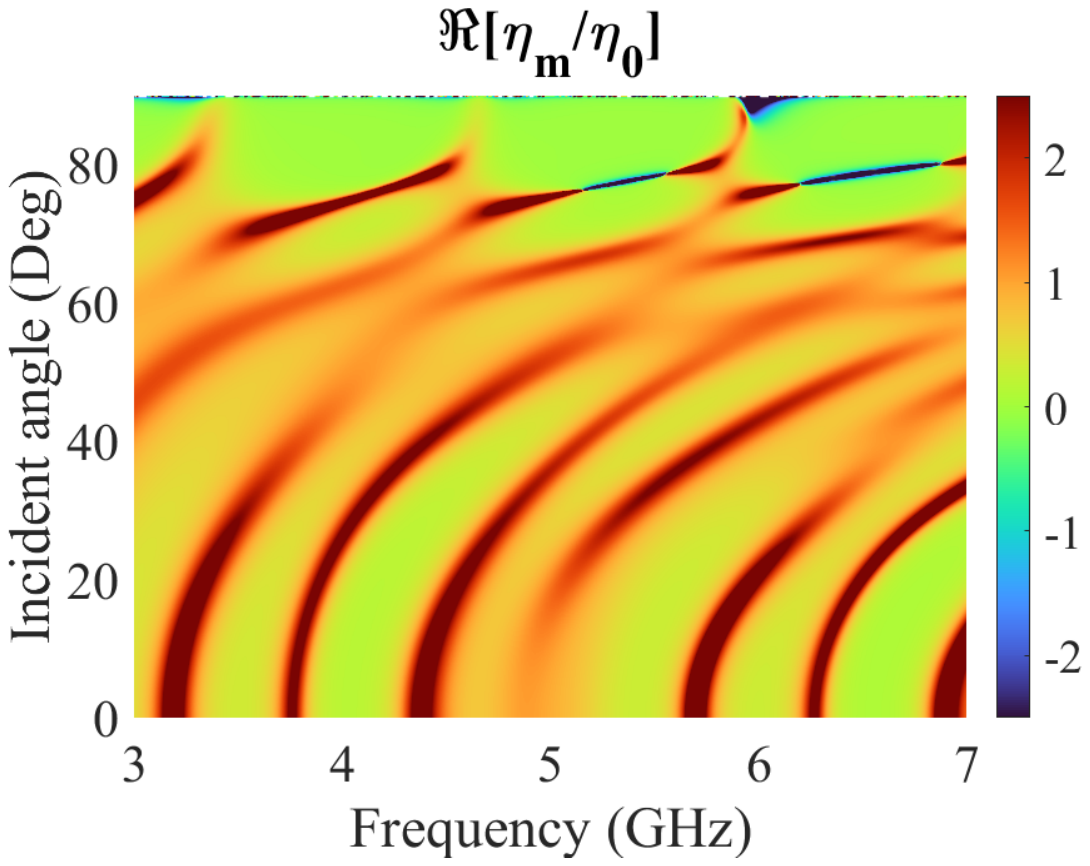}
    \includegraphics[trim={1cm 6cm 1cm 6cm},clip,width=43mm]{ 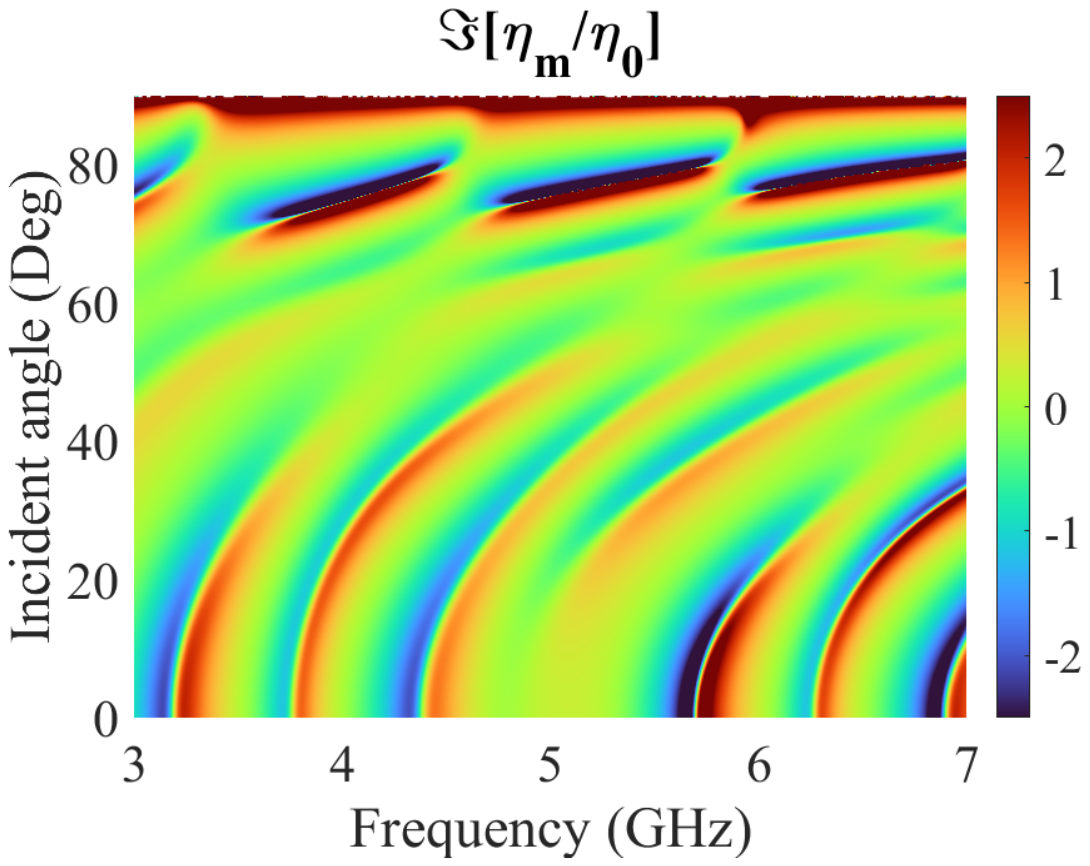}
  \caption{This figure illustrates the reflection coefficient and surface impedance characteristics essential for crafting the illusion with an engineered metasurface. It comprises four subfigures: two (left) display the absolute value and phase of the reflection coefficient for the engineered metasurface, while the other two (right) present the real and imaginary components of the normalized surface impedance. Each subfigure is structured versus the incident angle of the incoming wave, providing a comprehensive view of how the metasurface interacts with varying frequencies and angles of incidence.}
  \label{fig:msimp}
\end{figure}

By setting $\Gamma_m$ to be equivalent to $\Gamma_i$, it becomes possible to identify the required surface impedance of the metasurface to mimic the desired scattering properties. The critical reflection coefficient, which characterizes the behaviour of an impenetrable metasurface for achieving this illusion, is given by
\begin{equation}
  \label{eq:rho4m}
\rho_{4m}=\frac{A-B}{C-D}
\end{equation}
The values of $A$, $B$, $C$, and $D$ are determined as detailed in Appendix A. The reflection coefficients and the associated normalized impedance for a metasurface located at the PEC position $z=300mm$ are depicted in Fig. \ref{fig:msimp}.
\begin{equation}
  \label{eq:etam}
\eta_{m}=\eta_0\frac{1+\rho_{4m}}{1-\rho_{4m}}
\end{equation}
The real component of the normalized impedance predominantly resides in the positive domain, indicating the requirement for elements with loss characteristics. The reflection amplitude profiles corroborate this observation, suggesting that achieving the illusion predominantly relies on passive components, except at very steep incident angles where active elements may be necessary.

\begin{figure}[htbp]
  \centering
  \includegraphics[trim={1cm 6cm 1cm 6cm},clip,width=43mm]{ 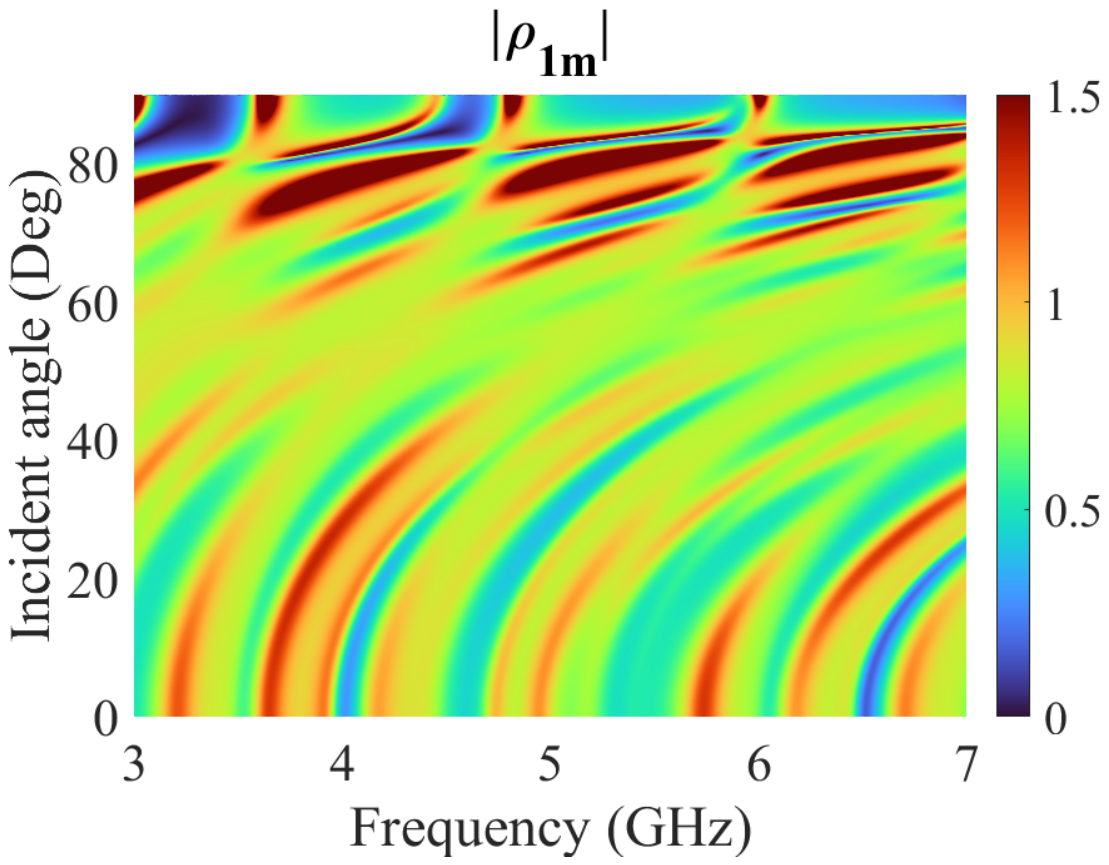}
    \includegraphics[trim={1cm 6cm 1cm 6cm},clip,width=43mm]{ 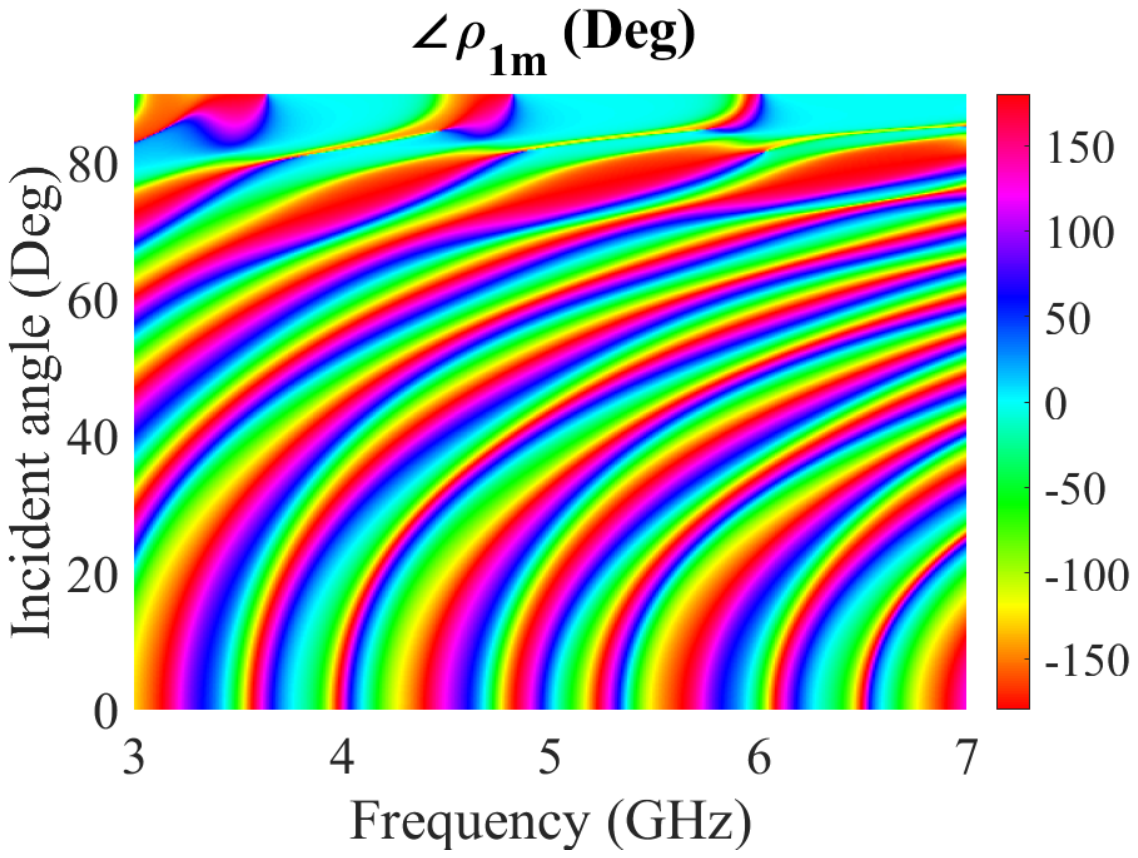}
  \includegraphics[trim={1cm 6cm 1cm 6cm},clip,width=43mm]{ 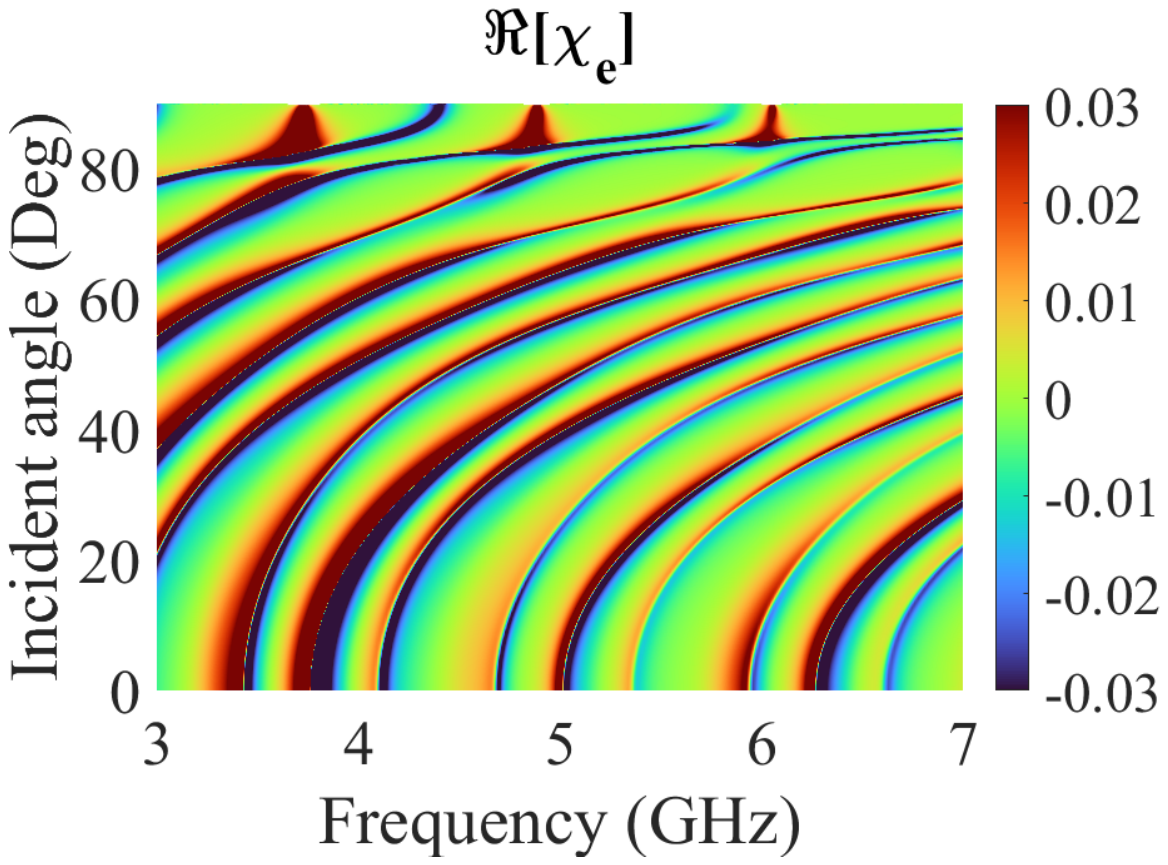}
    \includegraphics[trim={1cm 6cm 1cm 6cm},clip,width=43mm]{ 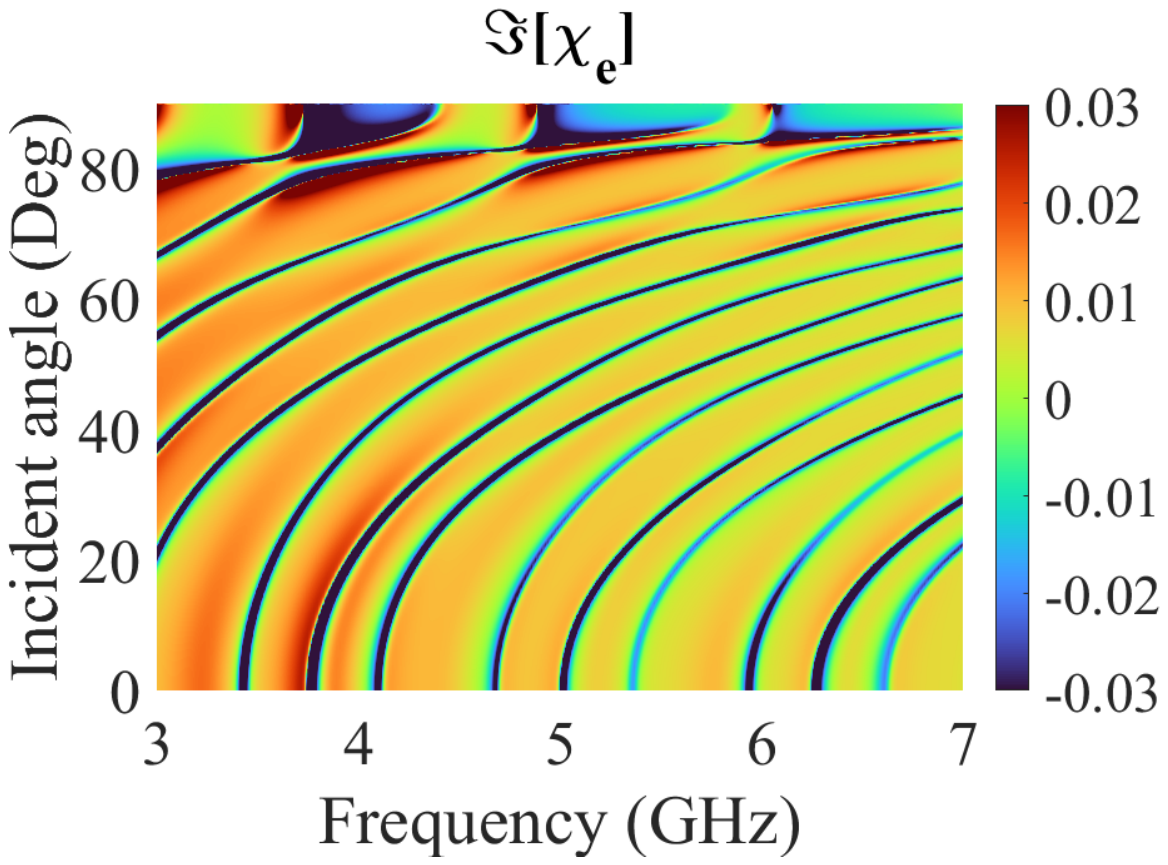}
  \caption{This figure showcases the reflection coefficient and surface susceptibility of an engineered metasurface designed for illusion creation. It is composed of four subfigures: two (left) detail the absolute value and phase of the reflection coefficient for the engineered metasurface, and the other two (right) depict the real and imaginary components of the surface susceptibility. Each subfigure is arranged as a plot versus the incident angle of the wave, providing a comprehensive view of the metasurface's interaction with different frequencies and angles of incident waves.}
  \label{fig:mssus}
\end{figure}

In the second scenario, the focus shifts to designing a transmissive metasurface positioned at $z=0$, with the task of determining the necessary susceptibilities for the desired effect. The critical reflection coefficient, which describes the metasurface's behaviour for this particular illusion, is defined as follows:
\begin{equation}
  \label{eq:rho1m}
\rho_{1m}=\frac{a-b}{c-d}
\end{equation}
In this formulation, the coefficients $a$, $b$, $c$, and $d$ are elaborated upon in Appendix B. By substituting $\rho_{1f}$ with $\rho_{1m}$ as specified in Eq.~(\ref{eq:rho}) and focusing solely on the electric susceptibility ($\chi_m=0$), the following result is obtained:
\begin{equation}
  \label{eq:chi}
\chi_e=\frac{\rho_{1m}}{j\frac{k_0}{2cos\theta_1}(1-\rho_{1m})}
\end{equation}
The real and imaginary components of the calculated electric susceptibility, along with the corresponding reflection coefficients, are illustrated in Fig.~\ref{fig:mssus}. The analysis of the reflection amplitude from the transmissive metasurface indicates that active elements are often required to realize the intended illusion. Moreover, the reflection phases exhibit more rapid fluctuations in comparison to those of the reflective metasurface, suggesting that reflective metasurfaces may present a more viable option for achieving the desired outcomes.

The strategy outlined in this research is adaptable to various illusion effects, demonstrating that environmental engineering serves as a potent method for creating object illusions without direct manipulation of the object. It is projected that future advancements in EMI and camouflage will increasingly rely on RIS technology. The subsequent section will focus on this promising area, providing valuable insights to facilitate the development of reconfigurable metasurfaces for sophisticated and smart illusion applications.

\section{Reconfigurable metasurfaces for smart illusions}
As RIS technology matures, its integration into EMI systems is expected to revolutionize how we manipulate and control EM waves, leading to advancements in a range of scientific and practical applications. RIS can provide fine-grained spatial control over EM fields, enabling localized manipulation of EM waves. This level of control is vital for creating precise and convincing EMI, whether it's altering the apparent shape of an object or creating virtual objects in free space. The ability to integrate RIS into existing systems and scale them according to the application's needs makes RIS a versatile tool for realizing EMI in practical scenarios. Whether it's for enhancing privacy in wireless communications or creating adaptive camouflage in defence applications, RIS can be tailored to meet diverse EMI requirements. 

The initial phase of deploying an RIS focuses on the design of unit cells, with various designs enabling the reconfiguration process. These designs may incorporate relative phase shifts, amplitude adjustments, or timing changes \cite{9493185,10.1002.adom.202101699,YANG2022128234}. Tools such as analytical methods, simulations, and machine learning can aid in developing a model to design versatile unit cells \cite{10.1002/adts.201800132}. Metasurfaces are characterized by their complex, spatially modulated surface impedance. The integration of switching diodes offers a binary modulation approach, allowing unit cells to switch between two states of phase or amplitude \cite{HuangZhangZentgraf}. Using varactors, however, permits a continuous adjustment of the reactive (imaginary) component of surface impedance, with minimal changes to the resistive (real) component \cite{10.1002.adom.202000783}. For comprehensive control, both the real and imaginary parts of surface impedance should allow for continuous tuning \cite{PhysRevApplied.11.044024}, enabling complete manipulation in the complex plane for independent amplitude and phase control \cite{9769001,arxiv.2106.06789,9593176}. This level of manipulation necessitates incorporating a tunable chip within unit cells for continuously adjustable resistance and reactance \cite{9092996,9171580}. While individual adjustments in tunable chips are typically made with quantized voltage levels, these can be chosen to smoothly cover the impedance range as required \cite{9109701}.

\begin{figure}[htbp]
  \centering
  \includegraphics[trim={9cm 3cm 10cm 3cm},clip,width=86mm]{ 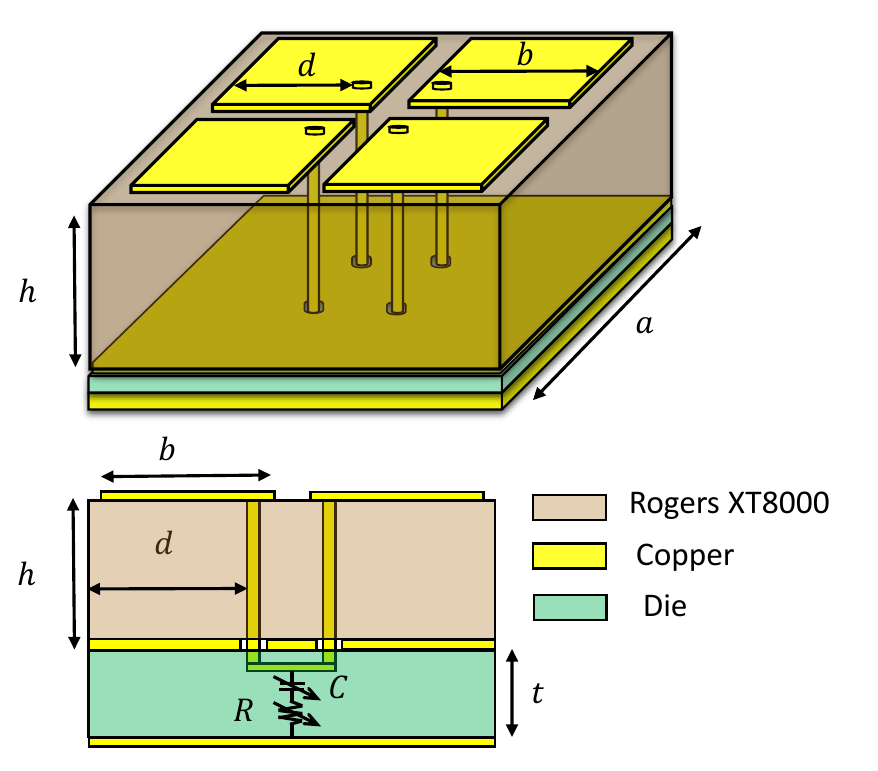}  
  \caption{This figure illustrates the layout of a unit cell design featuring four rectangular metallic patches situated on a Rogers XT8000 dielectric substrate. These patches are interconnected through vias to a die chip located in the backplane. Tunable elements such as varactors and resistors are incorporated in the die chip.}
  \label{fig:unit}
\end{figure}

Metasurface reconfiguration can be achieved by altering the composition or properties of materials or components, such as electronic elements, whose EM characteristics change with external stimuli. Methods to apply these stimuli include low-frequency electric or magnetic fields, optical pumps, voltage sources, or temperature changes. Modulating these stimuli varies the EM properties like permittivity or conductivity, thus tuning the metasurface's effective response and EM output. Global tuning affects the entire metasurface uniformly, while local tuning allows independent adjustment of each unit cell \cite{Taghvaee_2021}, supporting spatial impedance modulation for applications like wavefront control and dynamic functionality switching. Local tuning examples include liquid crystals for voltage-driven beam steering \cite{Zografopoulos2015}, nematic liquid crystals in fishnet metamaterials for tunable lenses \cite{6872806}, graphene for anomalous wave deflection \cite{Carrasco_2015,10.1002/adom.201500285}, and magnetic materials for lensing and beam steering through external magnetic fields \cite{Yang2017}. Optical local tuning has been explored with photodiode arrays behind metasurfaces, and biasing varactors to achieve cloaking and vortex beam generation \cite{Zhang2020,PhysRevLett.109.083902}. Each tuning method comes with specific attributes and limitations, influencing its appropriateness for applications and frequency ranges. Optical and thermal methods may face crosstalk issues due to the need for localized control signals. Modulation speeds can vary, with thermal control being slower and semiconductor processes faster. Reconfigurable functionalities, like wavefront manipulation, often require significant material parameter modulation \cite{Taghvaee_2021}. To demonstrate the proposed strategy's effectiveness, a practical unit cell design is proposed. Figure \ref{fig:unit} shows a unit cell layout with four rectangular metallic patches on a Rogers XT8000 dielectric substrate, connected through vias to a die chip in the backplane, emphasizing that the die thickness does not affect EM performance due to the isolating metallic sheet.

Adopting voltage-tunable lumped elements offers a pragmatic solution for achieving localized control over metasurfaces without interference or crosstalk between unit cells. This approach is particularly suited for integration with computer-based programmatic control, heralding advancements in software-defined metasurfaces. The advantages of employing voltage-tunable elements include:
\begin{enumerate}
  \item Compact Integration: Devices such as switch diodes and varactors are compact and can be easily integrated within the metasurface layout, preserving the unit cell structural integrity.

  \item These elements function within a low-voltage regime, facilitating efficient and manageable control mechanisms.

  \item Localized Impact: The effect of voltage-tunable elements is confined to individual meta-atoms, eliminating undesired interactions with adjacent cells.

\end{enumerate}
Voltage-tunable lumped elements, due to their compactness and precision, have become a cornerstone for the development of locally tunable metasurfaces, facilitating a wide spectrum of applications. For example, a study demonstrated the application of a voltage profile across varactors in a unidirectional manner to actively etch a specific 1D phase profile onto the metasurface \cite{10.1002/adom.201700485}. This technique has unlocked a variety of tunable functionalities, such as anomalous reflection for steering purposes, polarization control, and beam splitting, as evidenced by various research efforts \cite{1236089,4665714,10.1002/adom.201700485,9772314}. The ability to control each varactor independently within a 2D arrangement further enhances the metasurface's versatility, allowing for any desired bias distribution pattern. Such control has been instrumental in demonstrating dynamic imaging capabilities through focal point adjustment based on the Huygens principle \cite{10.1002/adma.201606422}, as well as enabling dynamic holography \cite{Li2017}.

The concept of coding metasurfaces, which employs locally switchable elements like pin diodes, varactors, or switches, transforms these surfaces into digitally controlled entities. Each metasurface can be depicted as a state matrix, programmable through reconfigurable logic devices such as FPGA \cite{Cui2014}. This digital representation allows for a broad range of functionalities to be realized across GHz and THz frequencies, demonstrating the effective implementation of reconfigurability \cite{9737695,sciadv.aao1749}. This paradigm shift towards software-defined control mechanisms exemplifies the fusion of traditional electromagnetic design with contemporary digital technologies, heralding new possibilities in dynamic wave manipulation and smart surface applications.

The concept of a beam-steering gradient metasurface realized through a reconfigurable gradient index, is an exemplary case of advanced metamaterial technology. A coding set, defining the potential states of each unit cell, is meticulously curated by identifying configurations where the unit cell showcases optimal reflection amplitude (RA) and a reflection phase (RP) that follows a progression of $2\pi/2^{N_\mathrm{bit}}$, where $N_\mathrm{bit}$ indicates the bit depth for control precision. This coding set is then applied via an FPGA to create the requisite phase profile, such as a linear phase gradient, facilitating precise beam direction adjustments.

Beyond serving as a natural platform for describing reconfigurable metasurfaces, the coding approach aligns well with modelling and optimization techniques used in signal processing and complex systems. Genetic algorithms (GA), for example, represent candidate solutions as arrays of bits, making coding metamaterials an ideal fit. This synergy has been explored not only with GA \cite{Li2016b, Sui:18, Sui_2018,1289250} but also with particle-swarm \cite{10.1002/adom.201500156,Gao2015}, simulated annealing \cite{Zhao2016, Jidi2018}, binary Bat \cite{Momeni2018}, and hybrid solutions \cite{Wang2014}. Recently, machine learning methods coupled with optimization algorithms have been proposed for programming metasurfaces to achieve complex functionalities, such as real-time imaging \cite{Li2019}.

The interplay between the Fourier transform and the domains of the physical structure and coding dramatically enhances the design capabilities for complex metasurfaces. The process of multiplying two spatial codes, as illustrated in research, leads to their scattering patterns undergoing convolution, enriching the scope of metasurface functionality \cite{10.1002/advs.201600156}. Furthermore, the Fourier transform's inherent linearity, as highlighted by Wu et al., means that adding two metasurface coding matrices directly translates to the superposition of their scattering patterns in phase-coded metasurfaces \cite{10.1002/adom.201701236}. This methodology has been extended to amplitude/phase-coded metasurfaces for practical applications, such as asymmetric power dividers, showcasing the method's versatility \cite{Rajabalipanah2019}. Such techniques facilitate the efficient encoding of multiple beams to meet specific requirements or the integration of diverse orthogonal functions within a single beam, opening new avenues in metasurface design \cite{10.1002/advs.201600156,10.1002/adom.201701236}. Additionally, analytical methods have been applied to establish scaling laws and design guidelines for RCS reduction in metasurfaces \cite{10.1002/adom.201700455}. Fault tolerance analysis further contributes to the robustness of metasurface programming by providing a comprehensive error model to assess the impact of inaccuracies \cite{8702080,8972463}.

Significant efficiency improvements are achieved in scenarios where unit cells are on the scale of the wavelength and thus have a pronounced impact on the scattering pattern. By adopting the 2D Inverse Fast Fourier Transform (2D-IFFT) method, the computational burden for a single simulation is drastically reduced from $(MN)^{2}$ to $MN \times \log(MN)$ \cite{Yang2016a}, streamlining the optimization process considerably. This methodology has been effectively applied in various applications within the GHz frequency spectrum, including polarization and focusing control \cite{Yang2016a}, beam manipulation \cite{Wan2016,Huang2017}, and the creation of holograms \cite{Li2017}. However, the THz spectrum has seen fewer implementations of these techniques, largely due to the challenges associated with tuning capabilities at these higher frequencies.

Expanding beyond spatial coding, which facilitates functionalities like wavefront manipulation \cite{Zhou2018, Zheng2017, Liu2016, Huang2017}, focusing \cite{Hosseininejad2019,PhysRevB.104.235409}, angular momentum conversion \cite{ROUHI2019125}, and polarization control \cite{10.1021/acsami.7b12468, 10.1002/adom.201700548}, space-time-coding digital metasurfaces have been introduced \cite{Zhang2018} to leverage the benefits of temporal modulation. The application of space-time modulation on coding metasurfaces enables control over the \emph{spectral} parameters of the electromagnetic wave response, providing a crucial alternative to nonlinear metasurfaces for generating new frequencies—a capability not feasible with conventional linear variants. This generalization positions RIS technology as an intriguing approach for implementing digital modulation systems with spatiotemporal control in wireless communications \cite{10.1093/nsr/nwy135, Cui2019}.

\begin{figure}[htbp]
  \centering
  \includegraphics[trim={1cm 6cm 1cm 6cm},clip,width=43mm]{ 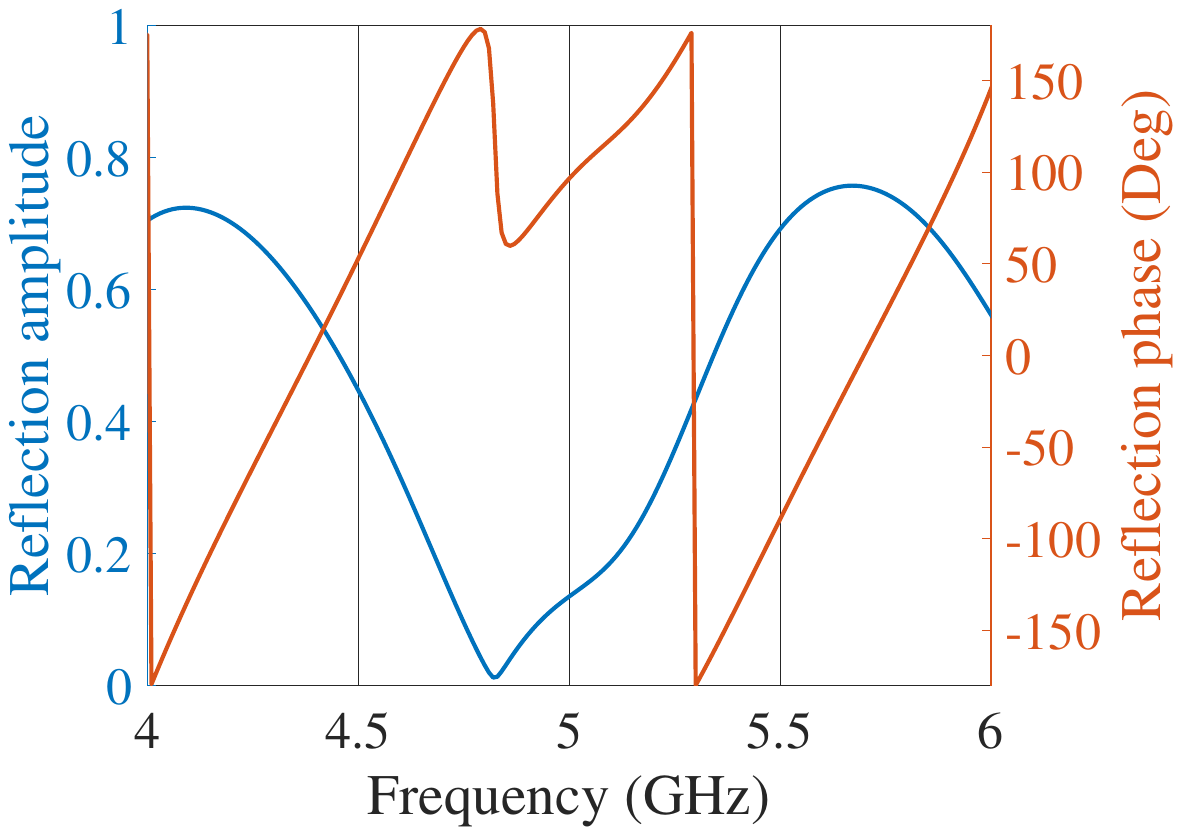}
  \includegraphics[trim={1cm 6cm 1cm 6cm},clip,width=43mm]{ 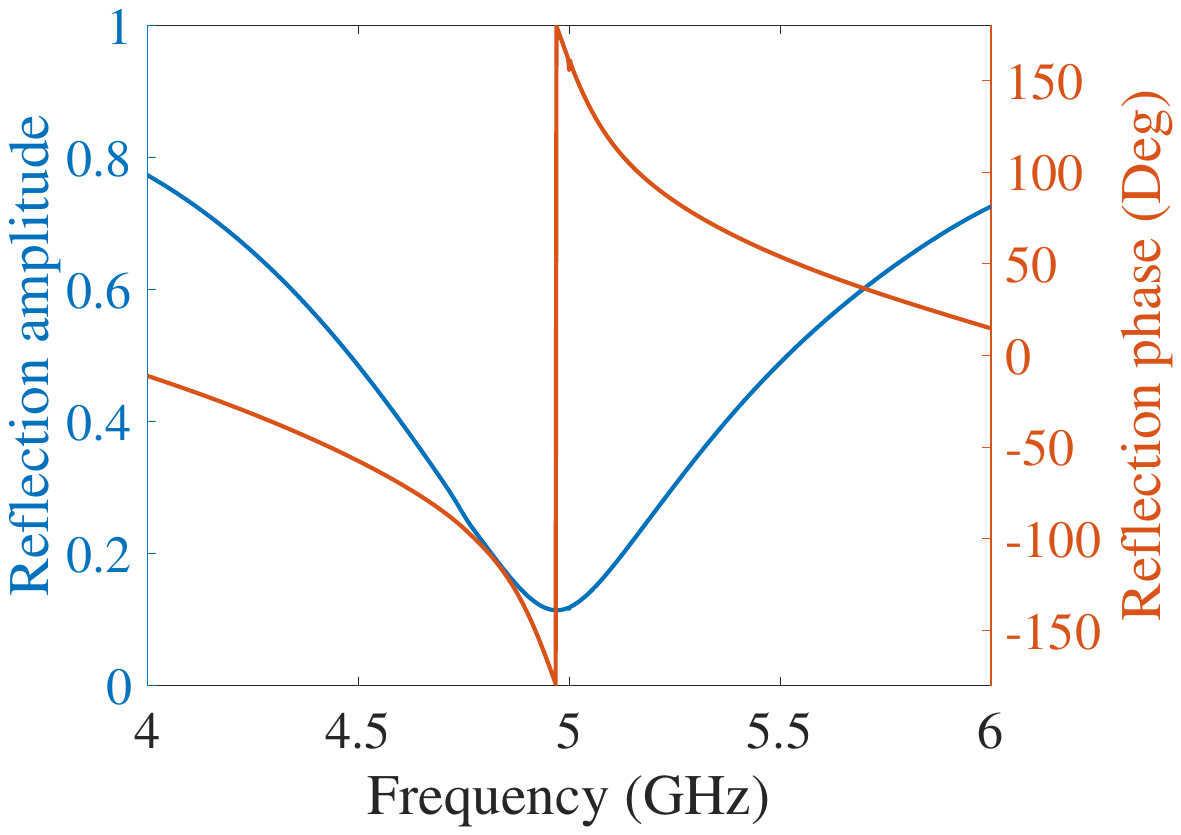}
  \caption{Reflective Characteristics for Illusion Creation and Unit Cell Design: The left subfigure illustrates the required reflective property ($\rho_{4m}$) of the RIS to achieve the illusion depicted in Fig. \ref{fig:cavity}. The right subfigure displays the reflective response ($\rho_m$) of the specifically designed unit cell, as seen in Fig. \ref{fig:unit}, demonstrating its capability to support the desired EM interplay for illusion realization.}
  \label{fig:rho4m1D}
\end{figure}

This study focuses on the 1D case, where global tuning (i.e., all unit cells are in the same state) is applicable. Within this context, Figure \ref{fig:rho4m1D} showcases the essential reflective characteristic ($\rho_{4m}$) of the RIS necessary to achieve the illusion depicted in Fig. \ref{fig:cavity}, under the condition of a TM polarized wave incident normally ($\theta_1=0$). It's noteworthy that the reflection from the specifically designed unit cell (illustrated in Fig. \ref{fig:unit}) manifests a singular resonance at 5 GHz, as detailed in Fig. \ref{fig:rho4m1D}. In the 1D case study, the symmetrical shape and dimensions of the unit cell pose challenges for discerning specific trends in reflection diagrams across the frequency spectrum. However, by meticulously adjusting the R/C values, it's feasible to synchronize with the performance outlined in Fig. \ref{fig:rho4m1D} for the envisioned illusion, thereby accommodating adjustments at a single frequency.

\begin{figure}[htbp]
  \centering
  \includegraphics[trim={1cm 6cm 1cm 6cm},clip,width=43mm]{ 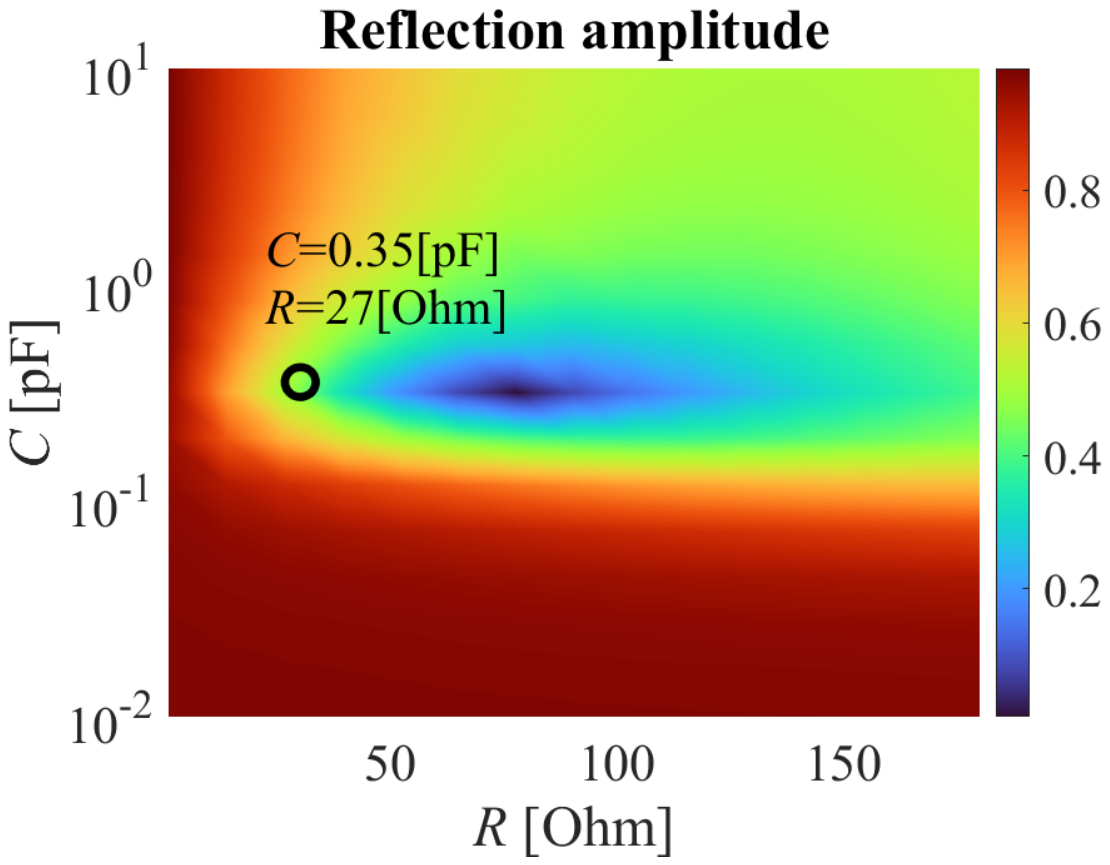}
  \includegraphics[trim={1cm 6cm 1cm 6cm},clip,width=43mm]{ 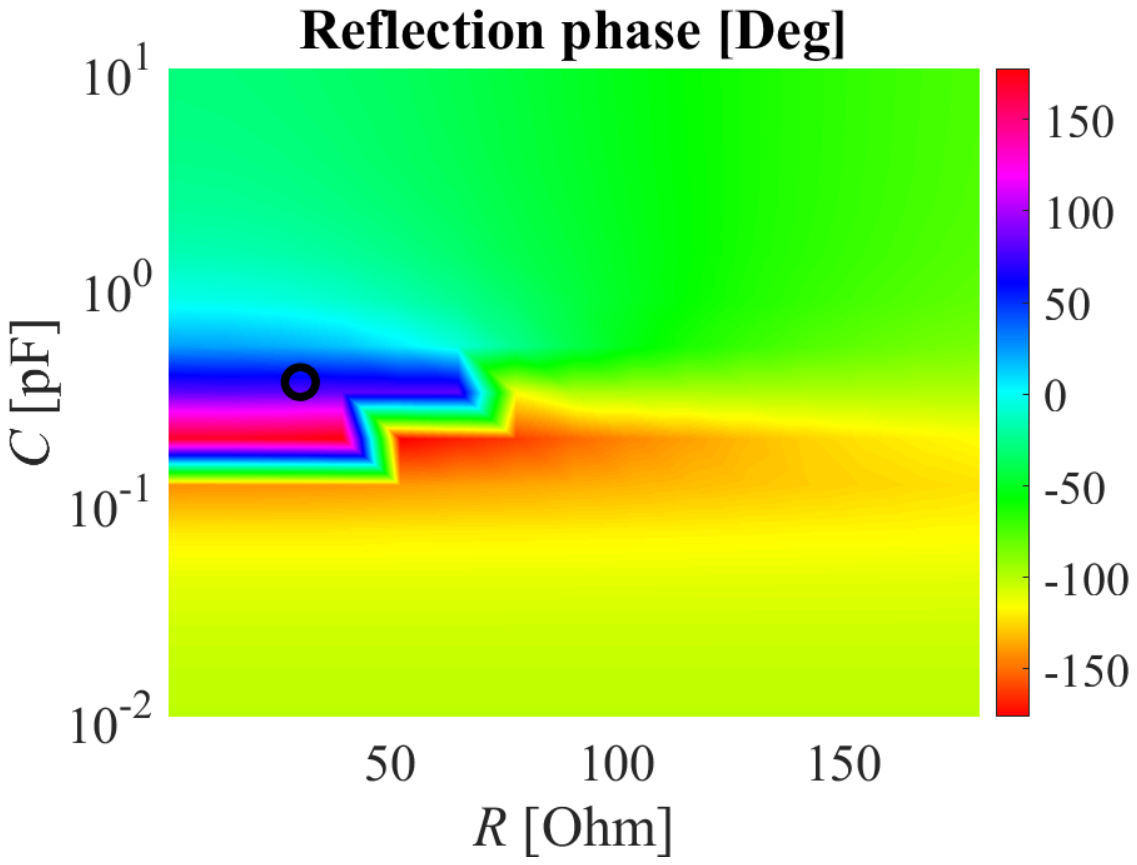}
  \includegraphics[trim={1cm 6cm 1cm 6cm},clip,width=43mm]{ 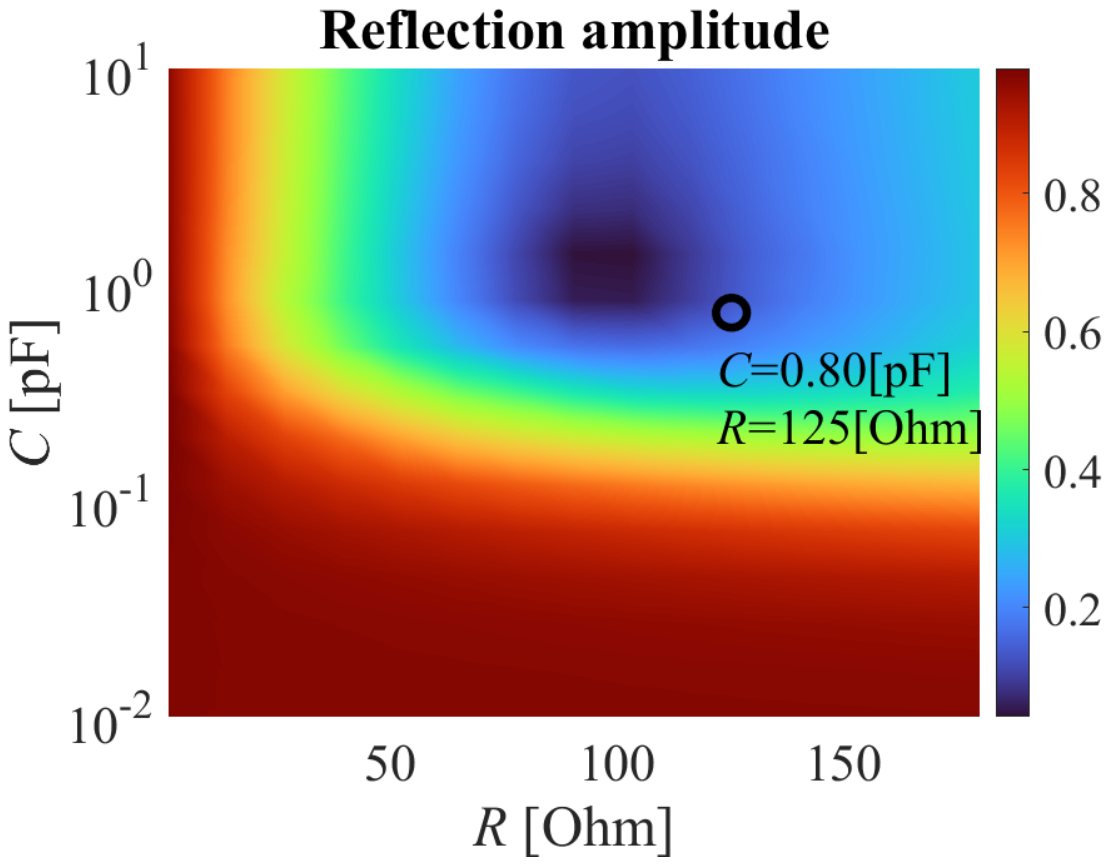}
  \includegraphics[trim={1cm 6cm 1cm 6cm},clip,width=43mm]{ 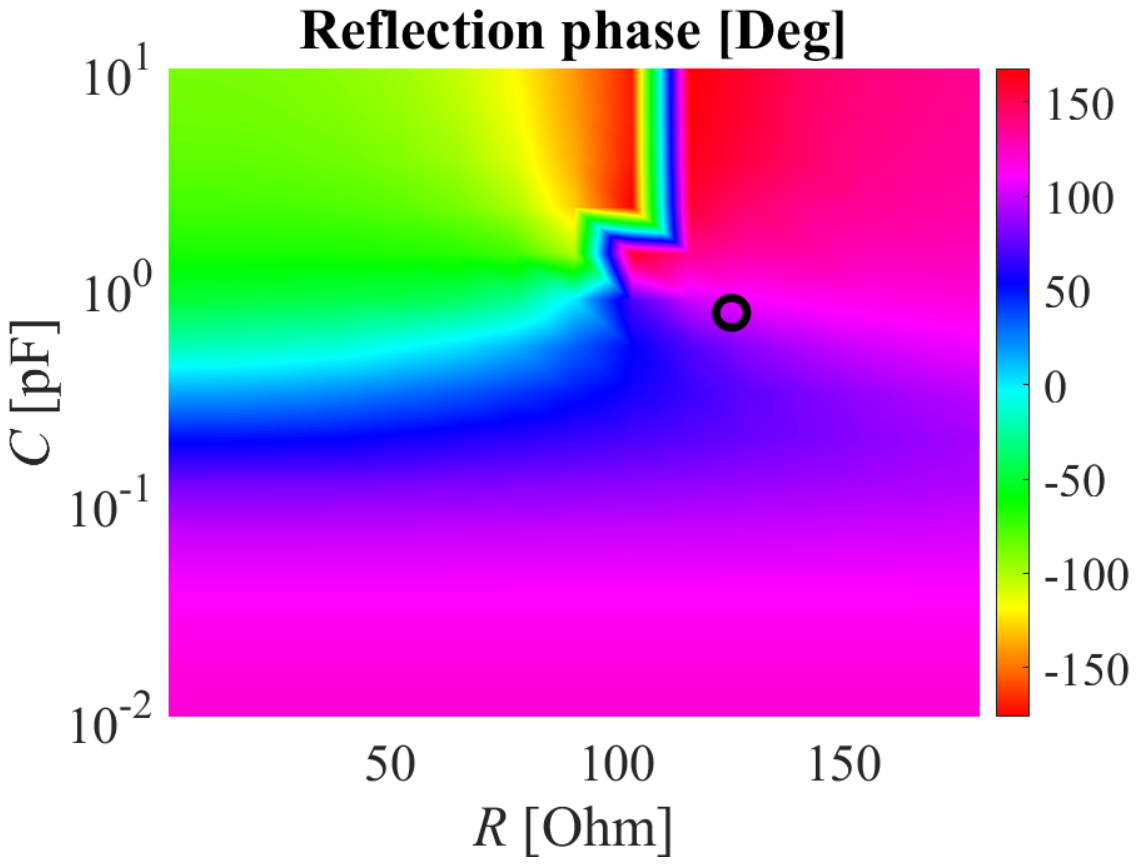}
  \includegraphics[trim={1cm 6cm 1cm 6cm},clip,width=43mm]{ 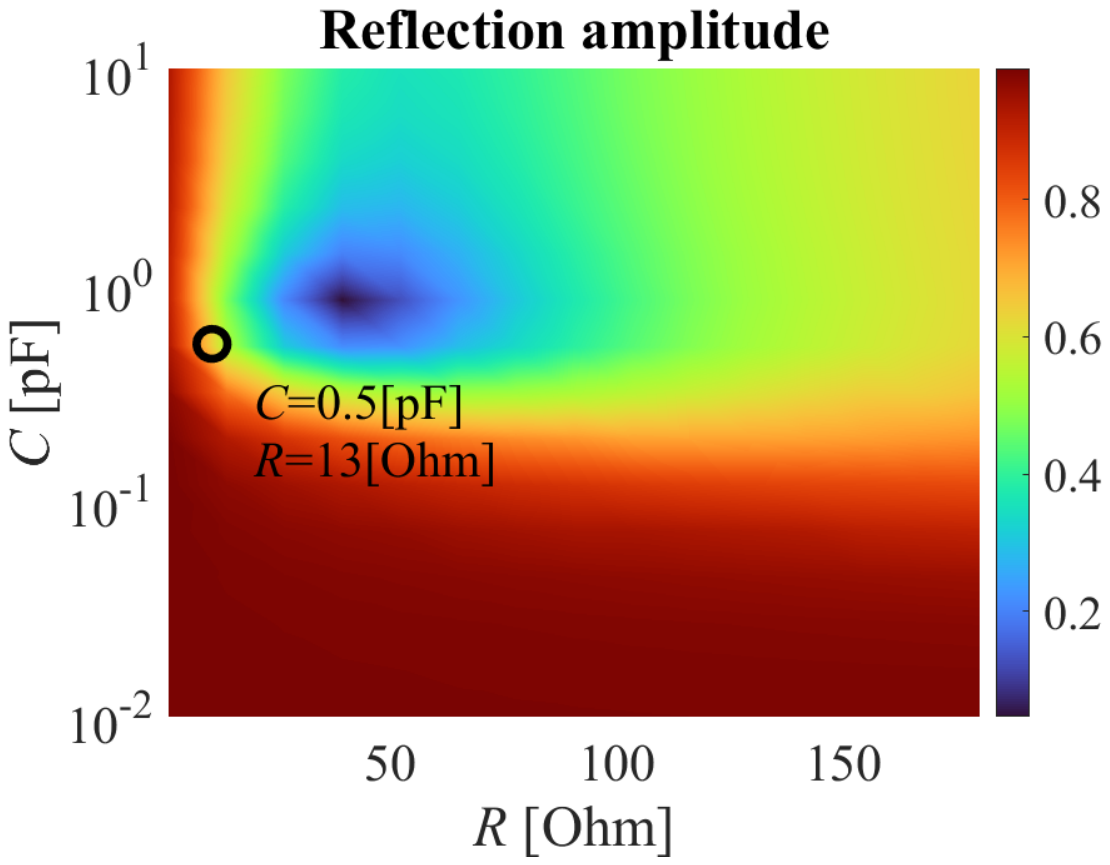}
  \includegraphics[trim={1cm 6cm 1cm 6cm},clip,width=43mm]{ 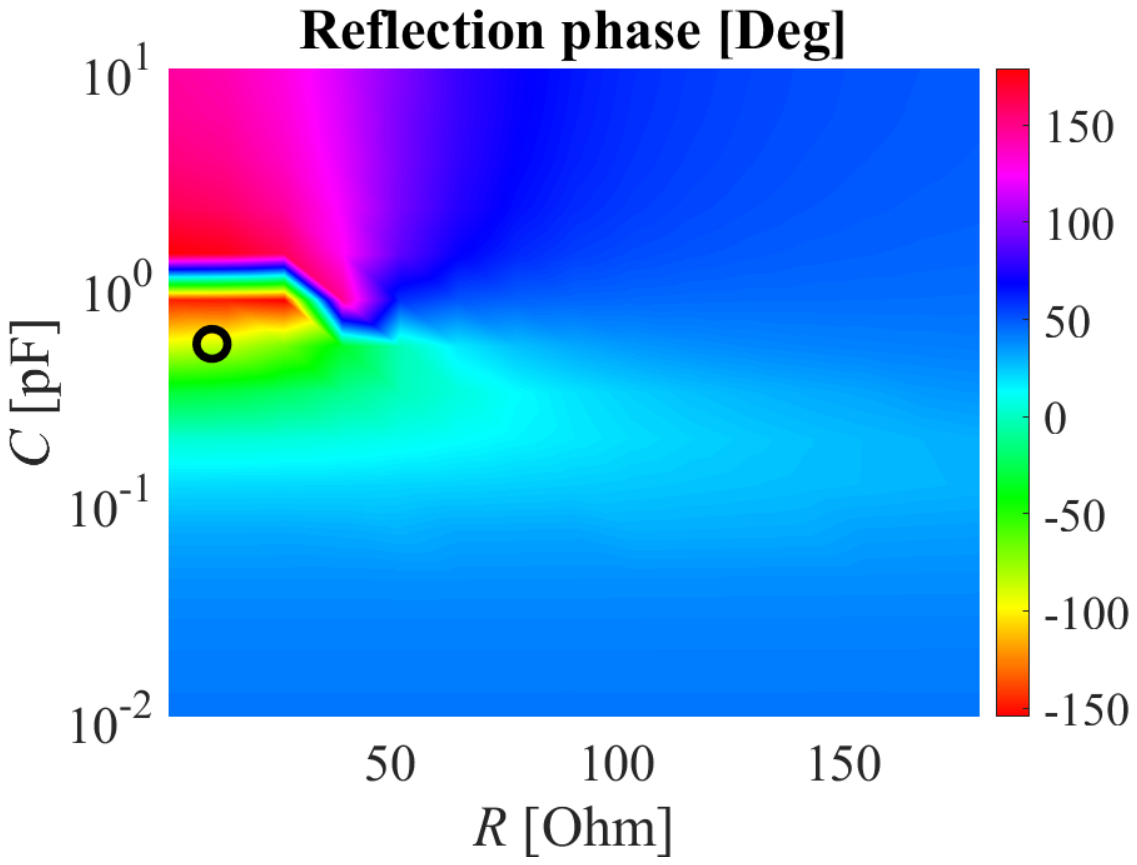}  
  \caption{Color Maps of Unit Cell Reflection Characteristics Across R/C Values: This figure presents six subfigures detailing the reflection amplitude and phase for the unit cell across varying R/C values at operation frequencies of $4.5$, $5.0$, and $5.5$ GHz. Black circles highlight the optimal states chosen to generate the illusion depicted in Fig. \ref{fig:cavity}, aligning with the necessary reflective properties of the RIS as outlined in Fig. \ref{fig:rho4m1D}.}
  \label{fig:ref}
\end{figure}

Fine-tuning the R/C values is an effective strategy for optimizing performance across various frequencies. However, it's crucial to recognize that deviations in resonance from its designed frequency lead to a reduction in the phase span available for modulation. To evaluate the influence of frequency on the reflective properties of the design, an exploration of the R/C design space is conducted on a logarithmic scale using a full-wave simulation at frequencies of $4.5$, $5.0$, and $5.5$ GHz, as depicted in Fig. \ref{fig:ref}. At the central frequency of $5$ GHz, the resonance exhibits a significant width, with the phase range comprehensively spanning from $0$ to $\pi$. At frequencies $4.5$ and $5.5$ GHz, some performance degradation is observed, yet the unit cell's functionality remains within acceptable bounds. This analysis indicates that the current unit cell design can effectively support the generation of illusions within the $4.5$ to $5.5$ GHz frequency range, maintaining the integrity of the physical structure.

For the specific illusion showcased in Fig. \ref{fig:cavity} and in alignment with the performance illustrated in Fig. \ref{fig:rho4m1D}, optimal R/C configurations are identified and marked with black circles. For example, at $4.5$ GHz, selecting $R=27$ ohms and $C=0.35$ pF achieves a RA of $0.5$ and a RP of $64$ degrees, closely matching the desired performance criteria detailed in Fig. \ref{fig:rho4m1D}. This methodology can be replicated for any target frequency within the specified range, facilitating the creation of diverse illusions without necessitating physical modifications to the metasurface.

\section{Use cases and technical requirements}
\begin{figure*}[htbp]
  \centering
  \includegraphics[trim={0cm 2cm 1cm 3cm},clip,width=160mm]{ 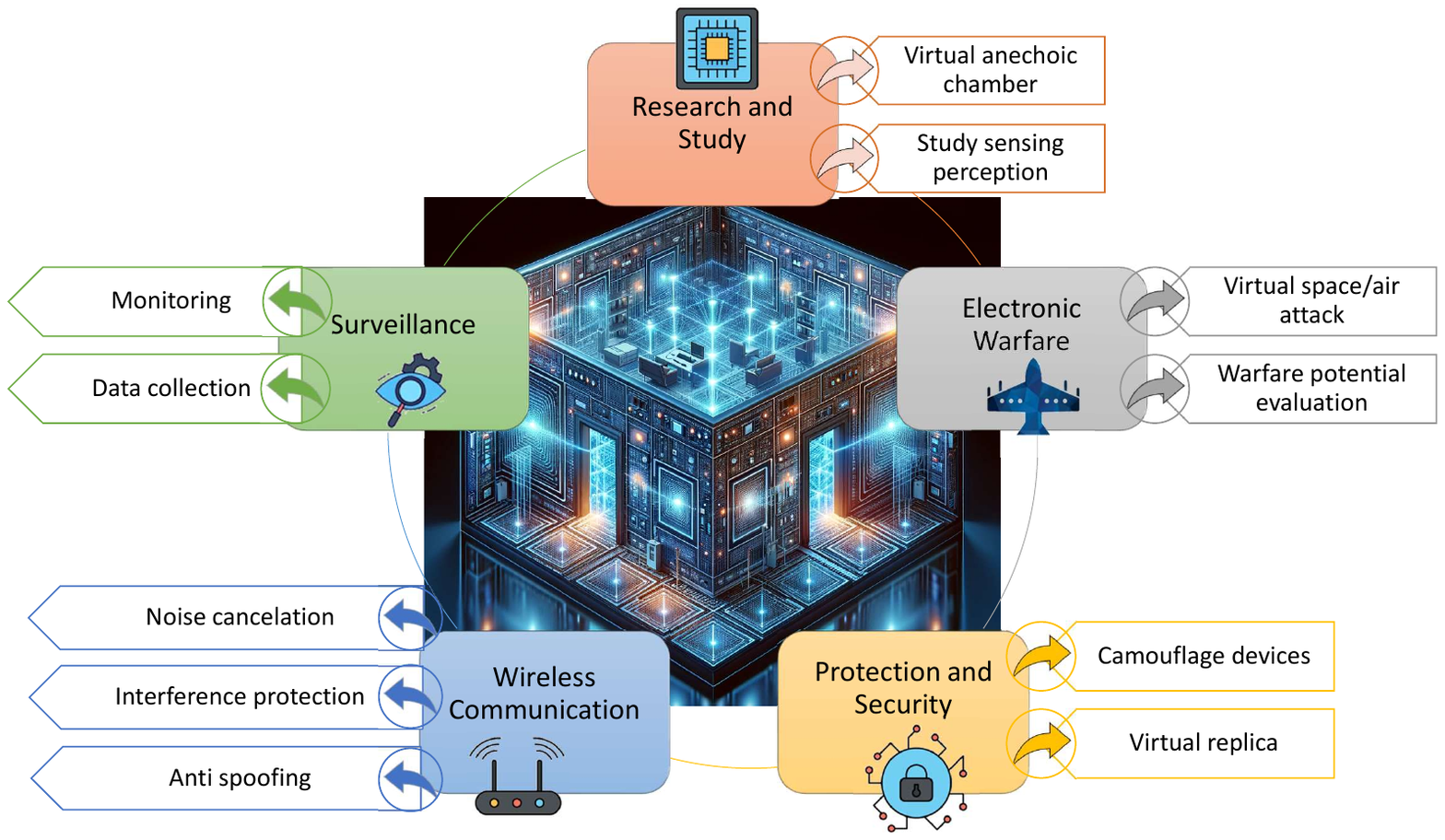}

  \caption{The figure highlights diverse applications of EMI, showcasing its impact on wireless communication, research and study, protection and security, surveillance, and electronic warfare. These applications underscore the versatility of EMI technology in addressing challenges and advancing capabilities across various domains.}
  \label{fig:use}
\end{figure*}

In the exploration of EMI, the practical application of this intriguing field necessitates a thorough understanding of use cases and the technical requirements that guide its implementation. This section delves into various scenarios and applications where EMI finds relevance, examining the specific technical considerations and prerequisites essential for successful deployment. Whether in the realms of communication, security, or advanced technologies, this section aims to elucidate the nuanced requirements that underscore the practicality and efficacy of EMI. Figure \ref{fig:use} is a graphical presentation of different use cases. In the following, each use case with relative technological requirements is discussed.

\subsection{Wireless communication}
As the evolution of wireless networks and the Internet of Things (IoT) usher in new technological landscapes \cite{8753447}, the imminent proliferation of countless interconnected devices underscores the critical need for effective EM compatibility. In this context, the imperative lies in noise control and device isolation, particularly considering the vulnerability of sensitive devices, such as sensors, to interfaces and disturbances. The potential ramifications of global systems, including mobile communications and global navigation satellite systems, facing vulnerabilities are substantial \cite{9187240}. Yet, the proactive implementation of EM control presents a promising solution. The dynamic manipulation of the EM field provides a mechanism to counteract interference, simultaneously rendering devices transparent to their environment. Here are some details on the potential impact of EMI on wireless communication:
\begin{itemize}

\item \textbf{Interference Mitigation:} EMI techniques can dynamically manipulate the EM field to cancel out interference, thereby mitigating the effects of unwanted signals and improving overall signal quality.
\item \textbf{Secure IoT:} Transmitters, and receivers can be hidden or disguised, making it challenging for unauthorized parties to detect, intercept, or interfere with communication signals. This enhances the security and privacy of wireless transmissions.

\item \textbf{Confidential Meetings and Communication:} EMI technologies can be employed to create secure zones for confidential meetings, making it more difficult for external entities to eavesdrop on sensitive discussions.

\item \textbf{Satellite Communication Security:} EMI technologies can enhance the security of satellite communication by implementing illusions that reduce the visibility of satellite signals, mitigating the risk of unauthorized access or interference.

\item \textbf{Resilience to Jamming:} EMI techniques can provide resilience against jamming by dynamically adapting to the EM environment, making it more challenging for adversaries to disrupt communication through interference.
\end{itemize}

In the realm of communication, network security stands as a foundational pillar. Threats from hackers and eavesdroppers pose significant risks to social networks, prompting governments to prioritize the privacy of individuals and businesses. Achieving a level of illusion or camouflage, where transmitters and receivers are concealed from unauthorized observers or are designed as decoys, enhances data security by making data sniffing arduous. 

\begin{itemize}
\item	The illusion/camouflage system shouldn't intervene with the operation of wireless networks.
\item	The design of the illusion/camouflage system should be compact to fit in any environment. 
\item	Since it is a commercial use the price of the product should be economical.
\end{itemize}





\subsection{Research and study}

Illusions, far beyond being captivating parlor tricks, hold tremendous potential for scientific exploration, particularly in the realms of vision, brain studies, and accurate measurement setups. In the domain of scientific research, EMI technology emerges as a transformative tool with implications for isolating environments and enhancing measurement accuracy.

Traditionally, achieving isolation in measurement setups involves enclosing rooms with bulky materials, creating a physical barrier against external signals and noises. While effective in mitigating external influences, this approach poses limitations by disconnecting devices inside the room from the external environment or canceling out back-scattering reflections. In contrast, EMI technologies offer a more dynamic solution by implementing virtual anechoic chambers without the need for physical barriers. Some potential impact of EMI on science are:

\begin{itemize}

\item \textbf{Precision Measurements:} EMI technologies can create virtual anechoic chambers, canceling out interrupting signals and minimizing interference, thereby enhancing the precision of measurements in fields such as physics, acoustics, and material science.
\item \textbf{Neuroscience and Brain Studies:} EMI can be applied to manipulate EM signals in neuroscience experiments, providing researchers with new insights into vision, brain processing, and cognitive functions.
\item \textbf{Earth and Environmental Sciences:} Illusion and camouflage techniques can aid in creating controlled environments for environmental sensors, minimizing interference and improving the accuracy of data collection in studies related to climate change, geophysics, and environmental monitoring.

\end{itemize}

Expectations centre around the ability of illusion devices to cancel out interrupting signals, minimizing interference with the sample under test without the need to block all signals. This nuanced approach aligns with the key requirements for effective illusion and camouflage techniques in scientific applications, promising to revolutionize the way scientists conduct experiments and measurements.

\begin{itemize}
\item	The system should be compatible with different devices.
\item	Easy to calibrate for measurements. 
\item	Specially designed for the frequency range of the device under test.

\end{itemize}

\subsection{Protection and security}
Inherently, the imperative to shield electronic devices from external attacks aimed at sabotaging their operation is ever more critical. EM camouflage emerges as a potent strategy to conceal electronic devices from detection, as demonstrated by recent studies \cite{smtd.202000918,9142383}. Beyond mere concealment, the concept of smart illusion introduces an innovative dimension by enabling the creation of a secure zone through the generation of a virtual replica of real devices. This virtual replica not only provides a visual illusion but also mimics the EM signatures, making it challenging for potential adversaries to distinguish between the real and virtual entities. Some potential impact of EMI on security are:
\begin{itemize}

\item \textbf{Critical Infrastructure Protection:} EMI technologies can enhance the security of critical infrastructure by minimizing the detectability of EM signals, making it more challenging for adversaries to target key assets.

\item \textbf{Concealment Efficacy:} EM camouflage should possess the capability to alter the EM signatures of devices, rendering them indistinguishable from the surrounding environment.

\item \textbf{Realistic Virtual Replication:} Advanced algorithms and technologies should be employed to generate virtual replicas that closely resemble the physical and EM attributes of the actual devices.

\end{itemize}

The key requirements for the successful implementation of this type of illusion and camouflage include:

\begin{itemize}
\item	The illusion/camouflage system shouldn't intervene with the operation of electronic devices.
\item	The transmitted power to create a safe zone must be below the standard limit.
\item	The performance should be reliable and resisting.
\end{itemize}

\subsection{Surveillance}
Effective surveillance hinges on the target under monitoring being unaware of detection. Hence, during data collection, the surveillance system must operate covertly to avoid exposure, as highlighted by recent studies \cite{PhysRevLett.102.233901}. In this context, camouflage or cloak techniques are categorized as either active or passive. In the active case, the emitter is surrounded by controllable sources that emit signals in the same frequency range, inducing illusions with deceitful shapes. This active camouflage involves manipulating the EM environment to create illusions and distort the perception of the surveillance target. The potential impact of EMI on surveillance is multifaceted and includes the following aspects: 

\begin{itemize}

\item \textbf{Covert Monitoring:} EMI allows for the creation of illusions that conceal surveillance emitters, enabling covert monitoring without the awareness of the target.

\item \textbf{Dynamic Adaptation:} EMI techniques provide the capability to dynamically adapt illusions in response to changing surveillance scenarios, ensuring continued concealment.

\item \textbf{Protection Against Countermeasures:} By actively manipulating EM signals, EMI offers protection against counter-surveillance measures, reducing the likelihood of detection by adversarial surveillance systems.

\item \textbf{Border and Perimeter Security:} EMI technologies can enhance border and perimeter security by making surveillance systems less susceptible to interference or tampering, improving the overall effectiveness of monitoring and detection.

\end{itemize}

The key requirements for this type of illusion/camouflage are as follows.

\begin{itemize}
\item	The design of the illusion/camouflage system should be disguised itself and cannot be detected.
\item	Surveillance systems are typically remote sensing devices and the attached illusion/camouflage should be below the power for a higher lifespan.
\end{itemize}

\subsection{Electronic warfare}
Electronic Warfare encompasses a broad range of activities involving the use of the EM spectrum to gain a strategic advantage in military operations. One specifc case is Virtual warfare, a fully automated virtual reality application where users control the simulated presence of warfare, has gained prominence in evaluating and enhancing military performance \cite{1611684}. This technology allows for the geographically distributed deployment of combat simulations, aiding in the assessment of warfare strategies and tactics \cite{1253409}. As virtual warfare continues to evolve, the integration of EMI technology introduces a new dimension to this domain. The potential impact of EMI on Electronic Warfare, particularly within the context of virtual warfare, includes the following aspects:
\begin{itemize}

\item \textbf{Electronic Attack:} aims to disrupt, deceive, or deny the adversary's use of the EM spectrum. This involves the use of jamming, spoofing, or other means to interfere with communication systems, radar, and other electronic devices.

\item \textbf{Innovation in Military Objectives:}
EMI within virtual warfare scenarios opens avenues for innovative military objectives, such as simulating electronic warfare strategies that involve the manipulation of enemy perceptions and responses through EMI.

\item \textbf{Simulation of Stealth and Camouflage:}
EMI technology allows the simulation of stealth and camouflage strategies within virtual warfare scenarios, enabling the testing and refinement of electronic countermeasures and stealth tactics.

\item \textbf{Radar Systems:} EMI technologies can enhance stealth capabilities by making radar cross-sections smaller, reducing the visibility of objects to radar systems, and making surveillance more challenging for potential adversaries.

\end{itemize}

 The key requirements for this type of illusion/camouflage are as follows.

\begin{itemize}
\item	Like any other military application the system should be resistant to harsh environments such as high/low temperature, high pressure, and mechanical forces.
\end{itemize}

\section{Conclusion}

This paper traverses the trajectory of EM cloaking technologies, from the advent of metamaterials and metasurfaces to the development of mantle cloaking and onward to environment-assisted illusions. It highlights the pivotal role of scattering suppression and the burgeoning interest in understanding object-environment interactions as a rich area for exploration. The need for novel methodologies—integrating statistical physics, random matrix theory, and applied mathematics—is emphasized to establish a coherent theoretical basis for metasurface reconfiguration. Offering a departure from traditional approaches, this study proposes a simple, yet effective, proof of concept within a 1D medium, avoiding object coatings and not limiting the object to free space conditions. It lays the groundwork for future 2D or 3D metasurface designs capable of creating powerful illusions through engineered wave scattering in object-environment interactions. This research initiates a novel paradigm of establishing smart environments for dynamic EMI, suggesting the use of reconfigurable metasurfaces with tunable elements to modify wave interactions for EMI purposes. It concludes by identifying use cases and technical prerequisites essential for making EMI viable and impactful in the context of advanced integrated sensing and communication technologies, looking beyond 5G.

\section*{Acknowledgments}
This work has been supported by the NSF-EPSRC collaborative scheme, grant number EP/X038491/1.


{\appendices
\section*{Appendix. A}
\label{appA}
To calculate the reflection coefficient from an impenetrable metasurface one needs to solve Eq.(\ref{eq:rho4m}). To this end, one needs to find  $A$, $B$, $C$, and $D$. Defining $\Gamma_m\equiv\Gamma_i$ and from Eq.(\ref{eq:refmatrix}), $\Gamma_m$ and $\Gamma_i$ are derived.
\begin{multline}  
\label{eq:A}
A=(((\rho_{1f}Z_{1f}^{-1}Z_{2f}^{-1}+Z_{1f}\rho_{2f}Z_{2f}^{-1})\rho_{3f}Z_{3f}+(\rho_{1f}Z_{1f}^{-1}\rho_{2f}\\Z_{2f}
+Z_{1f}Z_{2f})Z_{3f})\rho_{4f})(((Z_{1t}^{-1}Z_{2t}^{-1}+\rho_{1t}Z_{1t}\rho_{2t}Z_{2t}^{-1})Z_{3t}^{-1}+\\(Z_{1t}^{-1}\rho_{2t}Z_{2t}+\rho_{1t}Z_{1t}Z_{2t})\rho_{3t}Z_{3t}^{-1})+((Z_{1t}^{-1}Z_{2t}^{-1}+\rho_{1t}Z_{1t}\rho_{2t}\\Z_{2t}^{-1})\rho_{3t}Z_{3t}+(Z_{1t}^{-1}\rho_{2t}Z_{2t}+\rho{1t}Z_{1t}Z_{2t})Z_{3t})\rho_{4t})
\end{multline}
\begin{multline}  
\label{eq:B}
B=(((Z_{1f}^{-1}Z_{2f}^{-1}+\rho_{1f}Z_{1f}\rho_{2f}Z_{2f}^{-1})\rho_{3f}Z_{3f}+(Z_{1f}^{-1}\rho_{2f}Z_{2f}\\+\rho_{1f}Z_{1f}Z_{2f})Z_{3f})\rho_{4f})(((\rho_{1t}Z_{1t}^{-1}Z_{2t}^{-1}+Z_{1t}\rho_{2t}Z_{2t}^{-1})Z_{3t}^{-1}\\+(\rho_{1t}Z_{1t}^{-1}\rho_{2t}Z_{2t}+((\rho_{1t}Z_{1t}^{-1}Z_{2t}^{-1}+Z_{1t}\rho_{2t}Z_{2t}^{-1})\rho_{3t}Z_{3t}\\+((\rho_{1t}Z_{1t}^{-1}\rho_{2t}Z_{2t}+Z_{1t}Z_{2t})Z_{3t})\rho_{4t})
\end{multline}
\begin{multline}
\label{eq:C}
C=(((Z_{1f}^{-1}Z_{2f}^{-1}+\rho_{1f}Z_{1f}\rho_{2f}Z_{2f}^{-1})Z_{3f}^{-1}+(Z_{1f}^{-1}\rho_{2f}Z_{2f}\\+\rho_{1f}Z_{1f}Z_{2f})\rho_{3f}Z_{3f}^{-1}))(((\rho_{1t}Z_{1t}^{-1}Z_{2t}^{-1}+Z_{1t}\rho_{2t}Z_{2t}^{-1})Z_{3t}^{-1}\\+(\rho_{1t}Z_{1t}^{-1}\rho_{2t}Z_{2t}+((\rho_{1t}Z_{1t}^{-1}Z_{2t}^{-1}+Z_{1t}\rho_{2t}Z_{2t}^{-1})\rho_{3t}Z_{3t}\\+((\rho_{1t}Z_{1t}^{-1}\rho_{2t}Z_{2t}+Z_{1t}Z_{2t})Z_{3t})\rho_{4t})
\end{multline}
\begin{multline}  
\label{eq:D}
D=(((\rho_{1f}Z_{1f}^{-1}Z_{2f}^{-1}+Z_{1f}\rho_{2f}Z_{2f}^{-1})Z_{3f}^{-1}+(\rho_{1f}Z_{1f}^{-1}\rho_{2f}Z_{2f}\\+Z_{1f}Z_{2f})\rho_{3f}Z_{3f}^{-1}))(((Z_{1t}^{-1}Z_{2t}^{-1}+\rho_{1t}Z_{1t}\rho_{2t}Z_{2t}^{-1})Z_{3t}^{-1}+(Z_{1t}^{-1}\\\rho_{2t}Z_{2t}+\rho_{1t}Z_{1t}Z_{2t})\rho_{3t}Z_{3t}^{-1})+((Z_{1t}^{-1}Z_{2t}^{-1}+\rho_{1t}Z_{1t}\rho_{2t}Z_{2t}^{-1})\\\rho_{3t}Z_{3t}+(Z_{1t}^{-1}\rho_{2t}Z_{2t}+\rho_{1t}Z_{1t}Z_{2t})Z_{3t})\rho_{4t})
\end{multline}
\section*{Appendix. B}
\label{appB}
Same as Appendix. A, to calculate the reflection coefficient from a transmissive metasurface one needs to solve Eq.(\ref{eq:rho1m}). To this end, one needs to find  $a$, $b$, $c$, and $d$. Defining $\Gamma_m\equiv\Gamma_i$ and from Eq.(\ref{eq:refmatrix}), $\Gamma_m$ and $\Gamma_i$ are derived.
\begin{multline}  
\label{eq:a}
a=(Z_{3f}^{-1}Z_{1f}\rho_{2f}Z_{2f}^{-1}+\rho_{3f}Z_{3f}^{-1}Z_{1f}Z_{2f}+\rho_{4f}\rho_{3f}Z_{3f}Z_{1f}\\\rho_{2f}Z_{2f}^{-1}+\rho_{4f}Z_{3f}Z_{1f}Z_{2f})(((\rho_{1t}Z_{1t}^{-1}Z_{2t}^{-1}+Z_{1t}\rho_{2t}Z_{2t}^{-1})Z_{3t}^{-1}\\+(\rho_{1t}Z_{1t}^{-1}\rho_{2t}Z_{2t}+Z_{1t}Z_{2t})\rho_{3t}Z_{3t}^{-1})+((\rho_{1t}Z_{1t}^{-1}Z_{2t}^{-1}+Z_{1t}\rho_{2t}\\Z_{2t}^{-1})\rho_{3t}Z_{3t}+(\rho_{1t}Z_{1t}^{-1}\rho_{2t}Z_{2t}+Z_{1t}Z_{2t})Z_{3t})\rho_{4t})
\end{multline}
\begin{multline}  
\label{eq:b}
b=(Z_{1f}^{-1}Z_{2f}^{-1}Z_{3f}^{-1}+Z_{1f}^{-1}\rho_{2f}Z_{2f}\rho_{3f}Z_{3f}^{-1}+Z_{1f}^{-1}Z_{2f}^{-1}\rho_{3f}Z_{3f}\\\rho_{4f}+Z_{1f}^{-1}\rho_{2f}Z_{2f}Z_{3f}\rho_{4f})(((Z_{1t}^{-1}Z_{2t}^{-1}+\rho_{1t}Z_{1t}\rho_{2t}Z_{2t}^{-1})Z_{3t}^{-1}\\+(Z_{1t}^{-1}\rho_{2t}Z_{2t}+\rho_{1t}Z_{1t}Z_{2t})\rho_{3t}Z_{3t}^{-1})+((Z_{1t}^{-1}Z_{2t}^{-1}+\rho_{1t}Z_{1t}\\\rho_{2t}Z_{2t}^{-1})\rho_{3t}Z_{3t}+(Z_{1t}^{-1}\rho_{2t}Z_{2t}+\rho_{1t}Z_{1t}Z_{2t})Z_{3t})\rho_{4t})
\end{multline}
\begin{multline}  
\label{eq:c}
c=(Z_{1f}\rho_{2f}Z_{2f}^{-1}Z_{3f}^{-1}+\rho_{3f}Z_{3f}^{-1}Z_{1f}Z_{2f}+\rho_{3f}Z_{3f}Z_{1f}\rho_{2f}\\Z_{2f}^{-1}\rho_{4f}+\rho_{4f}Z_{3f}Z_{1f}Z_{2f})(((Z_{1t}^{-1}Z_{2t}^{-1}+\rho_{1t}Z_{1t}\rho_{2t}Z_{2t}^{-1})Z_{3t}^{-1}\\+(Z_{1t}^{-1}\rho_{2t}Z_{2t}+\rho_{1t}Z_{1t}Z_{2t})\rho_{3t}Z_{3t}^{-1})+((Z_{1t}^{-1}Z_{2t}^{-1}+\rho_{1t}Z_{1t}\rho_{2t}\\Z_{2t}^{-1})\rho_{3t}Z_{3t}+(Z_{1t}^{-1}\rho_{2t}Z_{2t}+\rho_{1t}Z_{1t}Z_{2t})Z_{3t})\rho_{4t})
\end{multline}
\begin{multline}  
\label{eq:d}
d=(Z_{1f}^{-1}Z_{2f}^{-1}Z_{3f}^{-1}+Z_{1f}^{-1}\rho_{2f}Z_{2f}\rho_{3f}Z_{3f}^{-1}+Z_{1f}^{-1}Z_{2f}^{-1}\rho_{3f}\\Z_{3f}\rho_{4f}+Z_{1f}^{-1}\rho_{2f}Z_{2f}Z_{3f}\rho_{4f})(((\rho_{1t}Z_{1t}^{-1}Z_{2t}^{-1}+Z_{1t}\rho_{2t}Z_{2t}^{-1})Z_{3t}^{-1}\\+(\rho_{1t}Z_{1t}^{-1}\rho_{2t}Z_{2t}+Z_{1t}Z_{2t})\rho_{3t}Z_{3t}^{-1})+((\rho_{1t}Z_{1t}^{-1}Z_{2t}^{-1}+Z_{1t}\rho_{2t}\\Z_{2t}^{-1})\rho_{3t}Z_{3t}+(\rho_{1t}Z_{1t}^{-1}\rho_{2t}Z_{2t}+Z_{1t}Z_{2t})Z_{3t})\rho_{4t})
\end{multline}
}


\newpage

 
\vspace{11pt}

\begin{IEEEbiography}[{\includegraphics[width=1in,height=1.25in,clip,keepaspectratio]{ 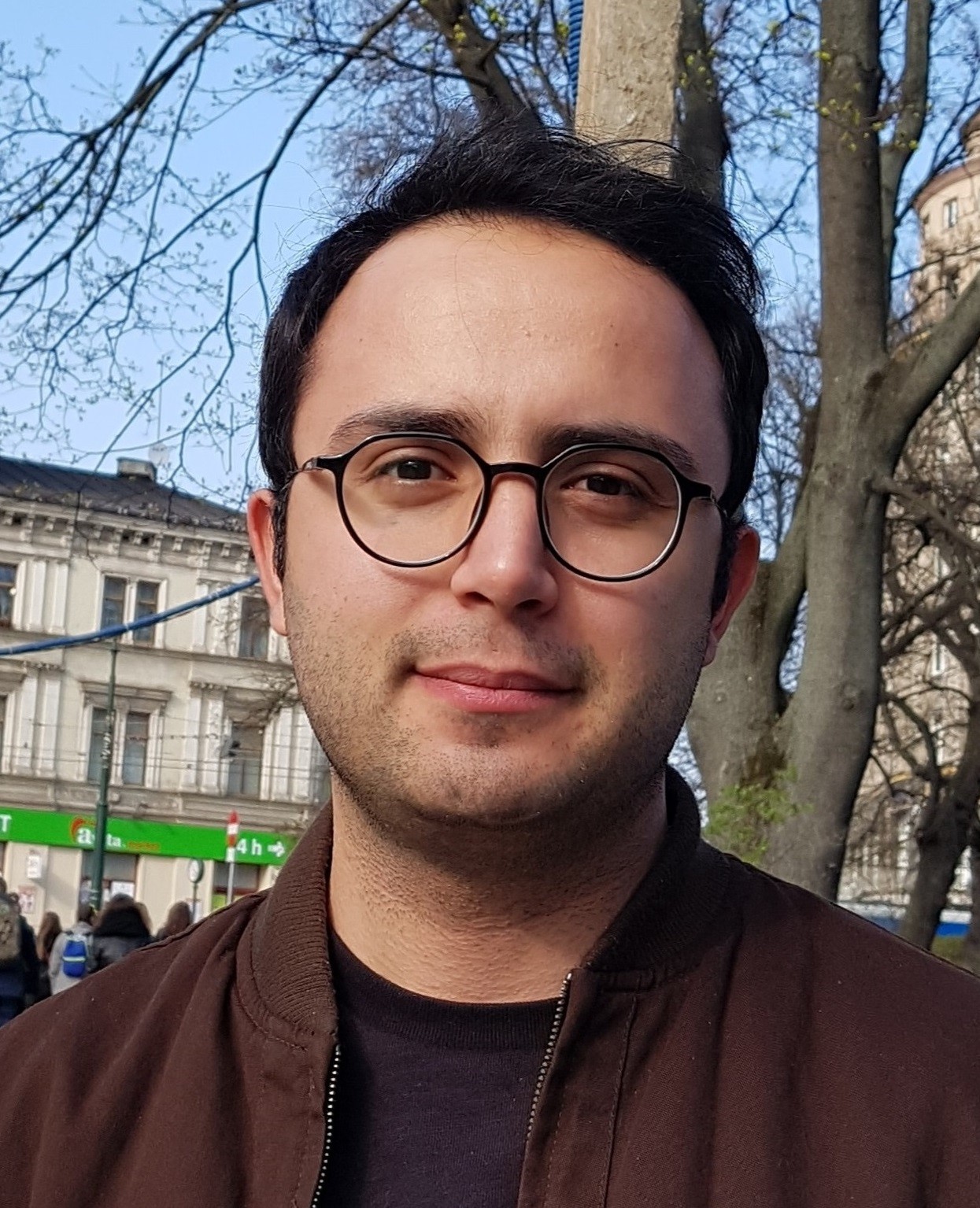}}]{Hamidreza Taghvaee}
(Member, IEEE) obtained his PhD in Computer Architecture from the Polytechnic University of Catalonia, Spain, with Cum Laude honors. His doctoral research garnered significant recognition, including a €7k grant each from the European Commission and the Academy of Finland for a visiting researcher stint at the Meta Group, Aalto University. His exceptional contributions during his PhD were further acknowledged with a special doctoral award for outstanding theses by the Universitat Politècnica de Catalunya in 2021. His doctoral research involved significant participation in the VISORSURF and WiPLASH FET-OPEN projects at N3Cat, Barcelona Tech. Progressing into postdoctoral research, Taghvaee contributed to the RISE-6G FET-OPEN and OBLICUE EPSRC projects at the George Green Institute, University of Nottingham. Currently, he holds a senior postdoctoral researcher position at the Institute for Communication Systems at the University of Surrey, home to the innovative 5G and 6G Innovation Centres. In this capacity, he leads the Flexi-DAS, CORE, and SCONDA projects, funded by DISIT. Moreover, he has successfully secured funding exceeding £200k from UK Innovate and over £500k from the Joint Lab initiative. Moreover, Taghvaee plays an active role in shaping the future of telecommunications technology as a member of the Industry Specification Group (ISG) on Reconfigurable Intelligent Surfaces within the European Telecommunications Standards Institute (ETSI), contributing to the development of standards that will underpin the next generation of communication systems.
\end{IEEEbiography}

\begin{IEEEbiography}[{\includegraphics[width=1.2in,height=1.5in,clip,keepaspectratio]{ 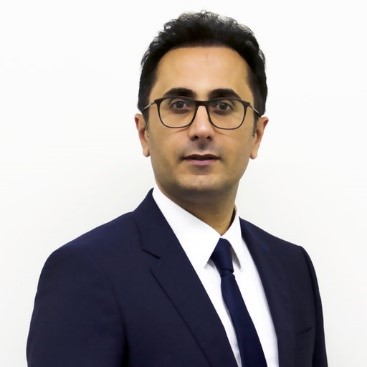}}]{Mohsen Khalily} (Senior Member, IEEE) is the esteemed Head of the Surface Electromagnetics Lab (SEMLAB) at the 5G and 6G Innovation Centres. With a robust background spanning over 12 years in both academia and industry focusing on antenna design, Dr. Khalily has emerged as a leading expert in his field. His prolific output includes holding 4 patents, authoring contributions to 4 book chapters, and publishing over 200 peer-reviewed journal and conference articles. His research primarily explores Metasurface Engineering, Reconfigurable Intelligent Surfaces, and mmWave and Terahertz channel propagation. As the principal investigator, Dr. Khalily has steered research grants totalling more than £4 million, significantly propelling forward the domain of surface electromagnetics. His leadership is also evident in his role at the European Telecommunications Standards Institute (ETSI), where he serves as the Rapporteur for the Industry Specification Group (ISG) on Reconfigurable Intelligent Surfaces, focusing on Implementation and Practical Considerations. In a notable achievement in 2020, Dr. Khalily demonstrated the potential of dynamic reflecting metasurfaces for continuous user tracking, a breakthrough ensuring uninterrupted wireless connectivity, which garnered him invitations to present at various esteemed international platforms. Managing a dynamic team comprising 6 academics and 15 researchers, Dr. Khalily's influence extends into editorial responsibilities, serving as an Associate Editor for prestigious publications like IEEE Antennas and Wireless Propagation Letters, and Scientific Reports. His accreditation as a Fellow of the U.K. Higher Education Academy further attests to his extensive contributions and commitment to the field of wireless communication and electromagnetics research.
\end{IEEEbiography}

\begin{IEEEbiography}[{\includegraphics[width=1in,height=1.25in,clip,keepaspectratio]{ 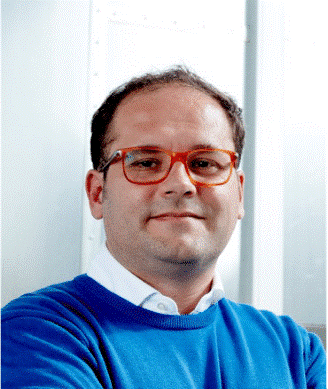}}]{Gabriele Gradoni} (Senior Member, IEEE) earned their Ph.D. in electromagnetics from Università Politecnica delle Marche, Ancona, Italy, in 2010. They further enriched their academic journey as a Visiting Researcher with the Time, Quantum, and Electromagnetics Team at the National Physical Laboratory, Teddington, U.K., in 2008. The period from 2010 to 2013 saw them contributing as a Research Associate at the Institute for Research in Electronics and Applied Physics, University of Maryland, College Park, MD, USA. From 2013 to 2016, they advanced their career as a Research Fellow at the School of Mathematical Sciences, University of Nottingham, U.K., culminating in their appointment as a Full Professor of Applied Mathematics and Electromagnetics Engineering in 2022. As of May 2023, they have been serving as Full Professor and Chair of Wireless and Satellite Communications at the 6G Innovation Centre, Institute for Communication Systems, University of Surrey, Guildford, U.K. Their role as a Royal Society Industry Fellow commenced in 2020 at British Telecom, U.K. Since December 2022, they have held positions as a Visiting Fellow at the Department of Computer Science and Technology, University of Cambridge, U.K., and as an Adjunct Professor at the Department of Electrical and Computer Engineering, University of Illinois at Urbana–Champaign, USA. Their research spans probabilistic and asymptotic methods for wave propagation in complex systems, metasurface modelling, wave chaos, and quantum computational electromagnetics. These areas of interest have significant implications for electromagnetic compatibility and the evolution of modern wireless communication systems. The individual is an esteemed member of both the IEEE and the Italian Electromagnetics Society. Their work has been recognized with several accolades, including the URSI Commission B. Young Scientist Award in 2010 and 2016, the Italian Electromagnetics Society Gaetano Latmiral Prize in 2015, and an Honorable Mention for the IEEE TEMC Richard B. Schulz Transactions Prize Paper Award in 2020. Furthermore, they have been honoured with multiple Best Paper awards at international forums, notably receiving the Best Electromagnetics Award at EuCAP 2022. Between 2014 and 2021, they represented early career professionals as the URSI Commission E Early Career Representative, highlighting their dedication to advancing the field of electromagnetic.

\end{IEEEbiography}

\begin{IEEEbiography}[{\includegraphics[width=1in,height=1.25in,clip,keepaspectratio]{ 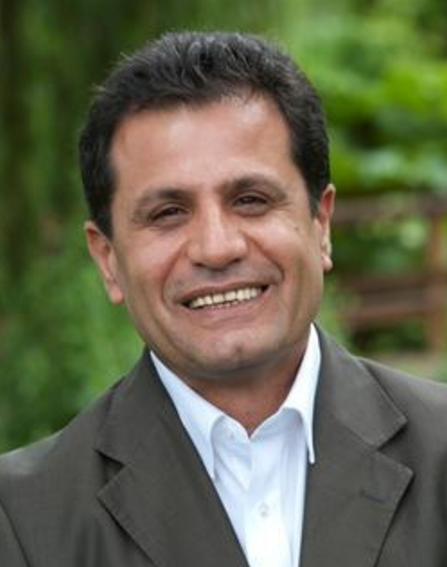}}]{Rahim Tafazolli} (Senior Member, IEEE) Regius Professor of Electronic Engineering and Professor of Mobile and Satellite Communications, Founder/Director of 5G and 6G Innovation Centre and Director of the Institute for Communication Systems, University of Surrey. Regius Professor Tafazolli is distinguished for his outstanding contribution to strategic research and development in future mobile and wireless communications including areas such as ad hoc networking, novel mobile architectures, massive MIMO, Green Communications, and the Internet of Things (IoT). He is the inventor of the NOMA technique for Mass connectivity in 5G and Beyond 5G. In Europe he pioneered clean slate research on Future Internet called “post-IP” era Internet. He is renowned for establishing the world’s first 5G Innovation Centre at Surrey University, which has galvanized national and international interest and funding for next-generation mobile communications. His scholarly contributions exceed 700 IEEE research papers across refereed journals, international conferences, and as an invited speaker. He has edited two volumes titled “Technologies for Wireless Future,” published by Wiley in 2004 and 2006, respectively. As a co-inventor, Tafazolli holds more than 30 granted patents, all within the realm of digital communications. His outstanding contributions to digital communications technologies over three decades were honoured in May 2018 when he was bestowed with the title of Regius Professor in electronic engineering. In 2020, his achievements led to his election as a fellow of the U.K. Royal Academy of Engineering. Furthermore, he is a fellow of the Institution of Engineering and Technology (IET) and the Wireless World Research Forum (WWRF), recognizing his significant impact and leadership in the field of communications technology.

\end{IEEEbiography}



\vfill

\end{document}